\documentclass[useAMS,usenatbib]{mn2e}

\usepackage{graphicx}
\usepackage{mathrsfs}
\usepackage{xcolor}
\usepackage{amssymb}
\usepackage{amsmath}
\usepackage{soul}     

\def\sun{{\odot}}

\title[Nuclear Dominated Accretion Flows]{Nuclear Dominated Accretion 
       Flows in Two Dimensions. II. Ejecta dynamics and nucleosynthesis for CO and ONe white dwarfs}
\author[Fern\'andez, Margalit, \& Metzger]{Rodrigo Fern\'andez$^{1}\thanks{E-mail: rafernan@ualberta.ca}$, 
Ben Margalit$^{2}$\thanks{NASA Einstein Fellow}, and Brian D. Metzger$^{3}$\\
$^1$ Department of Physics, University of Alberta, Edmonton, AB T6G 2E1, Canada\\
$^2$ Department of Astronomy and Theoretical Astrophysics Center, University of California, Berkeley, CA 94720, USA\\
$^3$ Department of Physics and Columbia Astrophysics Laboratory, Columbia University, New York, NY 10027, USA\\
}
\begin{document}

\date{Submitted to MNRAS}

\pagerange{\pageref{firstpage}--\pageref{lastpage}} \pubyear{2019}

\maketitle

\label{firstpage}

\begin{abstract}
We study mass ejection from accretion disks formed in the merger of a white
dwarf with a neutron star or black hole. These disks are mostly radiatively-inefficient and 
support nuclear fusion reactions, with ensuing outflows and electromagnetic transients. 
Here we perform time-dependent, axisymmetric hydrodynamic simulations of these disks including 
a physical equation of state, viscous angular momentum transport, a coupled $19$-isotope nuclear 
network, and self-gravity. We find no detonations in any of the configurations studied.
Our global models extend from the central object to radii much larger than the disk. We evolve these
global models for several orbits, as well as alternate versions with an excised inner boundary 
to much longer times. We obtain robust outflows, with a broad velocity distribution 
in the range $10^2-10^4$\,km\,s$^{-1}$. The outflow composition is mostly
that of the initial white dwarf, with burning products mixed in at the $\lesssim 10-30\%$
level by mass, including up to $\sim 10^{-2}M_\odot$ of ${}^{56}$Ni. These heavier
elements (plus ${}^{4}$He) are ejected within $\lesssim 40^\circ$ of the rotation axis, and should have
higher average velocities than the lighter elements that make up the white dwarf.
These results are in broad agreement with previous one- and two-dimensional studies, and point
to these systems as progenitors of rapidly-rising ($\sim $ few day) transients.
If accretion onto the central BH/NS powers a relativistic jet, these events could be accompanied by 
high energy transients with peak luminosities $\sim 10^{47}-10^{50}$\,erg\,s$^{-1}$ and peak durations 
of up to several minutes, possibly accounting for events like CDF-S XT2.
\end{abstract}


\section{Introduction}

Over the past decade, optical transient surveys have uncovered new types of 
relatively rare events with properties intermediate between supernovae
and classical novae (e.g., \citealt{kulkarni_2012}), and which have yet to be 
conclusively associated with a progenitor system. Examples include 
Ca-rich transients \citep{Perets+10,kasliwal_2012}, 
type Iax supernovae (e.g., \citealt{foley_2013}),  and rapidly-evolving 
blue transients (e.g., \citealt{drout_2014}).

Mergers of white dwarfs (WD) with neutron stars (NS) or stellar-mass black holes (BH) are expected
to generate transients, but theoretical predictions about their observational
signatures are not well developed yet. The key difficulty is the wide
range of scales and physical processes involved in the problem, in contrast
to mergers of similarly-sized objects (e.g., WD-WD, or NS-NS/BH) which
are more tractable and for which observational predictions are more mature
(e.g., \citealt{dan_2014,FM16}). 

Observationally, there are at least 20 confirmed galactic WD-NS binaries
plus a few dozen more candidates \citep{vankerkwijk_2005}. 
However, only a few of these systems will merge within a Hubble time (e.g., \citealt{lorimer_2008}).
Merger rates are predicted to be in the range $10^{-6}$-$10^{-4}$ per year in the Milky Way,
(e.g., \citealt{kim_2004,OShaughnessy&Kim10})
with the most frequent systems expected to contain CO and ONe WDs \citep{toonen_2018}. 
At present, only one candidate WD-BH binary is known in the Galaxy \citep{bahramian_2017}.

On the theoretical side, \citet{fryer1999} considered WD-BH mergers
as progenitors of long  gamma-ray bursts. They explored the dynamics of disk formation 
during unstable Roche lobe overflow in circular orbits around stellar-mass BHs
using Smooth Particle Hydrodynamc (SPH) simulations with nuclear burning, and predicted the accretion
power expected from the resulting disk using analytical arguments (see also \citealt{King+07}).
The disruption and disk formation process has also been explored by
\citet{Paschalidis+11} and \citet{bobrick_2017} with time-dependent simulations.
More extensive theoretical work exists in the context of tidal disruption of WDs on
parabolic orbits around BHs \citep{luminet_1989,rosswog_2009,macleod_2016,kawana_2018}. 
Thermonuclear burning due to tidal pinching
of the WD and/or tidal tail intersection are commonly found, although most existing work 
considers massive ($\geq 100M_\odot$) BHs with the exception of \citet{kawana_2018}, 
who obtains explosions with BHs of mass $10M_\odot$.

\citet{M12} explored the evolution of the torus formed during a 
quasi-circular WD-NS/BH merger using a steady-state, height-integrated model. Results
showed that nuclear reactions are important compared to viscous heating, and
that the radiatively-inefficient nature of these disks should result in
significant outflows. The importance of nuclear burning led \citet{M12}
to coin the term \emph{Nuclear-Dominated Accretion Flows (NuDAF)} for this regime.
The burning of increasingly heavier elements as they accrete deeper into the
gravitational potential generates an onion-shell-like stratification of composition
in radius, with non-trivial amounts of ${}^{56}$Ni production as a possible outcome.
A similar analysis has recently been applied to accretion disks 
in X-ray binaries and supermassive black holes \citep{ranjan_2019}.

The time-dependent evolution of height-integrated disks was carried out
by \citet[hereafter MM16]{margalit_2016} using a prescribed outflow model. A systematic
parameter exploration showed that disks from CO WDs evolve in a self-similar,
quasi steady-state fashion that is relatively robust to parameter variations.
Outflow velocities were found to be $\sim 10^4$\,km\,s$^{-1}$, with $\sim 10^{-3}M_\odot$
of radioactive ${}^{56}$Ni produced. At very late times,
these disks could in principle be a formation site of planets around the 
NS \citep{margalit_2017}.

\citet[hereafter Paper I]{FM12} performed global two-dimensional hydrodynamic 
simulations of the accretion disk using an ideal gas equation of state, parameterized 
nuclear burning, and viscous angular-momentum transport. Results
showed that turbulence-aided detonations of the disk are 
possible during the first few orbits if the nuclear energy release is significant compared to the local gravitational 
potential. Non-exploding cases yielded robust quasi-steady outflows, as expected from the radiatively 
inefficient character of the disk. A key uncertainty in these results was the robustness of
detonations when a more realistic equation of state that includes radiation
pressure is taken into account. Recently, \citet{zenati_2019} has reported
two-dimensional time-dependent simulations of the disk including a physical 
equation of state, a coupled 19-isotope nuclear reaction network, viscous angular
momentum transport, and self-gravity. Simulations are followed for a few
orbits at the initial disk radius ($\sim 100$\,s) and robust outflows are also found, with properties
consistent with previous 1D studies. Detonations are also reported.

Here we study the evolution of the disk with global two-dimensional axisymmetric simulations, focusing 
on the properties of the disk outflow in the case of the most common CO and ONe WDs. We improve upon 
Paper I by including a physical equation of state, a fully-coupled nuclear reaction network, 
and self-gravity. We also improve on previous work by resolving all spatial scales down to
the compact object surface over short times (few $100$\,s), and also follow the outer disk 
for much longer times ($\sim$\,hr) with an  excised boundary. Study of He-WD/NS binaries is
left for future work.

The paper is structured as follows. Section 2 describes the numerical method employed and
the parameter space surveyed. Section 3 presents our results. Section 4 summarizes
our conclusions and discusses observational implications. The Appendices contain a description
of the self-gravity implementation, the initial conditions for the disk, and a method to
obtain the accretion rate at the central object.

\section{Methods}

\subsection{Physical Model}

We consider accretion disks formed during the tidal disruption of
a white dwarf by a neutron star or a black hole via unstable
mass transfer. Following \citet{M12}, Paper I, and MM16, we
assume that nuclear burning is not dynamical during the
merger itself, and place the disk at the circularization
radius \citep{eggleton1983}
\begin{equation}
\label{eq:torus_radius}
R_{\rm t}  =  \frac{R_{\rm WD}}{(1+q)}\frac{0.6q^{2/3} + \ln (1 + q^{1/3})}{0.49 q^{2/3}}
\end{equation}
where $R_{\rm WD}$ is the radius of the white dwarf before disruption, which
depends on the white dwarf mass $M_{\rm WD}$ and composition (e.g., \citealt{nauenberg1972}) 
and $q = M_{\rm WD}/M_{\rm c}$ is the mass ratio of the binary, with $M_{\rm c}$ the
mass of the other compact object (neutron star or black hole). Mass transfer should
be unstable for most CO and ONe WDs (e.g., \citealt{bobrick_2017}). 

Outside the immediate vicinity of the central compact object, where neutrino 
or photodissociation losses from the disk\footnote{The neutron star is 
assumed to be old enough that its internal neutrino flux has no effect on the disk evolution.} 
can be important, the disk is radiatively inefficient.
Evolution occurs on a viscous timescale at the initial torus radius $R_{\rm t}$
by the action of angular momentum transport processes, which include magnetic
turbulence and perhaps also gravitational instabilities (MM16). 
While our models include self-gravity, they are axisymmetric and therefore
gravitational torques are not accounted for. Also, we do not include magnetic
fields, and instead parameterize angular momentum transport via a viscous shear
stress. See Paper I for a more extended discussion of the validity
of these approximations.

\subsection{Equations and Numerical Method}

We solve the equations of mass, momentum, energy, and
chemical species conservation in axisymmetric spherical polar 
coordinates $(r,\theta)$, with source terms due to gravity,
shear viscosity, nuclear reactions, and charged-current neutrino emission:
\begin{eqnarray}
\label{eq:mass_conservation}
\frac{\partial \rho}{\partial t} + \nabla \cdot (\rho\mathbf{v}_p) & = & 0\\
\label{eq:momentum_conservation}
\frac{d \mathbf{v}_p}{d t} + \frac{1}{\rho}\nabla p  & = &
-\nabla\Phi \\
\label{eq:angular_conservation}
\rho\frac{d j}{d t} & = & r\sin\theta\,(\nabla\cdot\mathbf{T})_\phi\\
\label{eq:energy_conservation}
\rho\frac{d e_{\rm int}}{d t} + p\nabla\cdot\mathbf{v}_p
& = & \frac{1}{\rho\nu}\mathbf{T}:\mathbf{T} + \rho\left(\dot{Q}_{\rm nuc} - \dot{Q}_{\rm cool}\right)\\
\label{eq:poisson}
\nabla^2\Phi & = & 4\pi G \rho + \nabla^2\Phi_{\rm c}\\
\label{eq:fuel_evolution}
\frac{\partial \mathbf X}{\partial t} & = & \mathbf{\Theta}(\rho,e_{\rm int},\mathbf{X}) + \mathbf{\Gamma}_{\rm cc}
\end{eqnarray}
where $d/dt\equiv \partial/\partial t + \mathbf{v_{\rm p}}\cdot\nabla$, and
$\rho$, $\mathbf{v}_{\rm p}$, $j$, $p$, $e_{\rm int}$, $\Phi$, and $\mathbf{X}$
are respectively the fluid density, poloidal velocity, specific angular momentum in the $z$-direction,
total pressure, specific internal energy, gravitational potential, 
and mass fractions of the isotopes considered ($\sum_i X_i =1$). Explicit source terms include
the viscous stress tensor for azimuthal shear, with non-vanishing components
\begin{eqnarray}
\label{eq:trphi_def}
T_{r\phi}      & = & \rho \nu\,\frac{r}{\sin\theta}\frac{\partial}{\partial r}\left(\frac{j}{r^2} \right)\\
T_{\theta\phi} & = & \rho \nu\,\frac{\sin\theta}{r^2}\frac{\partial}{\partial\theta}\left(\frac{j}{\sin^2\theta} \right),
\end{eqnarray}
the nuclear heating rate per unit volume $\dot{Q}_{\rm nuc}$, and the neutrino cooling rate 
per unit volume $\dot{Q}_{\rm cool}$. 
In equation~(\ref{eq:poisson}), we separate the gravitational potential of
the central object $\Phi_{\rm c}$ from that generated by the disk density field.
There is also an implicit source term in the expansion of $(\mathbf{v_{\rm p}}\cdot \nabla)\mathbf{v}_{\rm p}$
in spherical coordinates (left hand side of equation~\ref{eq:momentum_conservation}), which contains
a centrifugal acceleration\footnote{This acceleration is a subset of standard geometric source terms for finite-volume
hydrodynamic solvers in curvilinear coordinates (our reference frame is inertial).}
that depends on $j$:
\begin{equation}
\mathbf{f}_c = \frac{j^2}{(r\,\sin\theta)^3}\left[\sin\theta\hat r +\cos\theta\hat\theta\right].
\end{equation}

The system of equations~(\ref{eq:mass_conservation})-(\ref{eq:fuel_evolution})
is closed with the Helmholtz equation of state 
\citep{timmes2000}, the 19-isotope nuclear reaction network $\mathbf{\Theta}$ of \citet{weaver1978},
which provides a cost-effective description of energy generation by fusion and losses
from photodissociation and thermal neutrino emission ($\dot{Q}_{\rm nuc}$), 
and an alpha-viscosity prescription \citep{shakura1973}
\begin{equation}
\label{eq:viscosity_alpha}
\nu  =  \alpha \frac{p}{\rho\,\Omega_{\rm K}},
\end{equation}
where $\alpha$ is a free parameter and $\Omega_{\rm K}$ is the Keplerian frequency.  
In addition to the neutrino losses included in the nuclear network \citep{itoh1996}, 
neutrino emission via charged-current weak interactions is included (as described in 
\citealt{fernandez2019}), adding
a cooling term ($\dot{Q}_{\rm cool}$) and an extra source term for the mass
fraction of neutrons and protons ($\mathbf{\Gamma}_{\rm cc}$).

We use {\tt FLASH3} \citep{fryxell00,dubey2009} to evolve the system of
equations~(\ref{eq:mass_conservation})-(\ref{eq:fuel_evolution}) with the
dimensionally-split version of the Piecewise Parabolic Method (PPM, \citealt{colella84}).
The modifications to the code required to evolve accretion disks with a viscous
shear stress are described in \citet{FM13} and Paper I. Source terms are
applied in between updates by the hydrodynamic solver (operator-split). The
19-isotope nuclear reaction network is that included in {\tt FLASH3}; we use it with 
the MA28 sparse matrix solver and Bader-Deuflhard variable time stepping method 
(e.g., \citealt{timmes1999}). A time step limiter 
\begin{equation}
\Delta t_{\rm burn} < 0.1\frac{e_{\rm int}}{|\dot{Q}_{\rm nuc}|}
\end{equation}
is imposed for nuclear burning, in addition to the standard Courant, heating, and viscous 
time step restrictions.

Self-gravity is implemented using the algorithm of \citet{MuellerSteinmetz1995}, 
with a customized version for non-uniform spherical grids. The implementation and 
testing of this component is described in Appendix~\ref{s:self_gravity_appendix}.
The gravitational potential generated by the central object $\Phi_{\rm c}$ is modeled
as a pseudo-Newtonian point mass $M_{\rm c}$, with a spin-dependent 
event-horizon \citep{artemova1996,FKMQ14}.

The computational domain extends from an inner radius $R_{\rm in}$ to
an outer radius $R_{\rm max}$, with the grid logarithmically spaced
with $64$ cells per decade in radius. The full range of polar angles $[0,\pi]$
is covered with $56$ points equispaced in $\cos\theta$. The effective resolution
at the equatorial plane is 
$\Delta r /r \simeq 0.037\simeq \Delta\theta \simeq 2^\circ$. One model is
evolved at twice the resolution in radius and angle to test convergence (\S\ref{s:models_evolved}).

The boundary conditions are reflecting in polar angle and outflow at $r=R_{\rm max}$.
At $r=R_{\rm in}$, we use a reflecting boundary condition when resolving the
neutron star at the center, otherwise we set this boundary to outflow. 
The angular momentum is set to have a stress-free boundary condition at $r=R_{\rm in}$,
except in the case when the NS is assumed to be spinning, in which case a finite
stress is imposed. The boundary condition for the gravitational potential is 
such that it vanishes at $r\to \infty$.

\subsection{Initial Conditions}
\label{s:initial_conditions}

The initial condition is an equilibrium torus with constant
entropy, angular momentum, and composition. In a realistic
system, this initial state would be determined by the dynamics of
Roche lobe overflow, which requires a fully three-dimensional simulation
of the merger dynamics until the disk settles into a nearly axisymmetric state. 
In practice, the thermal time due to viscous heating in our disks is a few orbits, or
about $\sim 1/10$ of the viscous time, which means that for the timescales of
interest, the thermodynamics becomes quickly set by viscous heating
and nuclear burning. For systems in which a thermonuclear runaway is expected early 
on (i.e., ONe WD + NS mergers), the initial conditions are 
more important than in the systems we consider here. 

\begin{table*}
\centering
\begin{minipage}{18cm}
\caption{Parameters of models evolved. Columns from left to right show model name, WD mass, central
object mass, radius of initial torus density maximum, initial WD composition, initial
torus distortion parameter (equation~\ref{eq:pp_general}), viscosity coefficient 
(equation~\ref{eq:viscosity_alpha}),
radius of inner boundary, number of cells per decade in radius $n_{\rm r}$ and 
total number of cells in $\theta$-direction ($n_{\rm \theta}$), type of inner radial boundary 
condition (out: outflow, ref: reflecting), NS spin period $p_{\rm sp}$ or dimensionless
BH spin $\chi$, and maximum simulated time.}
\begin{tabular}{lcccccccccccc}
\hline
{Model}& $M_{\rm wd}$ & $M_{\rm c}$ & $R_{\rm t}$ & Mass Fractions & d &
  $\alpha$ & $R_{\rm in}$ & $(n_{\rm r}, n_{\rm \theta})$ & BC & $p_{\rm sp}/\chi$ & $t_{\rm max}$\\
       & $(M_\odot)$ & $(M_\odot)$  & ($10^9$\,cm)&   C/O/He/Ne          &   &
           & ($10^7$\,cm)         &                               &  & & (s) \\
\hline
{\tt CO+NS(l)}     & 0.6 & 1.4 & 2   & $0.50$/$0.50$/$0.0$/$0.0$ & 1.5 & 0.03  & 2   & (64,56)  & out & 0 & 4,123  \\
{\tt CO+NS(l-hr)}  &     &     &     &                           &     &       &     & (128,112) &     &   &        \\
{\tt CO/He+NS(l)}  &     &     &     & $0.45$/$0.45$/$0.1$/$0.0$ &     &       &     & (64,56)  &     &   &        \\
{\tt ONe+BH(l)}    & 1.2 & 5.0 & 1   & $0.00$/$0.60$/$0.0$/$0.4$ &     &       & 1   &           &     &   & 887  \\
\noalign{\smallskip}                                                           
{\tt CO+NS(s)}     & 0.6 & 1.4 & 2   & $0.50$/$0.50$/$0.0$/$0.0$ & 1.5 & 0.03  & 0.1 & (64,56) & ref & 0 & 245 \\
{\tt CO/He+NS(s)}  &     &     &     & $0.45$/$0.45$/$0.1$/$0.0$ &     &       &     &          &     &   & 215 \\
{\tt ONe+BH(s)}    & 1.2 & 5.0 & 1   & $0.00$/$0.60$/$0.0$/$0.4$ &     &       & 0.3 &          & out &   &  50 \\
\noalign{\smallskip}
{\tt CO+NS(s-vs)}  & 0.6 & 1.4 & 2   & $0.50$/$0.50$/$0.0$/$0.0$ & 1.5 & 0.10  & 0.1  & (64,56) & ref & 0     & 110 \\
{\tt CO+NS(s-sp)}  &     &     &     &                           &     & 0.03  &      &          &     & 2\,ms & 170 \\
{\tt ONe+BH(s-sp)} & 1.2 & 5.0 & 1   & $0.00$/$0.60$/$0.0$/$0.4$ &     &       & 0.17 &          & out & 0.8   & 31 \\
\label{t:models}
\end{tabular}
\end{minipage}
\end{table*}

The torus is initially constructed using only the gravity of the central object ($\Phi_{\rm c}$), 
for which a semi-analytic formulation is straightforward 
(e.g., \citealt{PP84,FM13}). This torus is then relaxed
with self-gravity, by evolving it without any other source
terms for 20 orbits at $r=R_{\rm t}$. A detailed description of this procedure
is provided in Appendix~\ref{s:initial_condition_appendix}.

Initial tori are described by their mass ($M_{\rm WD}$),
radius of maximum density ($R_{\rm t}$, equation~\ref{eq:torus_radius}), 
entropy (or $H/R = c_{\rm i}/\Omega_{\rm K}$ at density maximum, with $c_{\rm i}=\sqrt{p/\rho}$), 
angular momentum profile (constant), and composition.
The orbital time at $r=R_{\rm t}$ is given by 
\begin{equation}
\label{eq:torb_def}
t_{\rm orb} \simeq 40\,\textrm{s}\,\left(\frac{R_{\rm t}}{10^{9.3}\textrm{cm}}\right)^{3/2}
                                   \left(\frac{1.4M_\odot}{M_{\rm c}}\right)^{1/2},
\end{equation}
and the viscous time at the same location is
\begin{equation}
\label{eq:tvis_def}
t_{\rm vis} \simeq 900\,\textrm{s}\,\left(\frac{0.03}{\alpha}\right)\left(\frac{0.5}{\rm H/R} \right)^2
                                    \left(\frac{R_{\rm t}}{10^{9.3}\textrm{cm}}\right)^{3/2}                                                                       \left(\frac{1.4M_\odot}{M_{\rm c}}\right)^{1/2}.
\end{equation}

The torus is initially surrounded by a low-density adiabatic atmosphere with
hydrogen composition. This ambient density profile is set to $10^{-3}$\,g\,cm$^{-3}$ 
inside $r=4R_{\rm t}$ for most models, and decays as $r^{-2}$ outside this radius. 
In some cases we add a $r^{-0.5}$ dependence of this ambient inside $r=4R_{\rm t}$
whenever numerical problems at the inner radial boundary are encountered; this value is small enough it
does not affect the dynamics of the outflow. A density floor is set at 90\% of the 
initial ambient value. 
A constant floor of pressure ($10^8$\,erg\,cm$^{-3}$) about an order of magnitude lower than 
the lowest ambient value (near the outer boundary) is used to prevent numerical problems around 
the torus edges at early times and near the rotation axis at the inner boundary once an evacuated
funnel forms. A constant floor or temperature ($3\times 10^4$\,K) is also used to prevent the code from reaching
the lowest tabulated temperature in the Helmholtz EOS. The choice of floors is low enough that
results do not depend on it. When computing mass ejection and accretion, the
ambient matter is excluded, and material coming from the disk has densities much higher than this
initial ambient gas.

\subsection{Models Evolved}
\label{s:models_evolved}

All of our models are described in Table~\ref{t:models}. 
Given the large dynamic range in radius ($\gtrsim 10^3$) between the initial
disk radius $R_t$ and the characteristic size of the central compact object, 
and the fact that our time step is limited by the Courant condition at
the smallest radius in the simulation, we evolve two types of models. 

The first group sets the inner boundary radius at $R_{\rm in} = 10^{-2}R_{\rm t}\sim 10^7$\,cm, 
allowing evolution to long timescales ($\sim$ few $t_{\rm vis}$) but 
not resolving the regions from where the fastest outflows are launched and where
$^{56}$Ni is produced for a central NS. These models thus probe nucleosynthesis of intermediate-mass
elements and the time-dependence of mass ejection on long timescales (as in Paper I). 
These models have ``{\tt (l)}" appended to their names, for ``large inner boundary".

The group consists of our fiducial WD+NS model, {\tt CO+NS(l)}, a $0.6M_\odot$ CO 
WD around a $1.4M_\odot$ NS with a viscosity of $\alpha=0.03$ and an initial entropy of $3k_{\rm B}$ per
nucleon (c.f. Appendix~\ref{s:initial_condition_appendix}). 
To test the effect of a small admixture of helium, 
we include model {\tt CO/He+NS(l)}, which is identical to the fiducial case
except that the initial abundances of helium, carbon, and oxygen are $10\%$, $45\%$,
and $45\%$, respectively. Model {\tt ONe+BH(l)} probes a more massive ONe
WD ($1.2M_\odot$) around a $5M_\odot$ BH, with otherwise identical parameters.
Finally, model {\tt CO+NS(l-hr)} is identical to the fiducial case but with
twice the resolution in radius and angle, to probe the degree of convergence of our results.
All these models are evolved to $t=100t_{\rm orb}\simeq 4 t_{\rm vis}$.

The second group of models (``small inner boundary", or ``{\tt (s)}" for short)
resolve the central compact object but are evolved for
a shorter amount of time ($\sim$ several $t_{\rm orb}$). Model {\tt CO+NS(s)}
corresponds to the fiducial case but now with an inner reflecting radial boundary at $10$\,km. Likewise,
model {\tt CO/He+NS(s)} probes the hybrid WD while resolving the neutron star. In both
cases, the neutron star is assumed to be non-rotating.
Model {\tt ONe+BH(s)} extends the domain of the
corresponding large inner boundary BH model to a radius midway between the innermost stable
circular orbit (ISCO) and horizon. The BH is assumed to be non-spinning.

We include additional models that probe parameter variations among the small inner boundary
set. {\tt CO+NS(s-vs)} increases the alpha viscosity parameter from $\alpha=0.03$ to $0.1$.
Model {\tt CO+NS(s-sp)} adds a spin period of $2$\,ms to the NS and imposes a finite
viscous stress at this boundary, to probe energy release at the boundary layer. Finally,
model {\tt ONe+BH(s-sp)} adds a dimensionless spin of $\chi = 0.8$ to the BH, extending
accretion to smaller radii (the inner boundary is again placed midway between the new ISCO and
horizon radii).

For small boundary models, we first evolve the disk with a large inner
boundary to save computational time in this early phase, until the disk material
reaches this larger inner boundary. The result is then remapped into a grid that
extends the inner boundary further inward, with the process being repeated for each
additional order of magnitude that $R_{\rm in}$ decreases.

\section{Results}
\label{s:results}

\begin{figure}
\includegraphics*[width=\columnwidth]{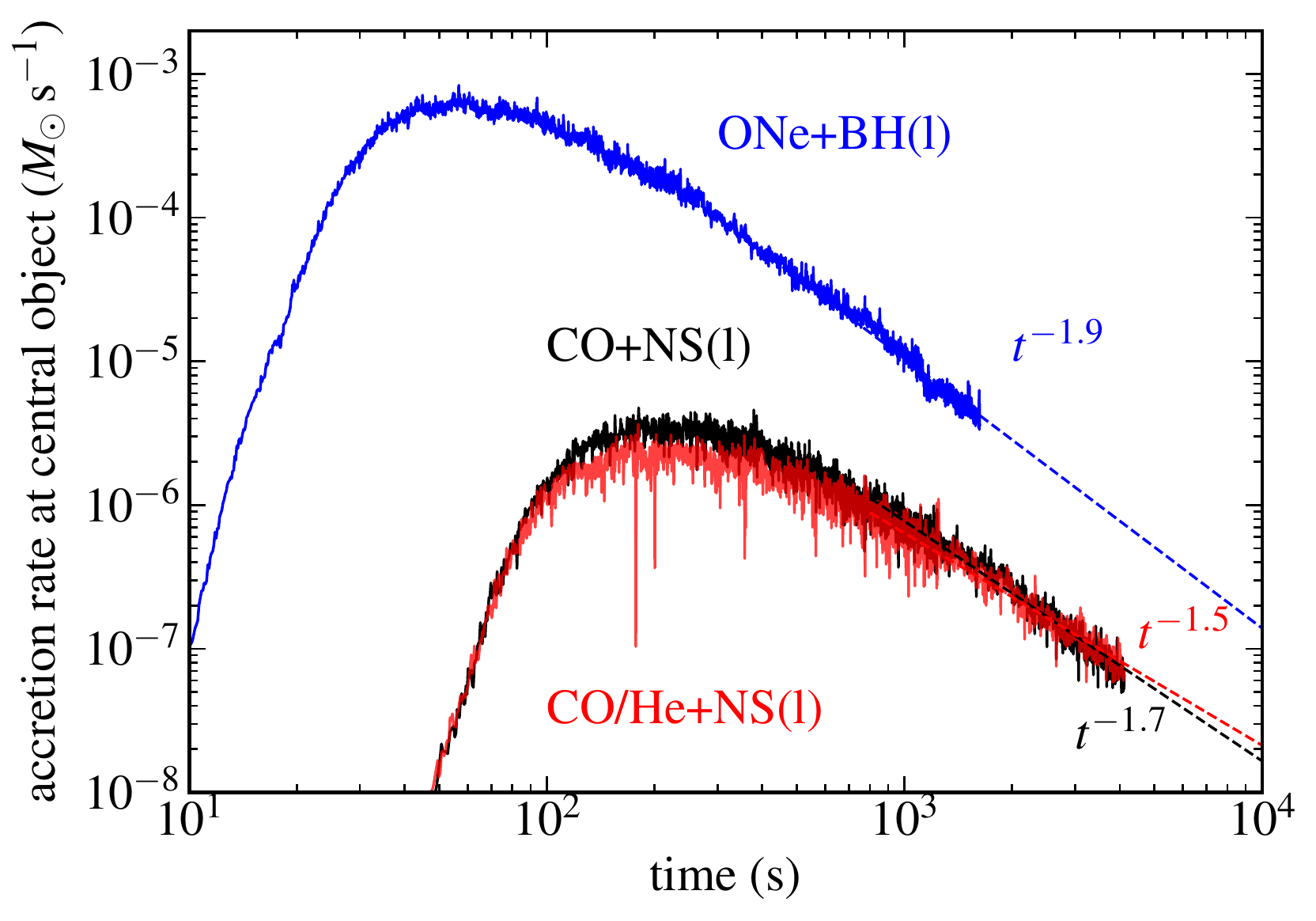}
\caption{Mass accretion rate at the central object inferred from the three large boundary
models by following the procedure described in Appendix~\ref{s:accretion_central}.
The dashed lines show power-law fits in time.}
\label{f:mdot_LB}
\end{figure}

\begin{figure*}
\includegraphics*[width=\textwidth]{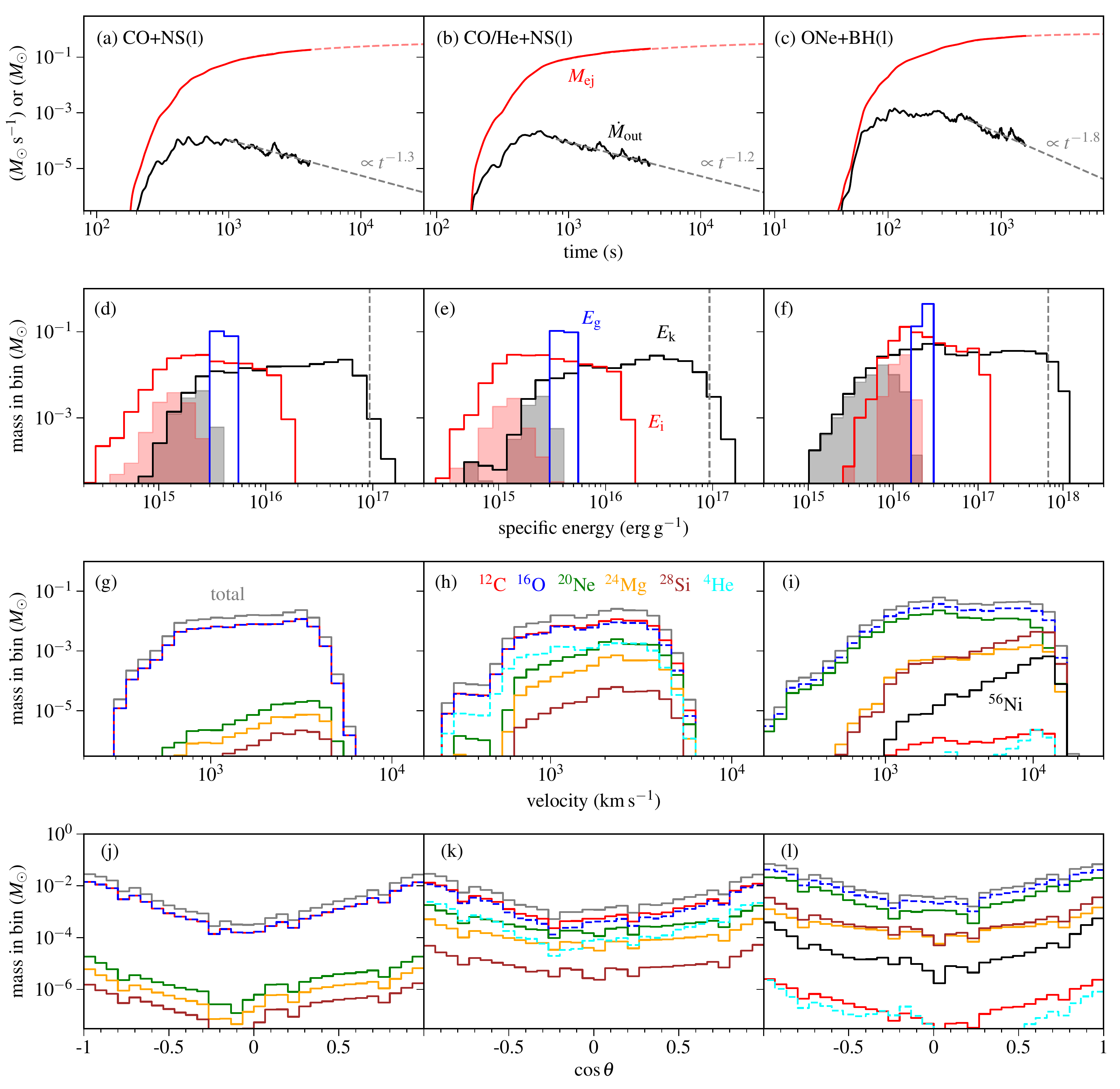}
\caption{Mass ejection properties of models with a large inner boundary ($R_{\rm in}=0.01R_{\rm t}$), as labeled.
\emph{Top:} Mass loss rate (solid black) in unbound material as a
function of time, measured at a radius $r_{\rm out} = 30R_{\rm t}$ 
($\simeq 6\times 10^{10}$\,cm for the fiducial {\tt CO+NS} model) and cumulative unbound mass ejected at 
this radius (solid red). The gray dashed line shows a power-law fit to the mass loss rate, which
when extrapolated yields a prediction for the total mass ejected (red dashed, equation~\ref{eq:mass_ejection_power-law}). 
\emph{Second row:}
Final mass histograms of unbound material as a function of specific kinetic energy (black), 
internal energy (red), and gravitational energy (blue) at $r=r_{\rm out}$. The vertical dashed line shows
the specific gravitational binding energy at the initial torus density peak, ignoring self-gravity ($GM_{\rm c}/R_{\rm t}$),
and the shaded histograms correspond to material with positive Bernoulli parameter but negative total
energy (i.e., marginally unbound). \emph{Third row:} Final mass histograms in unbound material 
at $r=r_{\rm out}$ as a function of radial velocity, for various species, as labeled. \emph{Bottom:} 
mass histograms as a function of polar angle, for various species. Note that these
models do not account for the mass ejected from  $r < 0.01R_{\rm t}$, 
and therefore the results shown correspond to lower limits to the ejected mass and its radial 
velocity (see \S\ref{s:small_bnd_evolution}).}
\label{f:evolution_LB}
\end{figure*}

\subsection{Mass ejection on long timescales}

Over the first few orbits at $r=R_{\rm t}$, the equilibrium initial tori begin accreting to
small radius while simultaneously transporting angular momentum outward. The absence of 
cooling during this initial stage leads to vigorous 
convection. This early phase of evolution is nearly identical to 
the quiescent models of Paper I, with quantitative details that depend on the parameters of the system.
Figure~\ref{f:mdot_LB} shows the accretion rate at the central object for the three large inner boundary 
models that remove the regions with $r < 0.01R_{\rm t}$ to 
allow for a longer evolution [{\tt CO+NS(l), CO/He+NS(l)}, and {\tt ONe+BH(l)}]. 
The accretion rate reaches a peak around 5-6 orbits at $r=R_{\rm t}$ ($\simeq 200$\,s for the NS
models, and $\simeq 35$\,s for the BH model).

None of our large boundary models detonate during the initial viscous spreading 
of the equilibrium torus nor at later stages. This stands in contrast to some of the
results of Paper I, which used parametric nuclear reactions, an ideal equation of
state, and point mass gravity. At higher temperatures, the increasing contribution of radiation pressure
results in more moderate increases in the temperature at small disk radii, 
preventing nuclear burning from ever causing a thermonuclear runaway (see \S\ref{s:comparison_1d}
for a more detailed discussion of this effect). Inclusion of
self-gravity only increases the density by a factor of $\sim 2$ and moves the radius of the torus density
peak inward by a few percent relative to using only point mass gravity (Appendix~\ref{s:initial_condition_appendix}).
The quantitative difference in the evolution once source terms are included is minor.
As a more extreme example, we evolved a test fiducial model in which the initial condition obtained 
with point mass gravity is not relaxed for self-gravity.
While stronger nuclear burning is obtained in some regions of the disk due
to radial oscillations, a detonation is not obtained within $2$ orbits.

The onset of convection is also accompanied by outflows from the disk, which continue
until the end of all simulations. We consider matter to be unbound from the disk
when its Bernoulli parameter
\begin{equation}
b = \frac{1}{2}\left[v_r^2+v_\theta^2+\frac{j^2}{(r\sin\theta)^2}\right] 
                         + e_{\rm int} + \frac{p}{\rho} + \Phi
\end{equation}
is positive. This criterion considers the conversion of thermal energy into
kinetic energy by adiabatic expansion, and is useful when measuring
the outflow at radii not much larger than the disk. 
The unbound mass outflow rate at a radius\footnote{We choose
this radius for sampling as a trade-off between being far enough away from the disk
to avoid including convective eddies, while also sampling enough outflow given the finite
simulation time.} $r_{\rm out}=30R_{\rm t}$
is shown in Figure~\ref{f:evolution_LB} for the three large inner boundary models. 
Peak outflow is reached around 15 orbits at $r=R_{\rm t}$, with a subsequent 
decay with time (after $t\simeq t_{\rm vis}$) that follows an approximate power-law.

By the time we stop our large inner boundary models
($100$ orbits at $r=R_{\rm t}$, see Table~\ref{t:models} for values in s), 
mass ejection is not yet complete.
Nevertheless, given the power-law dependence with time of the outflow rate, we can estimate the final ejecta mass
by extrapolating forward in time assuming that the same power law continues without changes
(see also MM16).
If this assumption holds, the extrapolation is a \emph{lower limit} on the total 
ejecta mass, because it does not include the contribution from $r < 0.01R_{\rm t}$, which is 
quite significant when a NS sits at the center (\S\ref{s:comparison_sb_lb}).
If the mass outflow rate at some radius is $\dot{M}_{\rm out}\propto t^{-\delta}$ ($\delta>0$), 
then we can write for $t > t_0$ 
\begin{equation}
\label{eq:mass_ejection_power-law}
M_{\rm ej}(t) = M_{\rm ej,0} + \frac{1}{\delta-1}\dot{M}_{\rm out,0}t_0
  \left[1 - \left(\frac{t}{t_0} \right)^{1-\delta}\right]
\end{equation}
with $M_{\rm ej,0}$ and $\dot{M}_{\rm out,0}$ the ejected mass and outflow rate at $t=t_0\sim t_{\rm vis}$, after which
the power-law dependence holds. For a finite value at $t\to \infty$, we need $\delta > 1$.
Figure~\ref{f:evolution_LB} shows $M_{\rm ej}$ as a function of time. The 
resulting exponents are $\delta \simeq \{1.3,1.2,1.8\}$ for models {\tt CO+NS(l)}, {\tt CO/He+NS(l)}, 
and {\tt ONe+BH(l)}, respectively, leading to a finite asymptotic value in all three cases.
All mass ejection results are shown in Table~\ref{t:results}. The asymptotic ejecta
masses for the large boundary models are $60-70\%$ of the initial WD mass before
including the contribution from $r < 0.01R_{\rm t}$. Over the short timescales that small
boundary models run, they eject about twice more mass when counted to the same radius and time 
than large boundary models (\S\ref{s:small_bnd_evolution}). A simple scaling of the asymptotic ejecta 
by this factor would exceed the initial WD mass, which indicates that
(1) most of the WD mass is indeed likely to be ejected, but that (2) the time exponents of the
outflow are also likely to change with time and/or be different than those derived from the large boundary
models. Upper limits to the ejected mass can be obtained
by subtracting the asymptotic accreted mass at the compact object (Figure~\ref{f:mdot_LB}) from the
WD mass. These accreted masses are $\simeq 3\times 10^{-3}M_\odot$ for the WD+NS models, and $0.12M_\odot$
for the ONe+BH model. 

Figure~\ref{f:evolution_LB} also shows how the cumulative ejecta is distributed in specific energy
at the radius $r_{\rm out} = 30R_{\rm t}$ where we sample the outflow. In all three
large boundary models, the highest kinetic energies achieved correspond approximately to 
the gravitational potential energy at the initial torus radius $R_{\rm t}$. The resulting
maximum velocities are $\sim 6,000$\,km\,s$^{-1}$ for models {\tt CO+NS(l)} and {\tt CO/He+NS(l)},
and $\sim 15,000$\,km\,s$^{-1}$ for {\tt ONe+BH(l)}. 

The bulk of the  ejecta has not yet reached homology at this radius, as indicated by the significant 
internal energy component. Nonetheless, most of the ejecta with $b>0$ has more 
than sufficient energy to escape the gravitational field of the system. 
For the NS models, only a fraction $3-5$\% by mass  has negative
specific energy but positive Bernoulli parameter at $r_{\rm out}=30R_{\rm t}$, 
while for the BH model this fraction is $10\%$ (shown as a shaded area in 
Figure~\ref{f:evolution_LB}, representing marginally bound ejecta).
The ratio of total internal energy to kinetic energy in Figure~\ref{f:evolution_LB} is
$E_{\rm i}/E_{\rm k}\simeq 0.15$ for the NS models and $0.2$ for the BH model, while
the total internal energy is very close to the gravitational energy, $E_{\rm i}\simeq E_{\rm g}$,
in all cases.
Assuming that all of the internal energy is converted to kinetic energy upon adiabatic
expansion, the root-mean-square velocity would increase by a factor 
$\sqrt{1+E_{\rm i}/E_{\rm k}}\lesssim 1.1$. In practice, this is an upper limit, since
some of the internal energy will be used to escape the gravitational potential.
Therefore the kinetic energy distributions of Figure~\ref{f:evolution_LB} are close
to their values in homology.

The velocity distribution of the ejecta is broad, as shown in Figure~\ref{f:evolution_LB},
spanning about two orders of magnitude in radial velocity. Note that this distribution is incomplete, however,
as including the region close to the compact object will add even faster outflows (\S\ref{s:small_bnd_evolution}). 
The angular distributions at the end of the simulations are strongly peaked toward the
poles, with an excess of about two orders of magnitude relative to the equatorial direction.

\begin{table*}
\centering
\begin{minipage}{17cm}
\caption{Mass ejection results for all models. Columns from left to right show unbound
mass ejected at $r_{\rm out}=30R_{\rm t}$ by the end of the simulation at $t=t_{\rm max}$ (Table~\ref{t:models}),
extrapolated mass ejected for $t\to\infty$ using equation~(\ref{eq:mass_ejection_power-law}),
fiducial comparison time $t_{\rm cmp}$, unbound mass ejected at 
$r=R_{\rm t}$ by $t=t_{\rm cmp}$, and selected mass fractions in this ejecta, computed according 
to equation~(\ref{eq:average_abundance}). Mass fractions smaller than $10^{-9}$ are not shown, and values larger than $0.1$
are rounded to two significant digits. Note that large boundary models {\tt (l)} do not include the 
contribution from inside $r<0.01R_{\rm t}$ and thus generate significantly less burning products than fully global models.}
\begin{tabular}{lcc|cc|ccccccccc}
\hline
{Model}& $M_{\rm ej}(t_{\rm max})$ & $M^\infty_{\rm ej}$ & $t_{\rm cmp}$ & $M_{\rm ej}(t_{\rm cmp})$ & 
\multicolumn{9}{c}{Mass Fractions @$R_{\rm t}$ and $t=t_{\rm cmp}$} \\
       & \multicolumn{2}{c}{@$30R_{\rm t}$ $(M_\odot)$} & (s)            & @$R_{\rm t}$ $(M_\odot)$ & 
 ${}^{12}$C & ${}^{16}$O & ${}^{4}$He & ${}^{20}$Ne & ${}^{24}$Mg & ${}^{28}$Si & 
 ${}^{32}$S & ${}^{40}$Ca & ${}^{56}$Ni\\
\hline
{\tt CO+NS(l)}       & 0.18 & 0.44 & 170 & 3.4E-3 & 0.50 & 0.50 & 5E-8 & 4E-3 & 1E-3 & 3E-4 & 6E-6 & 2E-9 & ...  \\ 
{\tt CO+NS(l-hr)}    & 0.21 & 0.42 &     & 3.5E-3 & 0.49 & 0.50 & 6E-8 & 6E-3 & 2E-3 & 4E-4 & 6E-6 & ...  & ...  \\
{\tt CO/He+NS(l)}    & 0.20 & 0.49 &     & 7.2E-3 & 0.45 & 0.21 & 2E-2 & 0.23 & 8E-2 & 7E-3 & 3E-6 & ...  & ...  \\
{\tt ONe+BH(l)}      & 0.59 & 0.72 & 30  & 1.2E-3 & 7E-5 & 0.60 & 3E-7 & 0.16 & 5E-2 & 0.11 & 5E-2 & 9E-3 & 3E-3 \\
\noalign{\smallskip}             
{\tt CO+NS(s)}       & 8.7E-3 & ... & 170 & 8.4E-3 & 0.34 & 0.45 & 4E-2 & 2E-2 & 3E-2 & 6E-2 & 2E-2 & 9E-3 & 2E-2\\  
{\tt CO/He+NS(s)}    & 4.4E-3 & ... & 170 & 1.0E-2 & 0.38 & 0.20 & 4E-2 & 0.21 & 9E-2 & 4E-2 & 8E-3 & 5E-3 & 1E-2 \\ 
{\tt ONe+BH(s)}      & 7.6E-4 & ... & 30  & 2.2E-3 & 8E-5 & 0.56 & 8E-4 & 0.10 & 5E-2 & 0.16 & 7E-2 & 2E-2 & 3E-2 \\ 
\noalign{\smallskip}             
{\tt CO+NS(s-vs)}    & 1.5E-2 & ... & 110 & 7.3E-2 & 0.40 & 0.46 & 3E-2 & 1E-2 & 2E-2 & 3E-2 & 1E-2 & 9E-3 & 2E-2\\  
{\tt CO+NS(s-sp)}    & 2.3E-3 & ... & 170 & 8.3E-3 & 0.33 & 0.44 & 7E-2 & 3E-2 & 3E-2 & 5E-2 & 1E-2 & 9E-3 & 2E-2 \\
{\tt ONe+BH(s-sp)}   & 9.7E-6 & ... & 30  & 2.4E-3 & 9E-5 & 0.46 & 3E-2 & 0.11 & 3E-2 & 0.15 & 8E-2 & 3E-2 & 9E-2 \\
\label{t:results}
\end{tabular}
\end{minipage}
\end{table*}

Table~\ref{t:results} shows that doubling the resolution in radius and angle results in enhanced mass ejection by
about $\sim 10\%$, which is consistent with other long-term hydrodynamic disk studies
carried out at similar resolution (e.g., \citealt{FM13}).

\subsection{Time-average behavior}
\label{s:time-average}

We average our results in time to remove the stochastic component of the flow, facilitating
structural analysis and comparison with previous one-dimensional work. We denote by angle
brackets the time- and angle average of a quantity \emph{per unit volume} $A(r,\theta\,t)$,
\begin{equation}
\langle A\rangle (r) = \frac{1}{(t_{\rm f}-t_{\rm i}) (\cos\theta_{\rm f}-\cos\theta_{\rm i})}
\int_{t_{\rm i}}^{t_{\rm f}}\int_{\cos\theta_{\rm i}}^{\cos\theta_{\rm f}} A\,dt\,d\cos\theta,
\end{equation}
where $[t_{\rm i},t_{\rm f}]$ and $[\theta_{\rm i},\theta_{\rm f}]$ are the time and 
polar-angle interval considered in the average. For quantities per unit mass 
$\tilde A = A/\rho$, we compute
the average as 
$\langle \rho \tilde A\rangle / \langle \rho\rangle$. For example, the average of the Bernoulli 
parameter is computed as
\begin{eqnarray}
\label{eq:bernoulli_average}
\langle b\rangle & = & \frac{1}{\langle\rho\rangle^2}\frac{1}{2}\left[\langle \rho v_r\rangle^2
     + \langle \rho v_\theta\rangle^2 + \frac{\langle \rho j\rangle^2}{(r\sin\theta)^2}\right]\nonumber\\
&& + \frac{1}{\langle\rho\rangle}\left[\langle \rho e_{\rm int}\rangle +\langle p\rangle + \langle\rho\Phi\rangle\right],
\end{eqnarray}
which we normalize with a local ``Keplerian" speed
\begin{equation}
\label{eq:vk_definition}
\langle v_K^2 \rangle = -\langle \rho\Phi\rangle / \langle\rho\rangle,
\end{equation}
which is simply the last term in equation~(\ref{eq:bernoulli_average}). Likewise, the root-mean-square
fluctuation of a quantity per unit mass is computed as
\begin{equation}
\textrm{r.m.s.}\, \left(\tilde A\right) \equiv \frac{\langle \rho \tilde A^2\rangle}{\langle\rho\rangle} - 
                                    \frac{\langle \rho \tilde A\rangle^2}{\langle\rho\rangle^2}.
\end{equation}

\begin{figure}
\includegraphics*[width=\columnwidth]{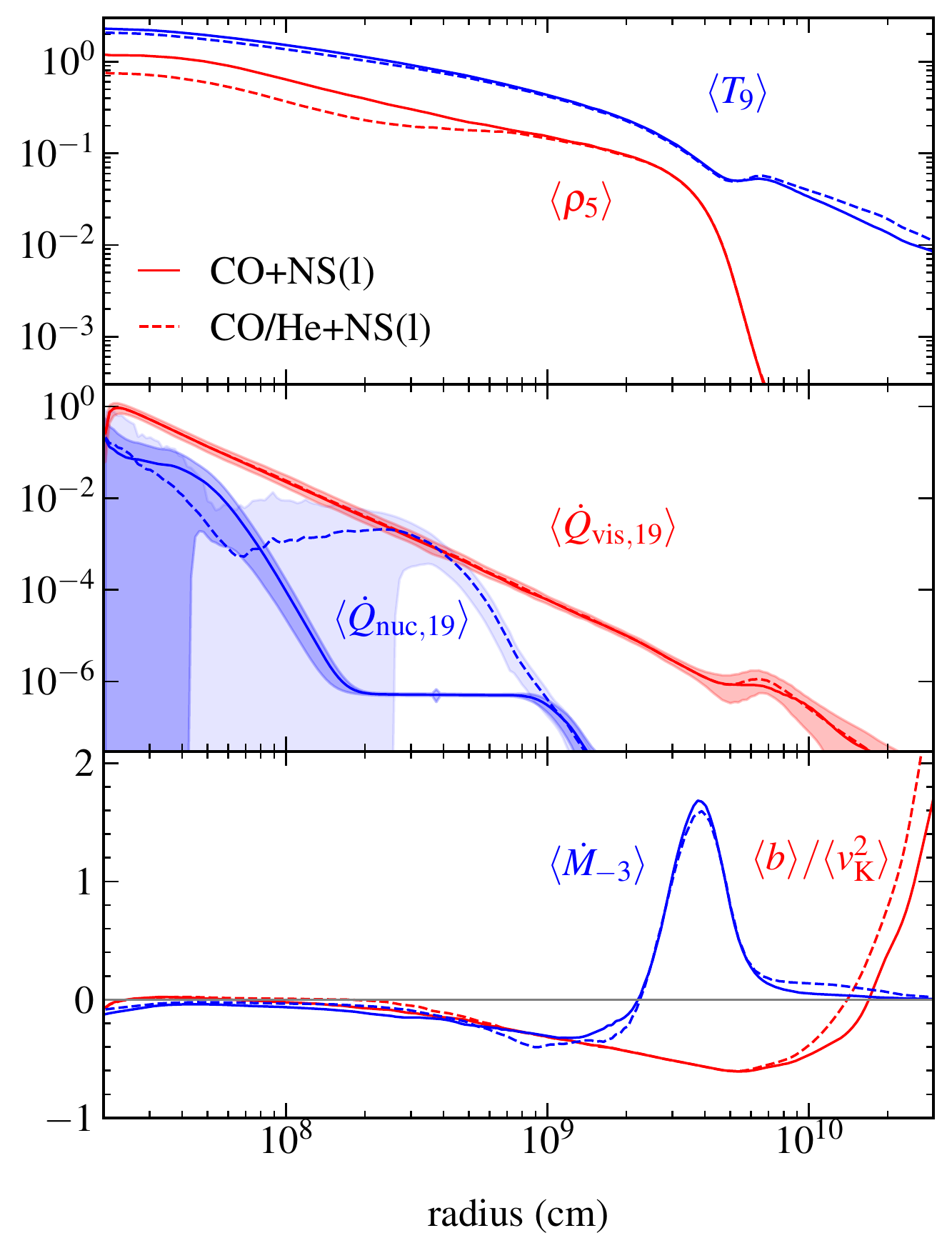}
\caption{Time- and angle-averaged profiles of structural quantities 
for models {\tt CO+NS(l)} and {\tt CO/He+NS(l)} as a function
of radius. The average is taken within $\pm 30^\circ$ of the equatorial plane, and
within $\pm 60$\,s (1.5 orbits at $r=R_{\rm t}$) of the time at which peak accretion is reached 
($t\simeq 200$\,s $\simeq$ 5 orbits). Quantities are defined as $\rho_5 = \rho/(10^5\,\textrm{g\,cm}^{-3})$, 
$T_9 = T/(10^9\,\textrm{K})$, 
$\dot{Q}_{\rm nuc,19} = \dot{Q}_{\rm nuc}/(10^{19}\,\textrm{erg\,[g\,s]}^{-1})$,
$\dot{Q}_{\rm vis,19} = \dot{Q}_{\rm vis}/(10^{19}\,\textrm{erg\,[g\,s]}^{-1})$,
and $\dot{M}_{-3} = \dot{M}/(10^{-3}\,M_\odot\textrm{s}^{-1})$. The average
Keplerian speed $v_{\rm K}$ is given by equation~(\ref{eq:vk_definition}), and the shaded
areas in the middle panel bracket root-mean-square fluctuations (values for $\dot{Q}_{\rm vis}$
in model {\tt CO+NS(l)} are not shown, for clarity).}
\label{f:timeave_profiles_LB}
\end{figure}

\begin{figure}
\includegraphics*[width=\columnwidth]{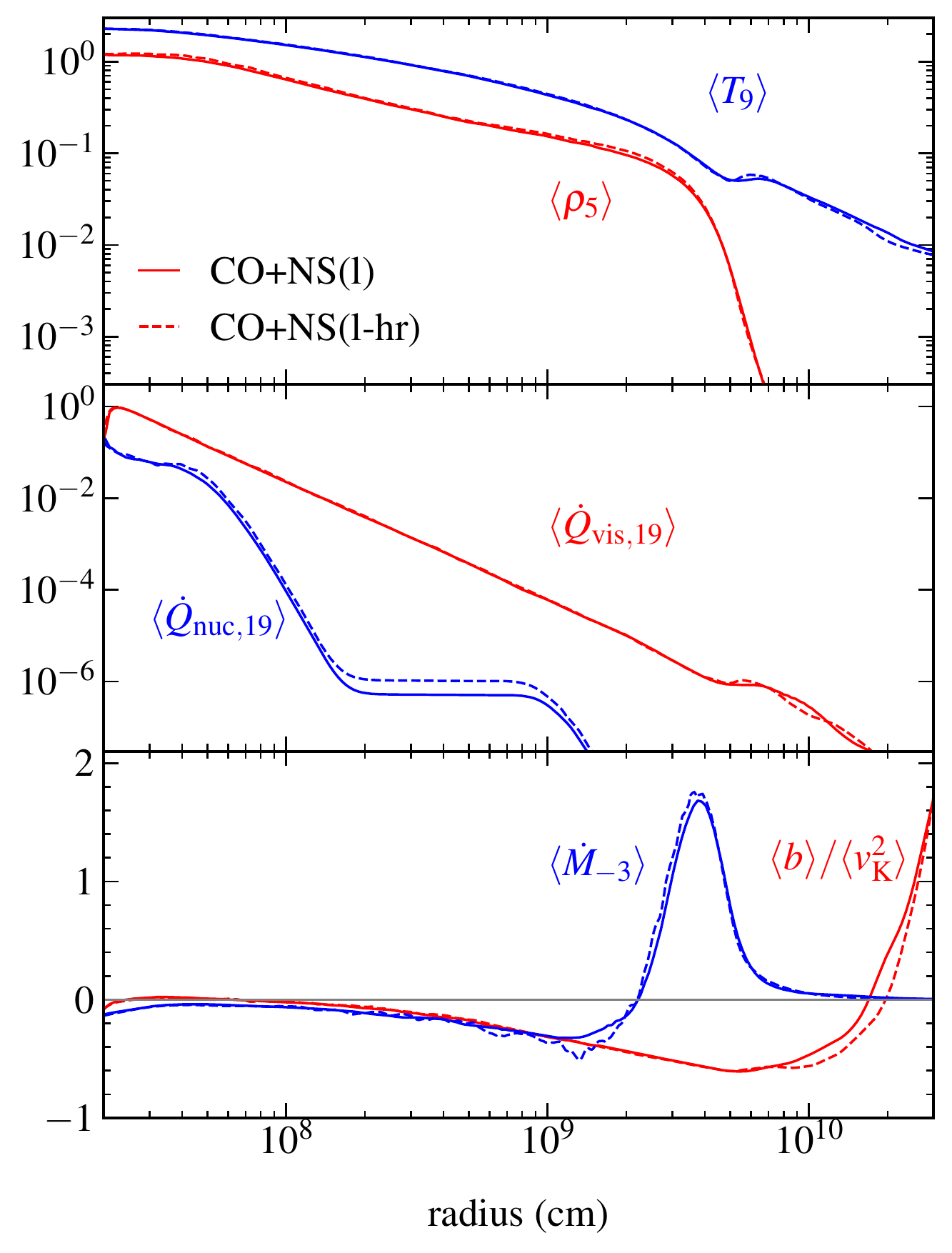}
\caption{Same as Figure~\ref{f:timeave_profiles_LB}, but now comparing model {\tt CO+NS(l)}
with its high-resolution counterpart, {\tt CO+NS(l-hr).} The structural profiles of the disk 
are essentially converged with resolution.
The time-averaged profiles
in the high-resolution model show more fluctuation because the data outputs were made
at larger intervals in time, hence fewer snapshots are involved in the average for the
same time period.
}
\label{f:timeave_profiles_resolution}
\end{figure}

Figure~\ref{f:timeave_profiles_LB} shows the average radial profiles of various 
quantities for the large inner boundary models {\tt CO+NS(l)} and {\tt CO/He+NS(l)}, with the average taken
within $30$\,deg of the equatorial plane and within $3$ orbits\footnote{The eddy turnover
time is of the order of the orbital time at each radius, as inferred from the r.m.s fluctuation 
of the meridional velocity.} 
at $r=R_{\rm t}$ from the time
of peak accretion at the inner boundary $r=0.01R_{\rm t}$ ($206\pm 62$\,s). The inner and outer
portions of the disk which are respectively accreting and expanding are separated by the
radius at which the accretion rate $\dot{M}=0$, and this radius moves outward in time.
The temperature and density profiles vary slowly with radius in both models, with a 
slight decrease in the density profile
for the hybrid WD model due to enhanced nuclear heating from He-burning reactions. 

\begin{figure*}
\includegraphics*[width=\textwidth]{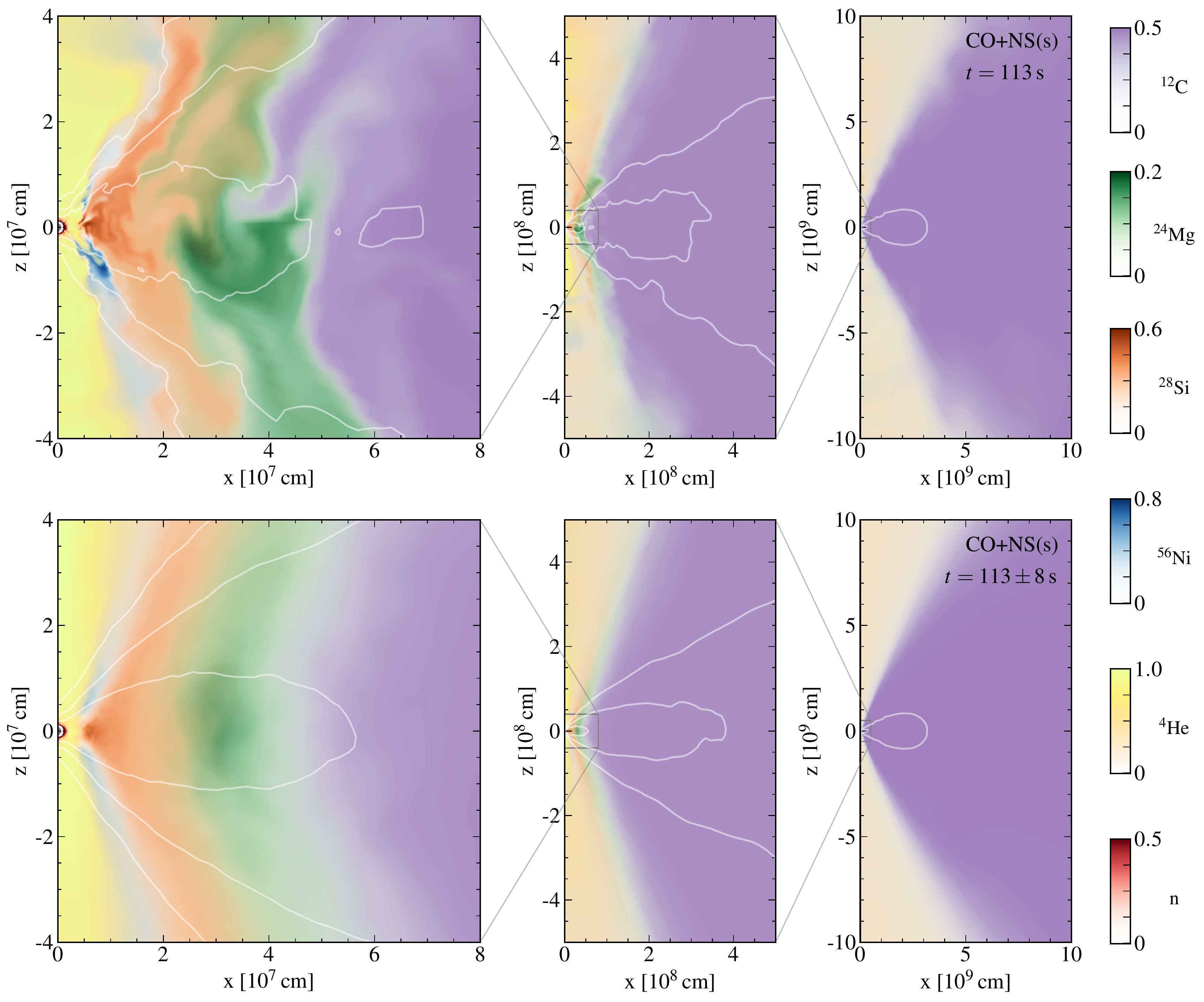}
\caption{Snapshot of the default WD+NS merger model with small
inner boundary [{\tt CO+NS(s)}], which resolves accretion onto the neutron star.
The top row shows the instantaneous mass fractions of various species, as labeled,
at a time $t=113$\,s ($2.7$ orbits at $r=R_{\rm t}$). The bottom row is a time average
of snapshots in the range $105-121$\,s. White contours correspond, from the inside out, 
to densities of \{$10^4$, $3\times 10^5$, and $10^5$\}g\,cm$^{-3}$, respectively. 
Panels on the left side are zoom-ins of panels on the right, as indicated by
the gray lines.
}
\label{f:timeave_abund_snapshots}
\end{figure*}

In both models, the mean viscous heating dominates at all radii, except in the region where most of the
He is burned in model {\tt CO/He+NS}, where the mean nuclear heating rate is at most comparable to the 
average viscous heating. This additional nuclear heating is associated with an enhancement of $\sim 10\%$
in the ejected mass in this hybrid model (Table~\ref{t:results}). While the fluctuations in the viscous 
heating term remain small over the entirety of the disk, nuclear burning becomes highly stochastic
inside radii where heavier elements start to be produced. In the case of the hybrid model, these fluctuations
can exceed the average viscous heating over a narrow range of radii, while for the fiducial model nuclear burning
never dominates (the steep decrease of the viscous heating at small radii is an artifact of the boundary condition
in the models shown in Figure~\ref{f:timeave_profiles_LB}). The relative weakness of nuclear burning helps 
explain why a thermonuclear runaway never takes place in our models.

Figure~\ref{f:timeave_profiles_LB} also shows the profile of averaged Bernoulli parameter 
on the disk equatorial plane. This quantity adjusts to negative values close to zero at
small radii. While the average Bernoulli parameter can be slightly positive near the inner boundary, 
this is a consequence of vertical alternations in sign in regions from which the outflow is launched. 
These average profiles of Bernoulli parameter are in broad agreement with the assumptions
of \citet{M12} and MM16.

Figure~\ref{f:timeave_profiles_resolution} compares average radial profiles in the 
fiducial WD+NS model and a version at twice the resolution in radius and angle. The
profiles of all quantities are in excellent agreement
except for nuclear burning around
$r=R_{\rm t}$, which is slightly higher in the high-resolution model (but still sub-dominant
relative to viscous heating). Table~\ref{t:results}
shows that the overall mass ejection is higher by about $10\%$ in the high-resolution
model. While higher spatial resolution allows a better characterization of convective
turbulence in the disk, the modest increase in mass ejection indicates that this
convective activity is a sub-dominant factor in determining mass ejection compared
to other processes such as viscous heating and angular momentum transport.

%
\subsection{Evolution near the central object}
\label{s:small_bnd_evolution}

\begin{figure}
\includegraphics*[width=\columnwidth]{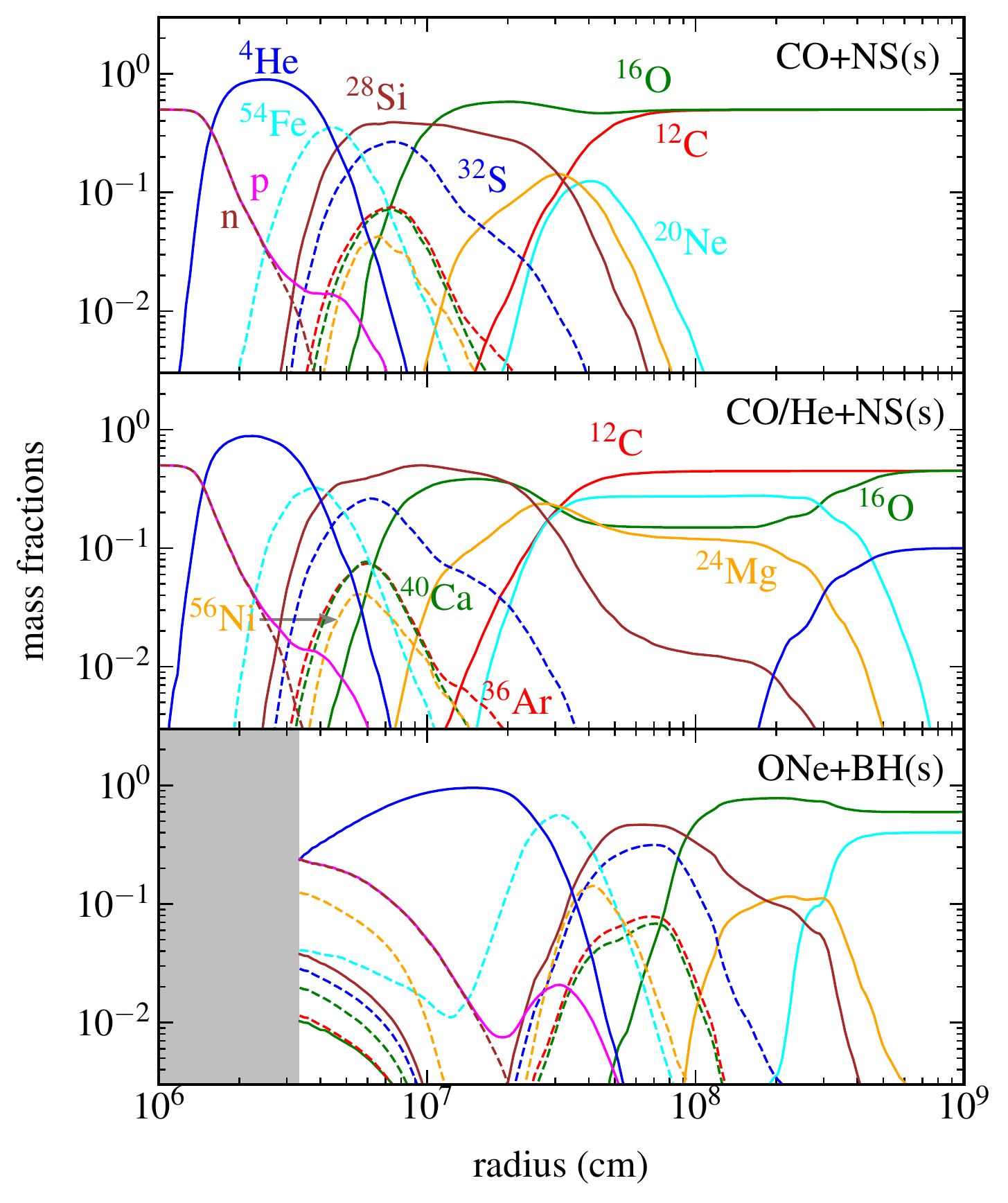}
\caption{Time- and angle-averaged mass fractions of various species, as labeled,
for small inner boundary models {\tt CO+NS(s)} (top), {\tt CO/He+NS(s)} (middle),
and {\tt ONe+BH(s)} (bottom), at $t=113\pm 3$\,s ($2.8\pm 0.2$ orbits at $r=R_{\rm t}$) for the
NS models and $t=39\pm 1.8$\,s ($4.4\pm 0.2$ orbits at $r=R_{\rm t}$) for the BH model. 
The gray shaded area markes the inner boundary of the BH model, midway 
between the horizon and the ISCO.}
\label{f:abund_profiles_SB}
\end{figure}

A key property of disks formed in WD-NS/BH mergers is that nuclear fusion
reactions of increasingly heavier elements take place as 
material accretes to smaller radii with higher temperatures and
densities \citep{M12}. Our small inner boundary models can resolve this
phenomenon in its entirety, at the expense of evolving for a short amount of time
relative to $t_{\rm vis}$ given the more restrictive Courant condition
at smaller radii.

None of our small-boundary models undergo a detonation. Given the deeper
gravitational potential than in the large boundary models, nuclear energy 
release at these radii is less dynamically important, so this outcome is to be expected if
detonations did not already occur at larger radii.

Figure~\ref{f:timeave_abund_snapshots} shows the spatial distribution of various
species in our fiducial WD+NS model that resolves the compact object [{\tt CO+NS(s)}].
Turbulence is associated with convection driven
mostly by viscous heating but also by the nuclear energy released in fusion
reactions. Species are launched from the same
radii of the disk in which they are produced, with only moderate radial mixing. This
stratification of mass ejection into different species becomes evident
when taking a time-average of the flow (also shown in Figure~\ref{f:timeave_abund_snapshots}),
yielding a characteristic onion-shell-like structure as envisioned by \citet{M12}.

Time-averaged radial profiles of different abundances in the disk are shown
in Figure~\ref{f:abund_profiles_SB} for the three baseline small boundary models.
During accretion, elements that initially made up the WD
are fused into heavier ones from the outside-in. 
The hybrid CO-He WD model
shows a larger fraction of intermediate mass elements at larger radius than the
fiducial CO WD, while the BH model completes all nucleosynthesis at
larger radii due to the higher disk temperatures.

Given that we resolve the compact object, we are also able to resolve the radius inside
which heavy elements undergo photodissociation into ${}^{4}$He nuclei and nucleons.
In the vicinity of the NS surface, the composition is almost entirely neutrons and protons
at the times shown. The BH model shows an increase in the heavy element abundance
as the inner boundary is approached. This phenomenon is associated with a decreasing
entropy given the net energy losses from nuclear reactions.

\begin{figure}
\includegraphics*[width=\columnwidth]{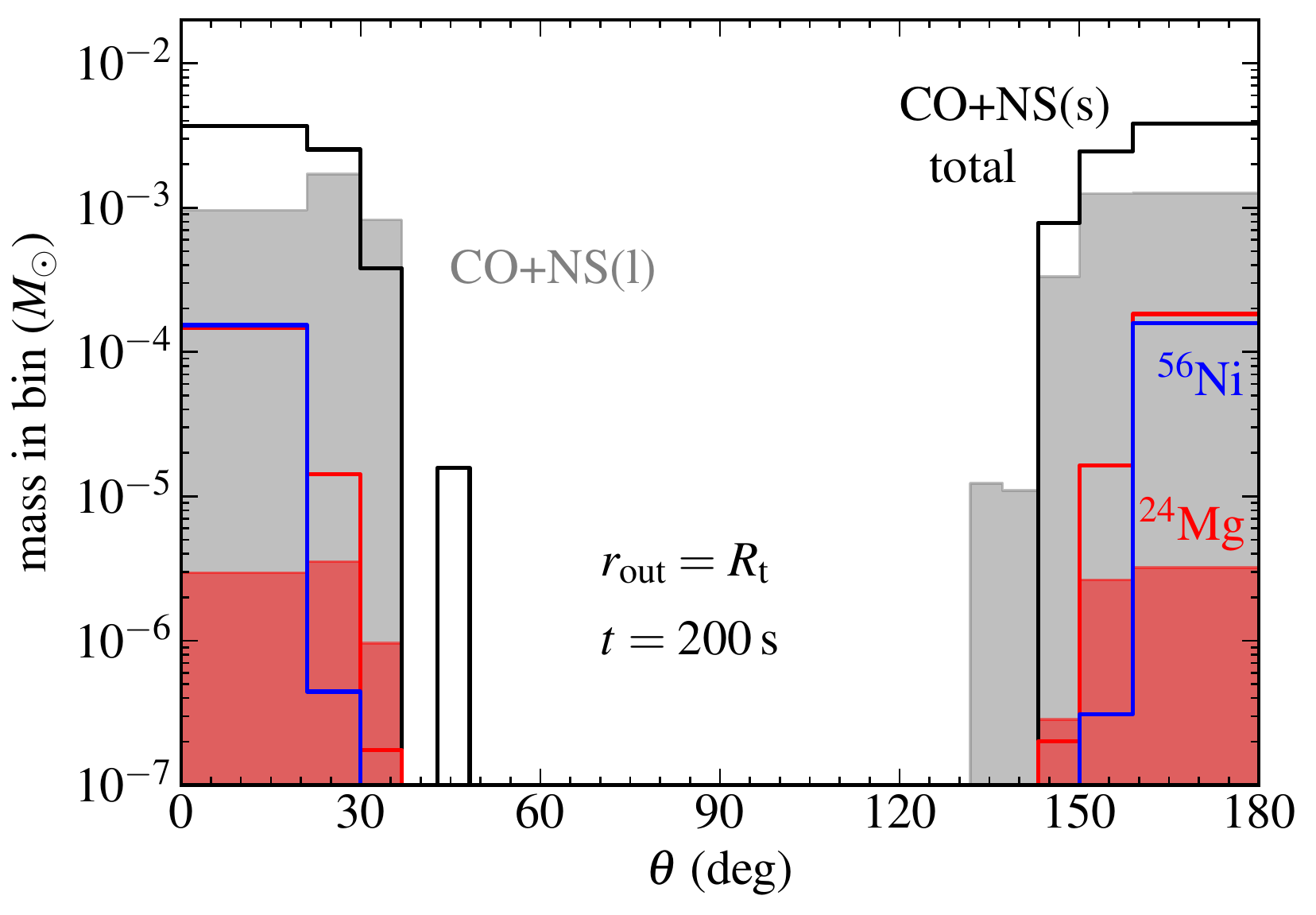}
\caption{Mass histogram of unbound ejecta beyond $r=R_{\rm t}$ by $t=200$\,s 
as a function of polar angle, for the default WD+NS model with small inner boundary 
[{\tt CO+NS(s)}, solid lines] and large inner boundary [{\tt CO+NS(l)}, shaded areas]. The black lines
and gray shaded areas show total mass, while red and blue correspond to
${}^{24}$Mg and ${}^{56}$Ni only (the large boundary model does not
make any ${}^{56}$Ni).}
\label{f:hist_ang_def_LB-SB}
\end{figure}

While the outflow composition is well stratified on spatial scales comparable to the disk thickness,
as shown by Figure~\ref{f:timeave_abund_snapshots},
significant mixing of the ejecta occurs as it expands outward, to the point where
individual species are not distinguishable on scales comparable to the initial 
circularization radius $R_{\rm t}\sim 10^9$\,cm. Note also that ejection of fusion products
is confined to a narrow cone in angle $\lesssim 40$\,deg from the rotation axis (Figure~\ref{f:hist_ang_def_LB-SB}),
which persists out to very large radii (Figure~\ref{f:timeave_abund_snapshots}). 
Figure~\ref{f:hist_ang_def_LB-SB} suggests that nucleosynthesis products produced at deeper radii have
narrower angular distributions around the rotation axis.

To characterize the composition of outflows, we compute an average ejecta mass
fraction for species $i$ as
\begin{equation}
\label{eq:average_abundance}
\bar X_i(r_{\rm out},t_{\rm cmp}) = \frac{\int dt\int d\Omega\, \rho v_r X_i}{\int dt \int d\Omega\, \rho v_r}
\end{equation}
where the time integrals are carried out from the beginning of the simulation out to
some fiducial comparison time $t_{\rm cmp}$, and
the angular integral covers all polar angles. Table~\ref{t:results} shows mass ejected
and abundances for all models, measured at $r_{\rm out} = R_{\rm t}$ and by a time that allows
to compare models with different durations ($170$\,s for most NS models, and $30$\,s
for the BH models). The outflow from the fiducial small-boundary CO WD is dominated 
by ${}^{12}$C and ${}^{16}$O at a combined $\sim 80\%$ by mass, with all nucleosynthesis products contributing
each at a few $\%$ level by mass. For the hybrid CO-He WD, the original
WD elements are preserved at a combined $\sim 60\%$ by mass, with ${}^{20}$Ne
and ${}^{24}$Mg being a significant secondary contribution at $21\%$ and $9\%$,
respectively. 

In the same way as with the large boundary models, the admixture of He in
the fiducial CO WD results in more energetic nuclear burning and enhanced
mass ejection. Table~\ref{t:results} shows that when integrated out to the same time, the total unbound 
mass ejection within $r=R_{\rm t}$ is higher by $\sim 10\%$ in
model {\tt CO/He+NS(s)} than in {\tt CO+NS(s)}.

The outflow from the ONe WD + BH model is 
qualitatively different from the fiducial CO + NS
case. The ejected mass is higher given the larger disk mass and a similar
overall fraction ejected. Regarding composition, the initial WD material
is preserved at a combined mass fraction of $66\%$, with ${}^{28}$Si
being the dominant nucleosynthetic product at $16\%$ by mass. While
other products have abundances at a few $\%$ level by mass, a key
property of this combination is the small amounts of ${}^{12}$C and ${}^{4}$He
in the ejecta, at less than $0.1\%$ by mass.

In the fiducial and hybrid small boundary WD+NS models, the mass fraction of ${}^{56}$Ni in the ejecta is 
$\sim 2\%$ at a time $170$\,s. If we assume that this fraction remains constant in all 
ejecta and that the fraction of the disk mass is at least that estimated for the large
boundary models ($0.4M_\odot$, which is a lower limit), we obtain a characteristic ${}^{56}$Ni yield 
in the range $10^{-3}-10^{-2}M_\odot$.  The non-spinning BH model with small boundary
makes a larger fraction of ${}^{56}$Ni which suggests a yield $\gtrsim 10^{-2}M_\odot$,
given the larger WD mass and asymptotic ejected fraction. These estimates are optimistic, 
given the fact that burning fronts recede with time as
the disk density decreases (MM16), implying that ${}^{56}$Ni production will eventually
stop. Most of the mass is ejected during peak accretion (Figure~\ref{f:mout_def_LB-SB}), however,
and the ${}^{56}$Ni fraction should remain approximately constant during this period. We thus expect that
the late-time recession of the burning fronts will introduce corrections of order unity to the
final ${}^{56}$Ni yield. Our range of ejected ${}^{56}$Ni is in agreement with previous
estimates (\citealt{M12}; MM16; \citealt{zenati_2019}). 

No significant $r$-process production is expected in our models.
Figure~\ref{f:abund_profiles_SB} shows that after photodissociation, the mass
fractions of neutrons and protons remain equal all the way to the surface of the NS, thus preserving
the initial $Y_e = 0.5$ of the WD. While our models include charged-current weak
interactions that modify $Y_e$, no appreciable neutronization
occurs. At the surface of the NS we have $T\simeq 10^{10}$\,K 
and $\rho\sim 10^6$\,g\,cm${}^{-3}$ (\S\ref{s:parameter_dependencies}), 
for which electrons are trans-relativistic. The
non-relativistic and relativistic Fermi energies are comparable and smaller than 
the thermal energy
\begin{eqnarray}
\frac{p_{\rm F}^2}{2m_e kT} & \simeq 0.2\rho_6^{2/3} T_{10}^{-1}&\\
\frac{p_{\rm F}c}{kT} & \simeq 0.5\rho_6^{1/3}T_{10}^{-1}&
\end{eqnarray}
where $T_{10}=T/(10^{10}\,\textrm{K})$ and $\rho_6 = \rho/(10^6\,\textrm{g}\,\textrm{cm}^{-3})$.
Thus electrons are essentially non-degenerate, and the equilibrium electron fraction is 
close to $Y_e = 0.5$. Even though neutrino cooling from electron-positron capture on nucleons is sub-dominant
relative to other heating and cooling processes (\S\ref{s:parameter_dependencies}), 
the timescale to change of $Y_e$ from charged-current weak interactions for
non-degenerate material (e.g., \citealt{fernandez2019})
\begin{equation}
\left(\frac{d\ln Y_e}{dt}\right)^{-1}\simeq 5T_{10}^{-5}\,\textrm{s}
\end{equation}
is shorter than the characteristic evolutionary times. The lack of neutronization is 
thus a consequence of the non-degeneracy of the material. Whether these systems
generate any $r$-process elements might depend on the details of angular momentum 
transport, which might result in higher accretion rates and densities at small radii, 
increasing the degeneracy of the material. This is not found for our choice of parameters.

\subsubsection{Comparison of small- and large inner boundary models}
\label{s:comparison_sb_lb}

\begin{figure}
\includegraphics*[width=\columnwidth]{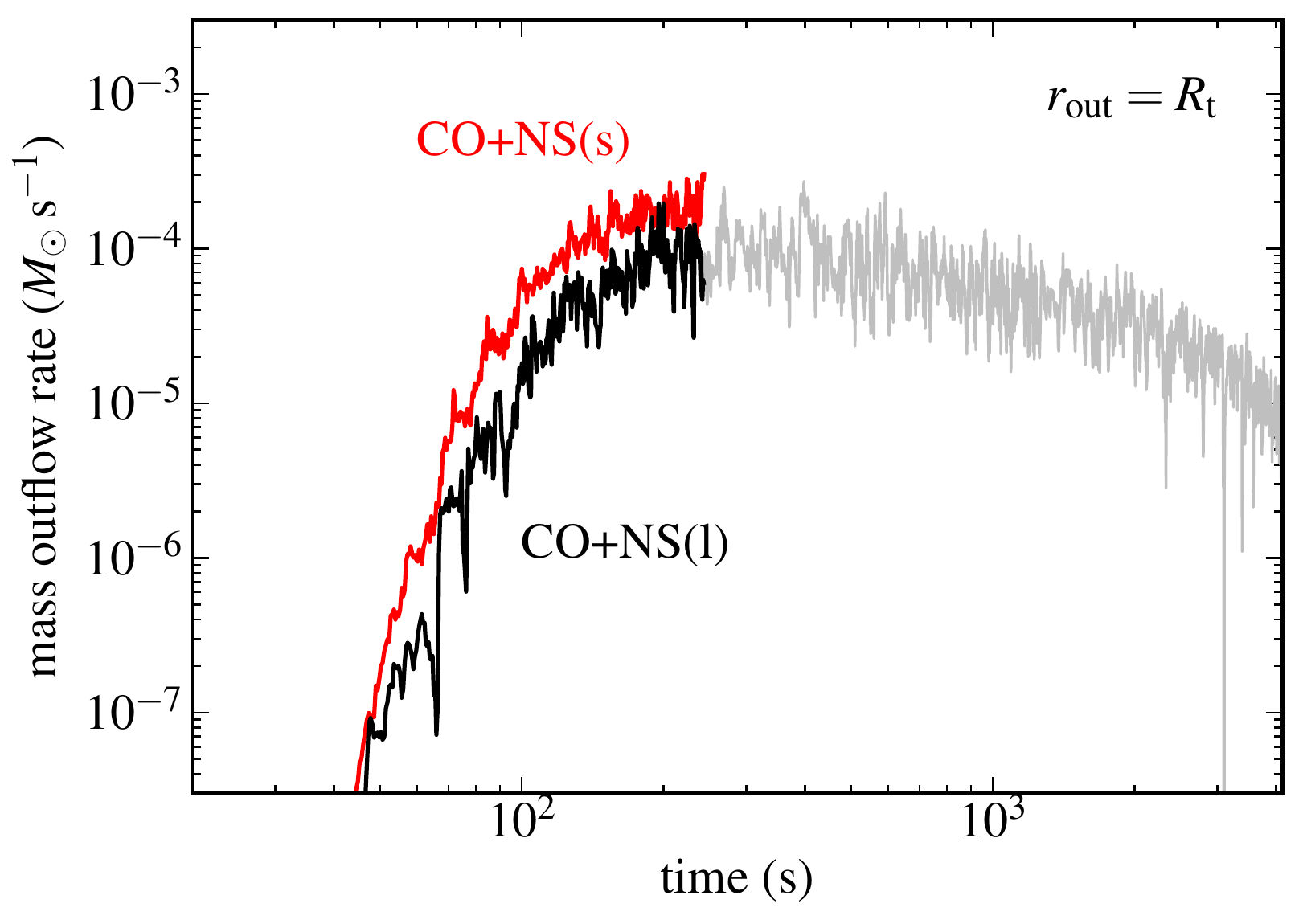}
\caption{Unbound mass ejected at $r=R_{\rm t}$ as a function of time
in the fiducial WD+NS models with large inner boundary (black and gray) 
and small inner boundary (red).}
\label{f:mout_def_LB-SB}
\end{figure}

Given our approach to disk evolution that separates large- from small
radius dynamics, it is important to make a connection between the two
run types and to quantify the ejecta missing from the large boundary runs.
Since the small boundary models cannot be evolved for nearly as long
as large boundary runs, most of the ejecta from the former does not
make it to a large enough radius to probe near-homologous expansion.
Instead, we need to make the comparison at smaller radius, which we
choose to be $r=R_{\rm t}$. By restricting the analysis to
material with positive Bernoulli parameter, we separate 
bound disk material from unbound ejecta.

\begin{figure}
\includegraphics*[width=\columnwidth]{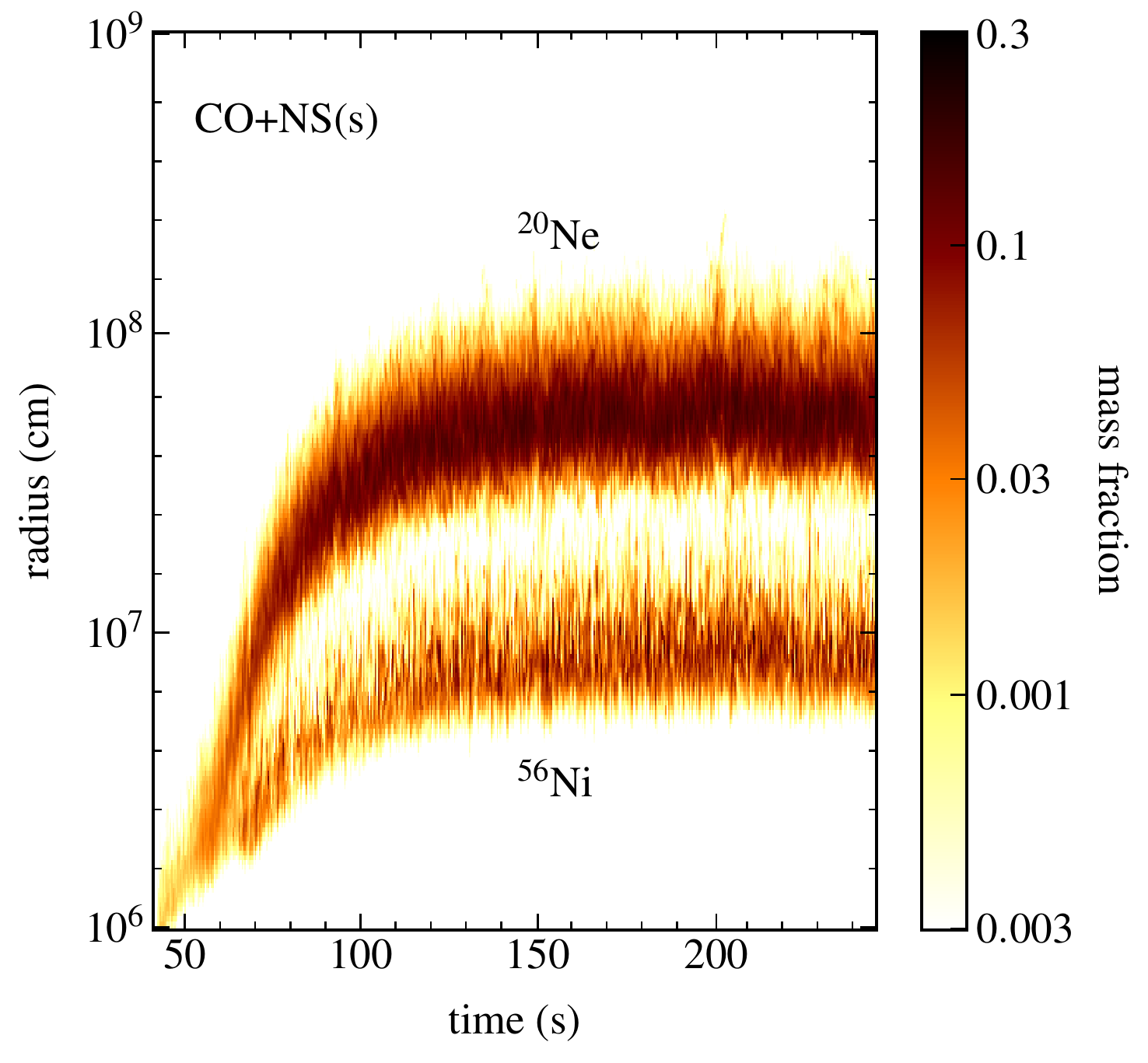}
\caption{Radial profiles of the ${}^{20}$Ne and ${}^{56}$Ni mass fractions as a function of
time for the fiducial WD+NS model with small inner boundary. See also Figure~\ref{f:abund_profiles_SB}.}
\label{f:profiles_spec_time}
\end{figure}

Figure~\ref{f:mout_def_LB-SB} compares the mass outflow rate at $r=R_{\rm t}$
from the default WD+NS with small- and large inner boundary. As expected,
the model that resolves the compact object ejects more mass (factor of $\sim 2$) than
the large boundary model at all times up to the end of the simulation at $t=245$\,s (Table~\ref{t:results}).
This time is in the range during which the mass accretion rate onto the compact object reaches its maximum value,
evolving slowly with time before entering the power-law decay regime at around $t\simeq t_{\rm vis}$.
The radial profiles of the ${}^{20}$Ne and ${}^{56}$Ni mass fractions 
as a function of time are shown in Figure~\ref{f:profiles_spec_time}, showing the
location of the burning fronts. On a linear scale in time, these burning fronts
are essentially at constant radii after $t\simeq 100$\,s for this value of the
viscosity parameter.

The angular distribution of material is very similar up to $t=200$\,s, with
both models ejecting the majority of the material within a funnel of $\lesssim 40$\,deg from
the rotation axis. The large
boundary simulation does not show a significant difference between the angular
distribution of the total ejecta (mostly C and O) and that of ${}^{24}$Mg (the
burning product with the largest mass fraction). In contrast, the small boundary
model shows a trend in which burning products that are generated at smaller
radii are ejected at narrower angles (on average) from the rotation axis. This is consistent
with the snapshots in Figure~\ref{f:timeave_abund_snapshots}, and suggests that
despite the mixing, this angular segregation can persist to large radii. 

\begin{figure}
\includegraphics*[width=\columnwidth]{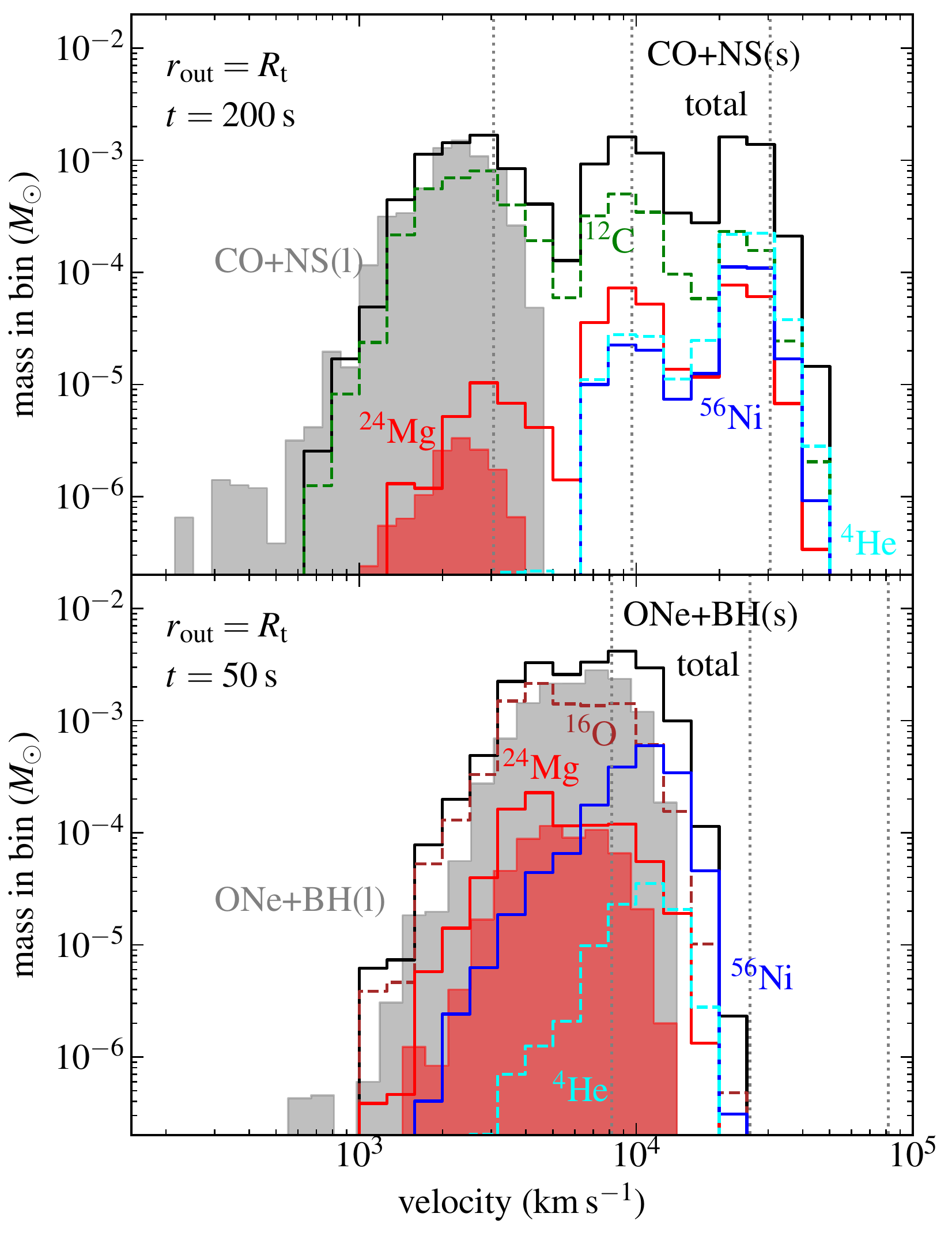}
\caption{Mass histograms as a function of radial velocity, for unbound ejecta
at $r=R_{\rm t}$ 
in the fiducial WD+NS models (top) and WD+BH models (bottom) at the times labeled,
with solid lines denoting small-boundary versions and shaded areas their large-boundary counterparts. 
Black/grey histograms correspond to total ejecta, while 
green, brown, red, blue,
and cyan histograms correspond respectively to 
${}^{12}$C, ${}^{16}$O,
${}^{24}$Mg, ${}^{56}$Ni, and ${}^{4}$He. The
vertical dashed lines correspond from left to right to the (point-mass) Keplerian velocity at radii 
$\{1,0.1,0.01\}R_{\rm t}$, respectively. Model {\tt CO+NS(s)} uses a reflecting boundary condition at the
surface of the NS, while all other models employ an outflow boundary condition at the smallest radius.}
\label{f:hist_vel_def_LB-SB}
\end{figure}

The importance of resolving small radii is illustrated in Figure~\ref{f:hist_vel_def_LB-SB},
which shows the velocity distribution of ejecta for both small- and large boundary
fiducial WD+NS. The velocity distribution of the large boundary model
cuts off at $v_{\rm max}\simeq \sqrt{GM_c/R_{\rm t}}\sim$ few $1,000$\,km\,s$^{-1}$, 
which persists up to the end of the simulation (c.f., Figure~\ref{f:evolution_LB}).
In contrast, the outflow from the small inner boundary model can reach maximum
velocities that are about $10$ times higher. These velocities correspond to
gravitational binding energies of radii as small as $0.01R_{\rm t}$, where
nuclear energy release is still significant (Figures~\ref{f:abund_profiles_SB} 
and \ref{f:profiles_spec_time}). Note also that the mean velocities of elements produced at smaller radii
are higher: ${}^{12}$C is on average slower than ${}^{24}$Mg (because it has more slow material), 
which in turn is slower than ${}^{56}$Ni and ${}^{4}$He (the latter two have comparable distributions).
This trend is consistent with the trend in the angular distribution of burning products.

The marked difference between the small- and large-boundary velocity distribution
for the fiducial WD+NS model is in part a consequence of the reflecting boundary 
condition at the NS surface for model {\tt CO+NS(s)}. The large boundary model has an outflow boundary 
condition, through which not only mass but also energy are lost. In contrast,
the small boundary model is such that energy from accretion has nowhere to go except into a wind, given 
the weakness of neutrino and photodissociation cooling (\S\ref{s:parameter_dependencies}).
This difference stands in contrast to the large- and small-boundary BH models (also shown
in Figure~\ref{f:hist_vel_def_LB-SB}),
both of which use an outflow inner radial boundary condition and
have a velocity distribution that 
differs only by a factor of $\sim 2$ in their maximum velocity.

Finally, the composition of the outflow between large- and small boundary models
is significantly different, as expected given the radii at which nucleosynthesis
occurs. The fiducial large boundary model preserves the original WD composition at
more than $99\%$ by mass, while this fraction drops to a combined $79\%$ 
(with different relative fractions) in the small boundary model. The large boundary
hybrid CO-He WD model consumes a substantial fraction of the original
${}^{16}$O and ${}^{4}$He, yet it does not manage to make any significant
${}^{56}$Ni. The large boundary BH model does make some ${}^{56}$Ni, but
the overall fractions of the heaviest elements are much smaller.

\begin{figure}
\includegraphics*[width=\columnwidth]{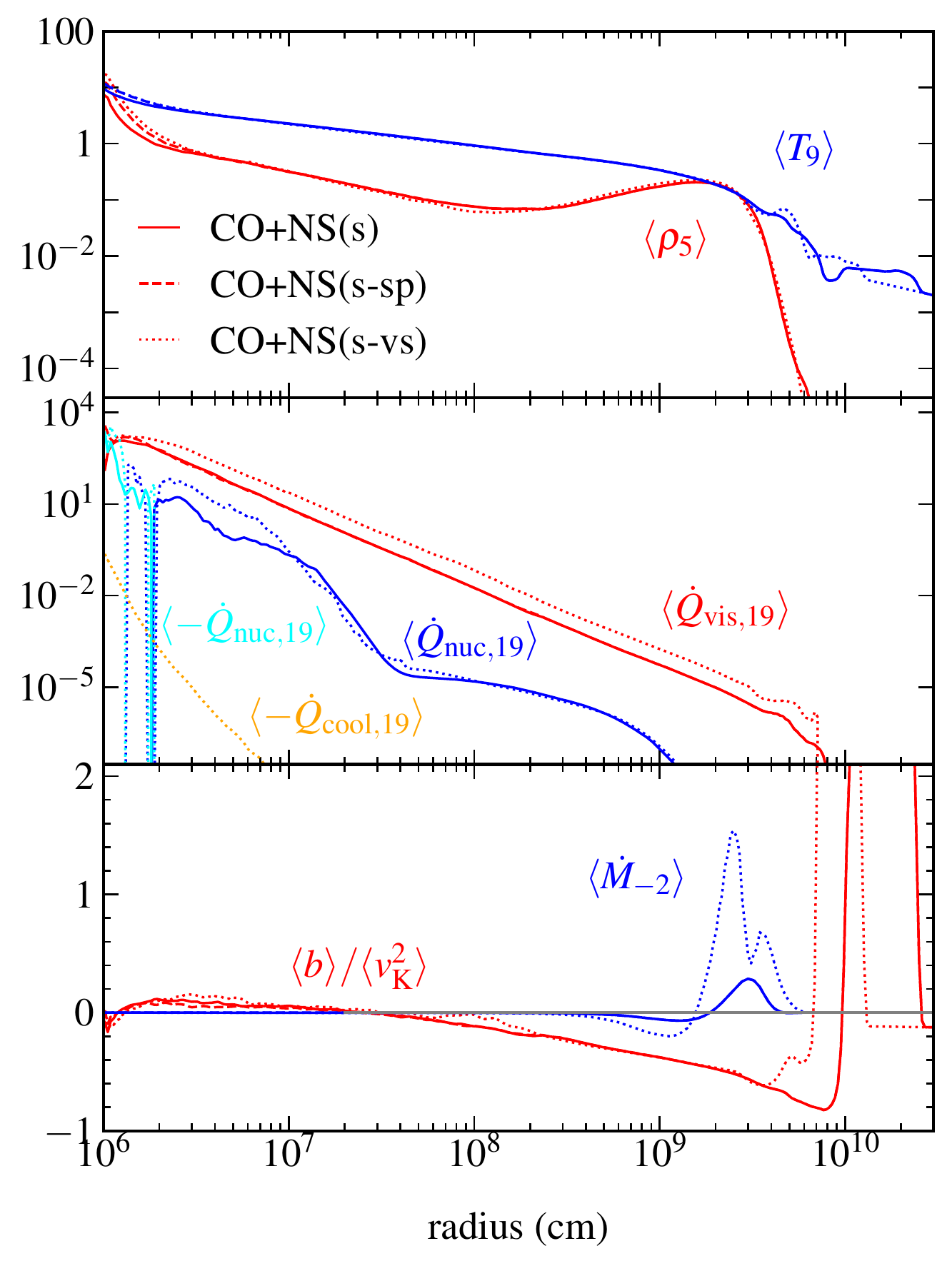}
\caption{Time- and angle-averaged profiles of structural quantities for small inner boundary models 
with a NS, comparing
the baseline model {\tt CO+NS(s)}, the model with a rotating NS {\tt CO+NS(s-sp)} and
a model with higher viscosity parameter {\tt CO+NS(s-vs)}. Quantities are the
same as in Figure~\ref{f:timeave_profiles_LB}, and the time range for the average
is $1.7\pm0.2$ orbits at $r=R_{\rm t}$ ($70\pm 8$\,s) for 
models {\tt CO+NS(s)} and {\tt CO+NS(s-sp)}, and 
$0.5\pm 0.02$ orbits ($20.2\pm 0.8$\,s) for model {\tt CO+NS(s-vs)}. 
Cyan curves show net energy loss from the nuclear reaction
network due to photodissiation and thermal neutrino losses (not shown for model {\tt CO+NS(s-sp)} for clarity, 
as it resembles that of {\tt CO+NS(s-vs)}), and the orange curve show charged-current neutrino
losses, shown only for model {\tt CO+NS(s-vs)}, for clarity.}
\label{f:timeave_profiles_rns}
\end{figure}

\subsection{Parameter dependencies}
\label{s:parameter_dependencies}

We now turn to addressing some of the parameter sensitivities of our results. 
Figure~\ref{f:timeave_profiles_rns} shows time- and angle averaged profiles of various quantities
for the baseline WD+NS model and variations of it with 
different viscosity parameter and spin of the neutron star. 
At a comparable evolutionary time, the model with higher viscosity differs in 
that (1) viscous heating is higher
throughout the disk, (2) the disk evolution is faster, as indicated by the larger
mass outflow rate at the disk outer edge, (3) nuclear energy release is
a factor of a few larger inside $0.01R_{\rm t}$, and (4) the transition from
positive to negative net energy generation by the reaction network moves inward in radius.
The model with a spinning NS, evaluated at the same time as the baseline model, 
differs only inside $\sim 30$\,km (3 NS radii),
where a boundary layer develops. The additional viscous heating in
this layer results in a somewhat higher temperature and an inward shift
of the transition where neutrino cooling dominates over nuclear energy release,
like in the high-viscosity model. In all three models, neutrino cooling
from electron/positron capture onto nucleons is sub-dominant.

\begin{figure}
\includegraphics*[width=\columnwidth]{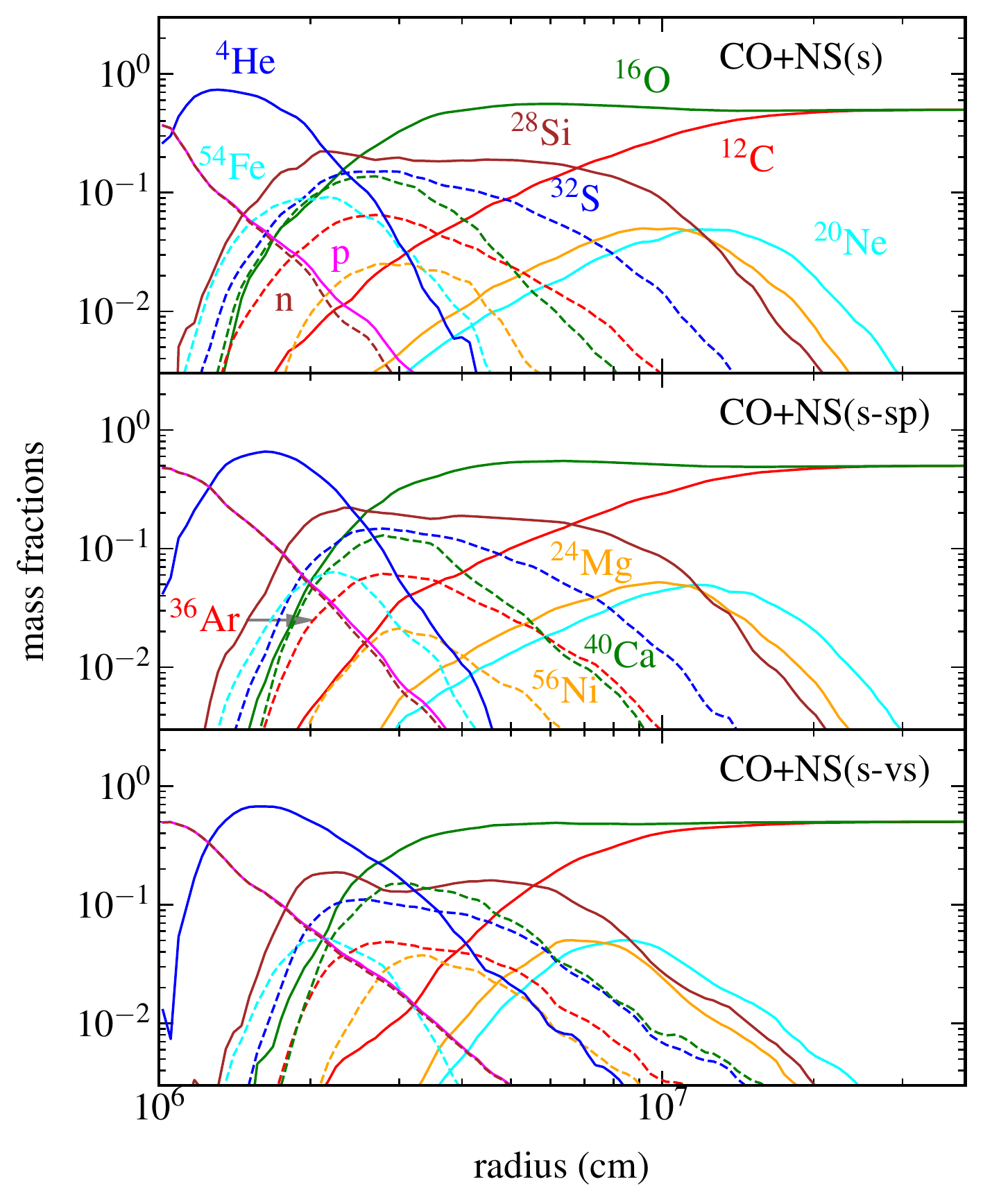}
\caption{Time- and angle-averaged abundance profiles within $30$\,deg of the equatorial plane over the 
same time period as in Figure~\ref{f:timeave_profiles_rns}, for
the three small boundary models that resolve the NS surface [{\tt CO+NS(s)}, {\tt CO+NS(s-sp)},
and {\tt CO+NS(s-vs)}]. Abundances have the same color coding as Figure~\ref{f:abund_profiles_SB}.}
\label{f:abund_profiles_rns}
\end{figure}

Figure~\ref{f:abund_profiles_rns} shows inner disk nucleosynthesis profiles for 
the three small boundary NS models. Abundances of all elements are very similar among models
except within a few NS radii of the stellar surface, where the high-viscosity and
spinning NS profiles both deviate from the baseline model in that photodissociation
of ${}^{4}$He moves further out in radius given the higher temperatures.
Table~\ref{t:results} shows that the mass fractions in the outflow are very similar
among all three models, with the possible exception of ${}^{12}$C and ${}^{4}$He, pointing to
a robustness in the composition of the wind to the details of how angular momentum
transport operates.

\begin{figure}
\includegraphics*[width=\columnwidth]{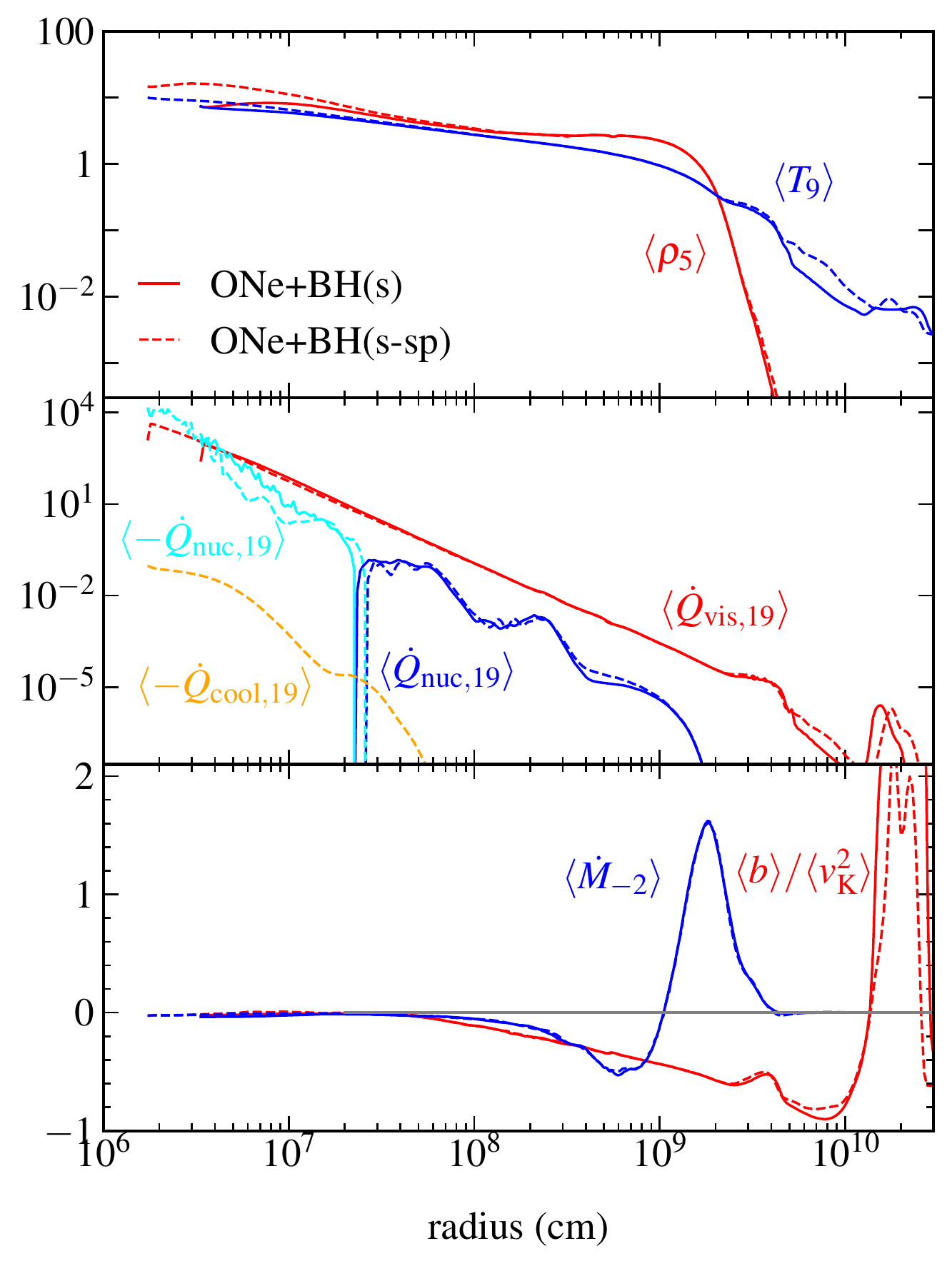}
\caption{Same as Figure~\ref{f:timeave_profiles_rns}, but for the two small boundary models
that resolve the BH: {\tt ONe+BH(s)} (non-spinning) and {\tt ONe+BH(s-sp)} (spin $\chi = 0.8$). 
The time interval is $3\pm 0.3$ orbits at $r=R_{\rm t}$ ($26.6\pm 2.7$\,s).}
\label{f:timeave_profiles_bh}
\end{figure}

The difference in profiles between the non-spinning and spinning BH models 
is shown in Figure~\ref{f:timeave_profiles_bh}. While the outer disk evolution is
nearly identical, differences arise near the inner boundary, where the spinning BH
model has slightly higher densities and temperatures. This bifurcation  does not
significantly affect the radius inside which photodissociation and thermal
neutrino cooling dominate over nuclear 
heating, although net energy loss is stronger in the spinning BH model, even exceeding
viscous heating near the inner boundary. Like the NS models, neutrino cooling
due to charged-current weak interactions is negligible.

\begin{figure}
\includegraphics*[width=\columnwidth]{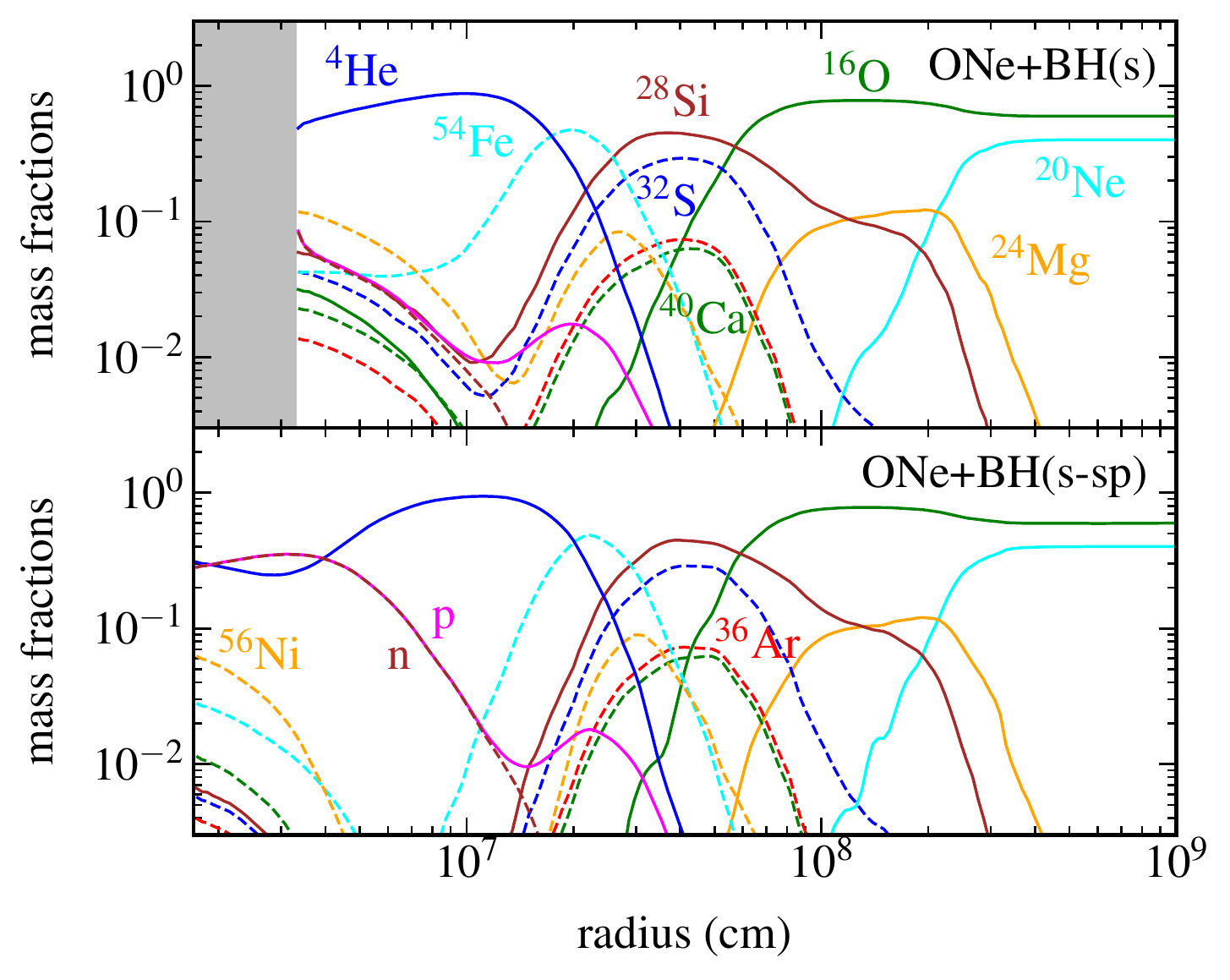}
\caption{Same as Figure~\ref{f:abund_profiles_rns}, but for the two small boundary models
that resolve the BH. The time intervals are the same as for Figure~\ref{f:timeave_profiles_bh}.}
\label{f:abund_profiles_bh}
\end{figure}

The nucleosynthesis profiles of the two BH models are shown in Figure~\ref{f:abund_profiles_bh}.
Differences become prominent inside the radius at which iron
group nuclei start undergoing photodissociation into ${}^{4}$He and nucleons. At smaller
radii, the spinning BH model has a lower abundance of heavy elements compared with the
non-spinning case. Table~\ref{t:results} shows however that the mass fractions in
in the wind are very similar in those models, with the exception of ${}^{16}$O and ${}^{4}$He,
indicating that radii close to the BH do not significantly contribute to the outflow.

\subsection{Comparison with previous work}
\label{s:comparison_1d}

Our results are a significant improvement relative to Paper I. First, our
new large boundary models, which cover a similar range as the simulations in Paper I,  
are evolved for a much longer time. Second, we also include more realistic 
microphysics, in particular an equation of state that accounts for radiation pressure,
and a realistic nuclear reaction network. Finally, we can resolve the dynamics at the 
surface of the central object. The key qualitative difference with the results of 
Paper I is the absence of any detonation in our current models. Accretion proceeds in 
a quasi-steady way, with secular mass ejection on the viscous time of the disk.

\begin{figure}
\includegraphics*[width=\columnwidth]{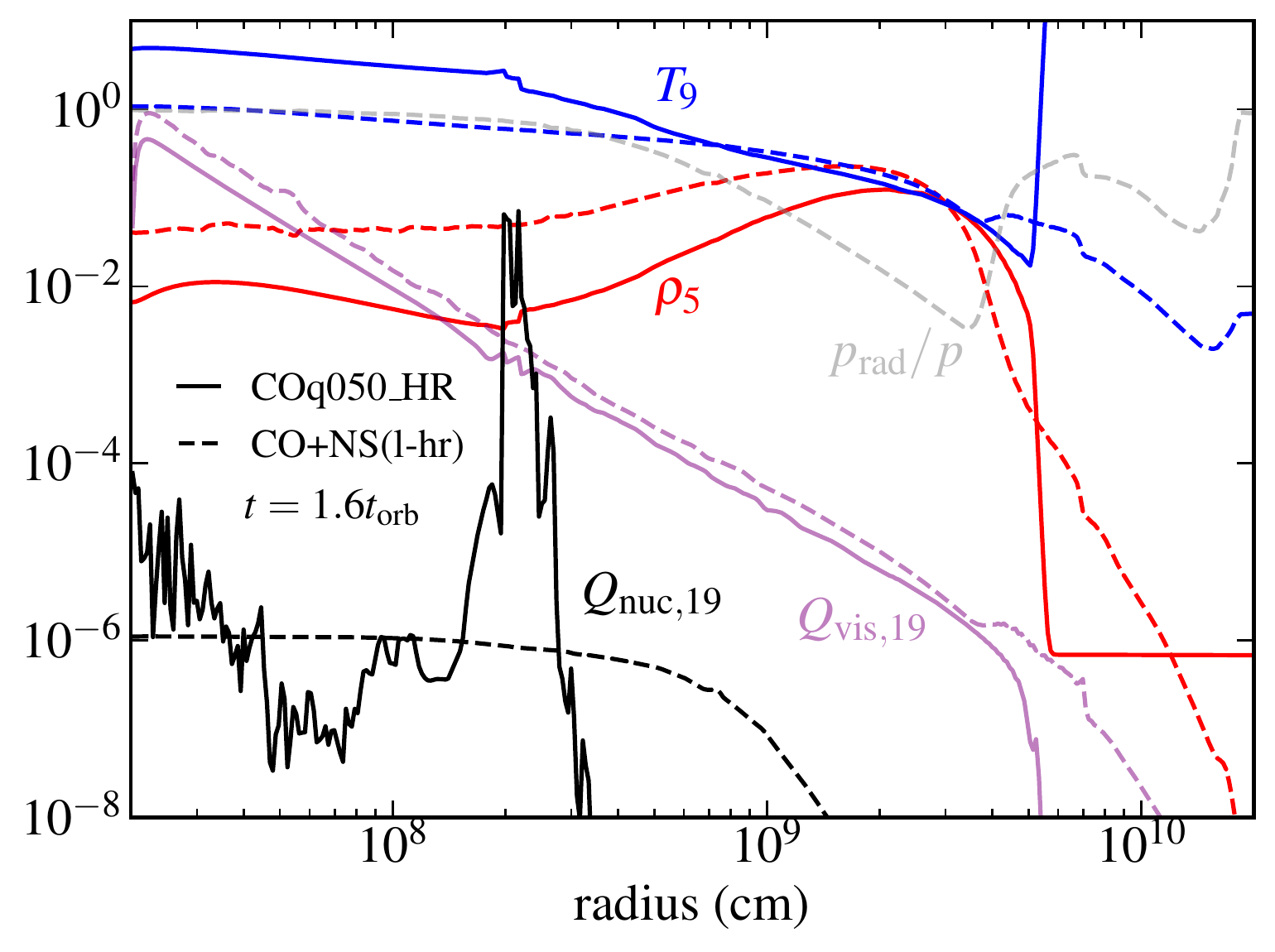}
\caption{Instantaneous equatorial profiles of structural quantities at $t=1.6t_{\rm orb}$
in the fiducial high-resolution model of Paper I ({\tt COq050\_HR}, solid lines) and our fiducial
high-resolution large boundary model ({\tt CO+NS(l-hr)}, dashed lines). 
The symbols have the same meaning as in Figure~\ref{f:timeave_profiles_LB}. The gray
dashed curve shows the ratio of radiation to total pressure in model {\tt CO+NS(l-hr)}.
The time is close to the onset of a detonation in model {\tt COq050\_HR}.}
\label{f:plot_old_nudaf}
\end{figure}

Figure~\ref{f:plot_old_nudaf} compares instantaneous equatorial profiles of key
quantities in the fiducial high-resolution model of Paper I ({\tt COq050\_HR}, used
in Figures 1-3 of that paper) and
in our high-resolution large-boundary WD+NS model {\tt CO+NS(l-hr)}, at a time
shortly before a detonation occurs in the former. Both models employ the same
equatorial resolution, domain size, torus parameters, central object mass, and boundary
conditions. The model from Paper I
assumes an ideal gas equation of state and point mass gravity, uses a different prescription
for the viscosity (proportional to density, as in \citealt{stone1999}), and
uses a single power-law nuclear reaction calibrated to match 
$^{12}$C($^{12}$C,$\gamma$)$^{24}$Mg. 
While our new model evolves somewhat faster due to the different 
viscosity, the profiles of viscous heating differ by less than a factor of $2$.
The key difference is the temperature profile, which differs by a factor
$4$ at the radius where most of the nuclear burning occurs in the model
of Paper I. The temperature profile in model {\tt CO+NS(l-hr)} is shallower
at small radius, which is a consequence of radiation pressure being dominant
at this location, as shown in Figure~\ref{f:plot_old_nudaf}. The burning rates 
correspondingly differ by several orders of magnitude.

A separate question is whether detonations that should be occurring
are not resolved in our current models. The mean accretion flow is such that burning
fronts are spread out over distances comparable to the local radius 
(e.g., Figure~\ref{f:timeave_abund_snapshots}),
so no sudden releases of energy occur given that the radial accretion speed is subsonic. 
The turbulent r.m.s. Mach number around $r\sim 10^8$\,cm is $\mathcal{M}_{\rm turb}\lesssim 0.3$ 
in model {\tt CO+NS(l-hr)}, which implies fractional temperature fluctuations of
$\mathcal{M}_{\rm turb}/3\lesssim 10\%$ 
if radiation pressure dominates. Figure~\ref{f:timeave_profiles_LB} 
shows that stochastic fluctuations in the burning 
rate are at most comparable to the viscous heating rate during peak accretion, when 
the density is the highest. The viscous heating timescale is itself a factor
$\sim 10$ lower than the sound crossing time, thus nuclear burning is far from being
able to increase the internal energy faster than the pressure can readjust the material.
Settling the question of whether detonations occur during the initial establishment
of the accretion flow to the central object will require simulations that employ
magnetic fields to transport angular momentum and that fully resolve turbulence.

\citet{zenati_2019} carried out 2D hydrodynamic simulations starting from an equilibrium
torus, and employing a nuclear reaction network, a realistic EOS, and self-gravity. 
They report weak detonations in all of their models excluding He WDs, followed 
by an outflow dominated by the initial WD composition, with an admixture of heavier elements. 
While we find the same type of outflow velocities and composition, our results 
differ in that we do not find any detonation in our models, weak or strong, even in the
case of a hybrid CO-He WD with an admixture of He. This difference might be in part 
due to resolution, as their finest grid size (in cylindrical geometry) is 1\,km. This 
is comparable to the resolution of our models at $r=10^7$\,cm, but 
coarser in the vicinity of the NS (our grid is logarithmic in radius, and on the midplane we have 
$\Delta r /r \simeq 0.037\simeq \Delta\theta \simeq 2^\circ$). 
Our results are consistent with 
those of \citet{fryer1999}, which found that nuclear burning was energetically
unimportant during disk formation.

\begin{figure}
\includegraphics*[width=\columnwidth]{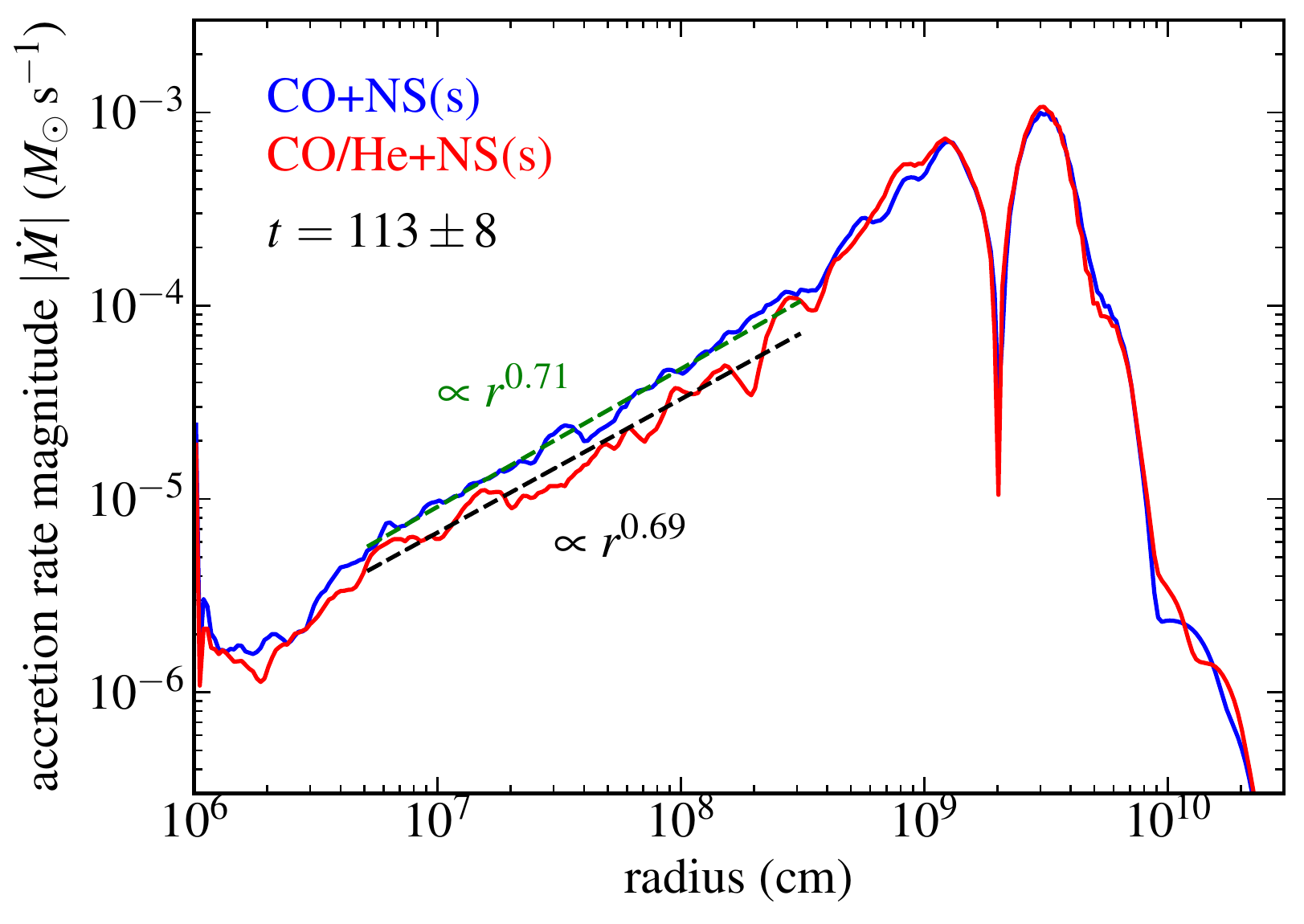}
\caption{Time- and angle-averaged mass flow rate (absolute value) for the fiducial small boundary
WD+NS model and its hybrid counterpart. The angle-average is taken within $30$\,deg of the equatorial
plane. The dashed lines show power-law fits to the radial dependence of the accretion rate.}
\label{f:accretion_exponents_rns}
\end{figure}

The time-averaged profiles in the disk equatorial plane show very close similarity
to the 1D results of \citet{M12} and MM16. Figure~\ref{f:accretion_exponents_rns}
shows the profile of absolute value of the radial mass flow rate for the fiducial and hybrid
small boundary models. A power-law fit to the radial dependence of the accretion rate
$\dot{M}\propto r^p$ yields $p\simeq 0.7$ except in the vicinity of the NS and where the disk has not yet
reached steady accretion. In the mass-loss model of MM16, this corresponds
to disk outflow velocities comparable to the local Keplerian speed $v_{\rm K}$. This is
consistent with the velocity distribution of the outflow (Figure~\ref{f:hist_vel_def_LB-SB}),
which shows an upper limit comparable to the Keplerian speed
at the innermost radius where ${}^{56}$Ni is produced (Figure~\ref{f:profiles_spec_time}).
We also find a characteristic power-law decline of the outflow rate with time after
peak accretion has been achieved (Figure~\ref{f:evolution_LB}). The radial dependence of the
mass fractions is remarkably similar to that of MM16 (c.f. their Figure~5),
although the radial position of our burning fronts evolves more slowly (compare our 
Figure~\ref{f:profiles_spec_time} with their Figure~6). This is a consequence of our fiducial
model using a lower viscosity parameter ($0.03$) than their fiducial case ($0.1$).

\section{Observational Implications}
\label{s:observations}

The outflow from the accretion disk should generate an
electromagnetic transient that rises over a few day timescale and reaches
a peak luminosity $\sim 10^{40}$\,erg\,s$^{-1}$ if powered only by radioactive decay. 
We can estimate this rise time and peak
luminosity from the velocity distribution of ejected ${}^{56}$Ni
and our estimate of the total ejecta from the disk 
(equation~\ref{eq:mass_ejection_power-law}, Table~\ref{t:results}).

\begin{figure}
\includegraphics*[width=\columnwidth]{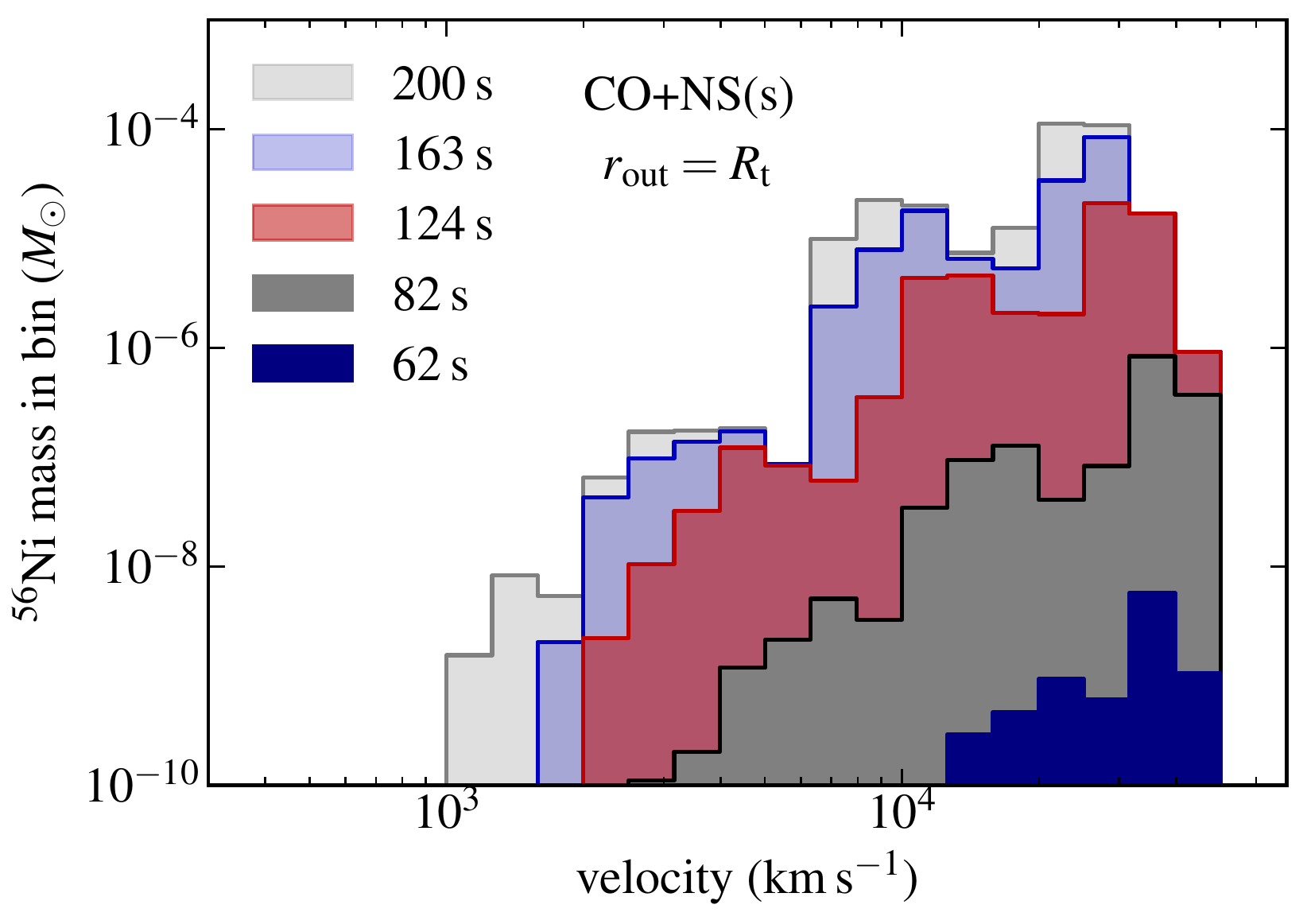}
\caption{Mass histograms of unbound ${}^{56}$Ni as a function of radial velocity for model
{\tt CO+NS(s)}, measured at $r_{\rm out}=R_t$ and at the labeled times.}
\label{f:hist_ni56_vel}
\end{figure}

Figure~\ref{f:hist_ni56_vel} shows the velocity distribution of
ejected ${}^{56}$Ni at various times in the fiducial small-boundary WD+NS model, 
measured at the initial torus radius, which is
$\sim 100$ times larger than the radius at which ${}^{56}$Ni is produced (Figure~\ref{f:profiles_spec_time}). The
average velocity decreases as a function of time, which means that on average,
faster material is ejected before slower material and therefore resides at larger radii, 
even if mixing takes place\footnote{At late times, the burning fronts are expected to recede 
(MM16) which should increase the average speed of ejecta again. However, this is 
not expected to be a dominant contribution to the total ${}^{56}$Ni mass ejected.}.
Ignoring corrections due to the geometric collimation 
of the outflow, this stratification in radius and velocity means that radiation escapes
from faster layers first. In our estimates, we therefore consider the cumulative mass 
starting from the highest velocity,
\begin{equation}
\label{eq:mass_velocity_cum}
M_i(>v) = \int_{v}^{v_{\rm max}}\frac{dM_i}{dv}\,dv.
\end{equation}
where the subscript $i$ stands for either total mass or ${}^{56}$Ni mass.
Note that this is a lower limit on the velocity, since thermal energy can be
converted to kinetic via adiabatic expansion.

The time for radiation to escape from a layer with total mass $M_{\rm tot}(>v)$ is
given by \citep{arnett_1979}
\begin{equation}
\label{eq:tpeak_def}
t_{\rm pk}(>v) = \left[\frac{3\kappa M_{\rm tot}(>v)}{4\pi c v} \right]^{1/2}.
\end{equation}
In evaluating equation~(\ref{eq:tpeak_def}), we adopt $\kappa = 0.05$\,g\,cm$^{-3}$
for a Fe-poor mixture (MM16), and obtain $M_{\rm tot}(>v)$ by 
renormalizing the ${}^{56}$Ni mass distribution by a conservative total ejecta mass 
of $0.4M_\odot$ (Table~\ref{t:results}). To estimate uncertainties, we also compute
this mass by re-normalizing the total (not just ${}^{56}$Ni) velocity distribution
to the same total ejected mass.

The luminosity of a layer with ${}^{56}$Ni mass $M_{\rm Ni}(>v)$ at time $t=t_{\rm pk}(>v)$ is
\begin{eqnarray}
\label{eq:luminosity_pk}
L_{\rm pk}(>v) & = &M_{\rm Ni}(>v)\left[\dot{Q}_{\rm Ni}(t_{\rm pk})+\dot{Q}_{\rm Co}(t_{\rm pk})\right]
\end{eqnarray}
where the specific nuclear heating rates from ${}^{56}$Ni and ${}^{56}$Co decay are
\begin{eqnarray}
\dot{Q}_{\rm Ni}(t) & = & \frac{\Delta E_{\rm Ni}}{m_{\rm Ni}\tau_{\rm Ni}}\,e^{-t/\tau_{\rm Ni}}\nonumber\\
& \simeq & 4.8\times 10^{10}\textrm{\,[erg\,g}^{-1}\textrm{\,s}^{-1}]\,e^{-t/\tau_{\rm Ni}}\\
\dot{Q}_{\rm Co}(t) & = & \frac{\Delta E_{\rm Co}}{m_{\rm Co}}
                          \frac{(\tau_{\rm Ni}\tau_{\rm Co})^{-1}}{(1/\tau_{\rm Ni}-1/\tau_{\rm Co})}\nonumber\\
& \simeq & 8.9\times 10^{9}\textrm{\,[erg\,g}^{-1}\textrm{\,s}^{-1}]\,\left(e^{-t_{\rm pk}/\tau_{\rm Co}} - e^{-t_{\rm pk}/\tau_{\rm Ni}}\right),\nonumber\\
\end{eqnarray}
with $\left\{\tau_{\rm Ni},\tau_{\rm Co}\right\}\simeq \left\{8.8,111\right\}$\,d the mean lifetimes
and $\left\{\Delta E_{\rm Ni},\Delta E_{\rm Co}\right\}\simeq \left\{2.1,4,6\right\}$\,MeV the decay
energies of ${}^{56}$Ni and ${}^{56}$Co, respectively. Equation~(\ref{eq:luminosity_pk}) assumes that
the gamma-rays from radioactive decay are thermalized with $100\%$ efficiency. The total ${}^{56}$Ni
mass is obtained by scaling the ejected distribution to a (conservative) estimate of $10^{-3}M_\odot$,
which is somewhat smaller than the ejected fraction times total ejected mass for this model (Table~\ref{t:results}).

\begin{figure}
\includegraphics*[width=\columnwidth]{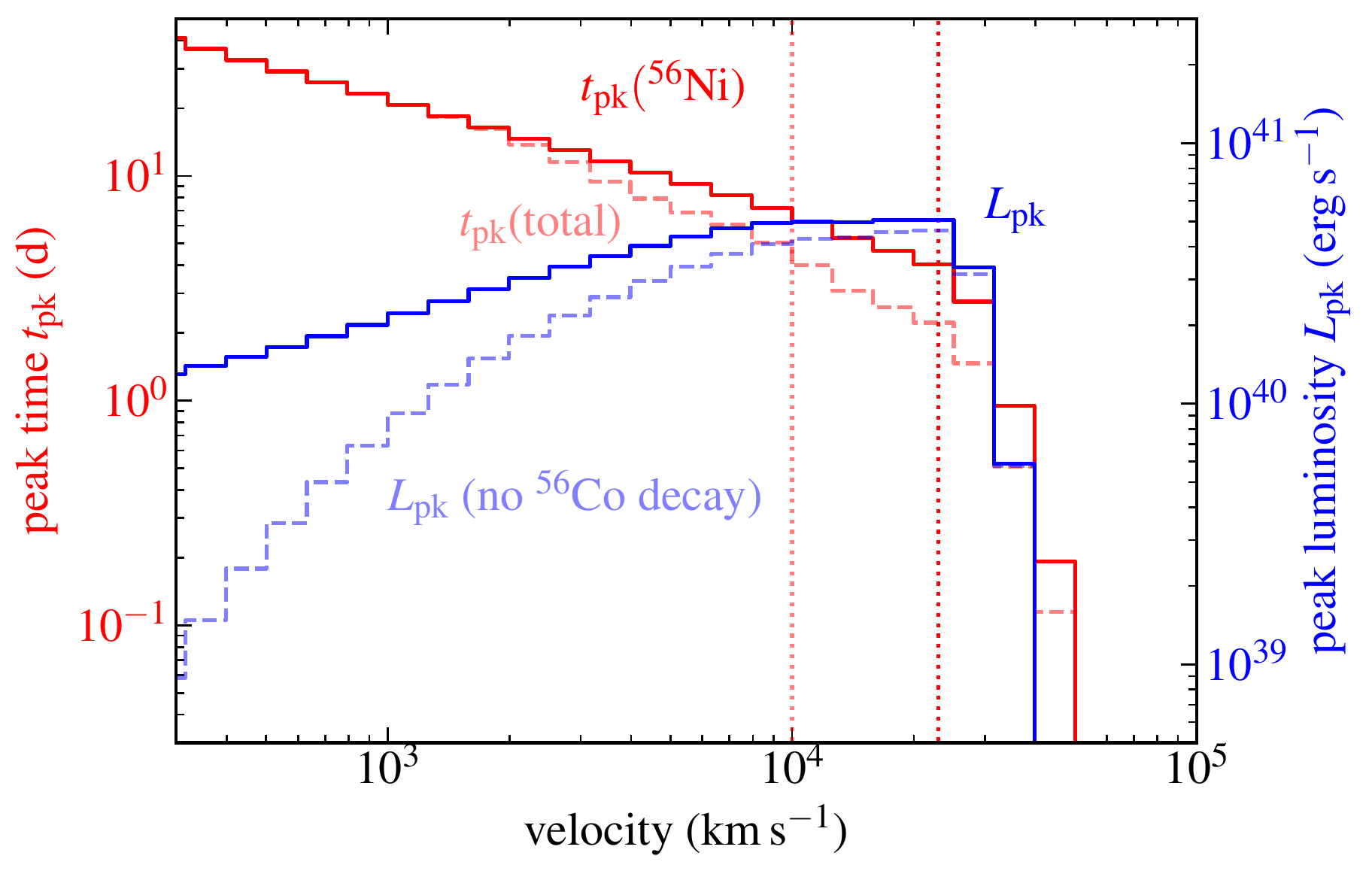}
\caption{Time to reach peak emission (equation~\ref{eq:tpeak_def}, red) and 
peak luminosity (equation~\ref{eq:luminosity_pk}, blue) for material with velocity larger
than a given value (equation~\ref{eq:mass_velocity_cum}) in model {\tt CO+NS(s)}. Solid red shows $t_{\rm pk}$
computed using the ${}^{56}$Ni velocity distribution at $t=200$\,s (Figure~\ref{f:hist_ni56_vel}) renormalized
to $0.4M_\odot$, while the dashed red line shows the same calculation but with the (renormalized) total
velocity distribution. The vertical dotted lines show the corresponding mass-weighted average velocities.
The dashed blue line shows $L_{\rm pk}$ without the contribution from ${}^{56}$Co heating 
in equation~(\ref{eq:luminosity_pk}). For the peak luminosity, we assume an initial ${}^{56}$Ni mass of $10^{-3}M_\odot$.}
\label{f:rise_time_ni56}
\end{figure}

\begin{figure}
\includegraphics*[width=\columnwidth]{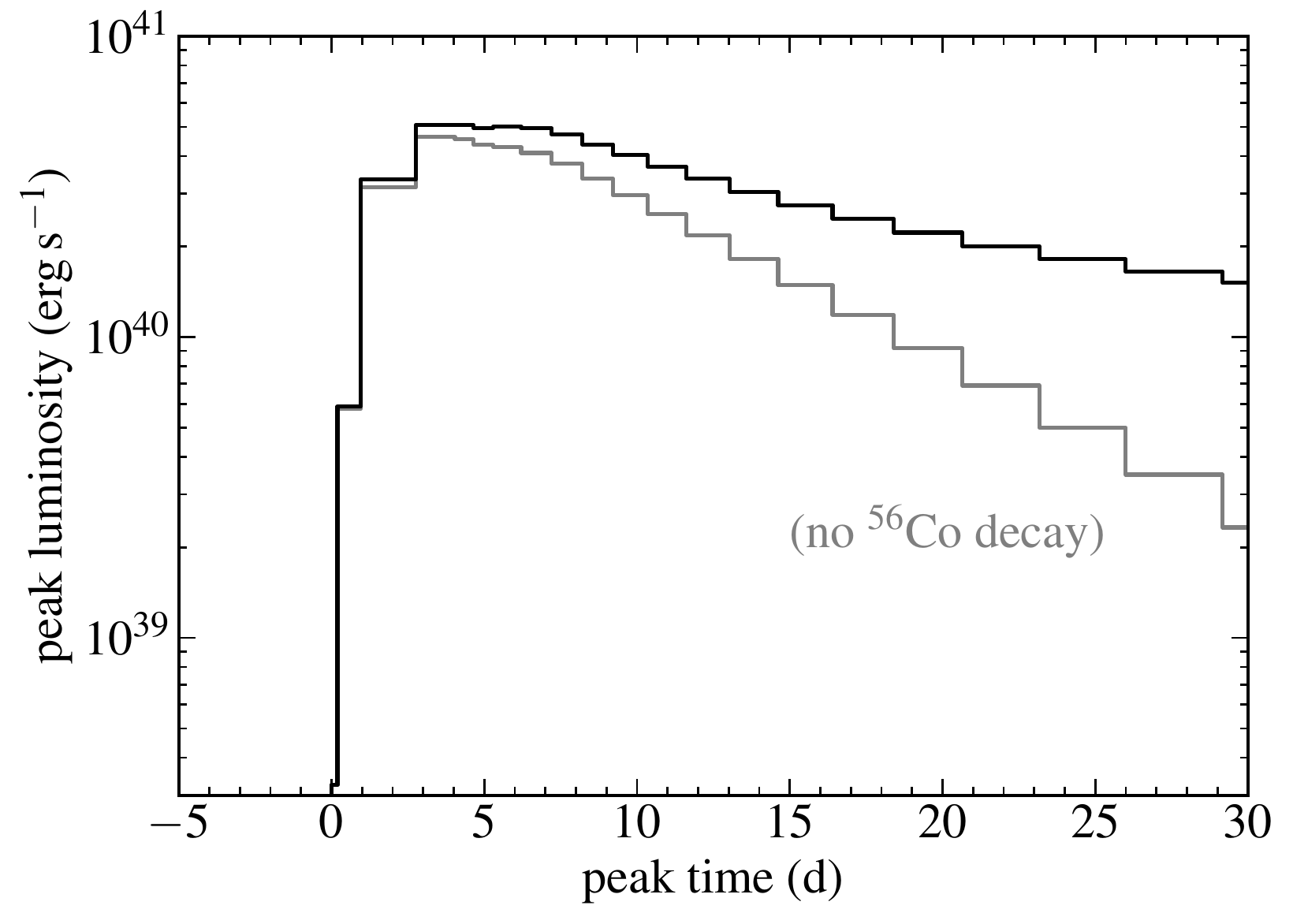}
\caption{Peak luminosity (equation~\ref{eq:luminosity_pk}) as a function of peak time 
(equation~\ref{eq:tpeak_def}) for material with velocity larger than a given value 
(equation~\ref{eq:mass_velocity_cum}) for model {\tt CO+NS(s)}. This is a rough approximation to the light curve
of the radioactively-powered transient expected from the disk outflow. The reverse-cumulative
mass distribution for the peak time is obtained by integrating the ${}^{56}$Ni mass distribution 
(Figure~\ref{f:hist_ni56_vel}) and re-normalizing by the asymptotic total mass ejected ($\sim 0.4M_\odot$),
while the luminosity is obtained by renormalizing the distribution to $10^{-3}M_\odot$ of ${}^{56}$Ni.
The gray line shows the peak luminosity without the contribution of ${}^{56}$Co to estimate the 
uncertainty range from our assumption of complete gamma-ray thermalization.}
\label{f:lpk_tpk_ni56}
\end{figure}

Figure~\ref{f:rise_time_ni56} shows $t_{\rm pk}(>v)$ and $L_{\rm pk}(>v)$ as a function
of outflow velocity for the fiducial small-boundary WD+NS model. The rise time to peak from half
maximum is in the range $2-3$\,d depending on whether the ${}^{56}$Ni or total velocity
distributions are used, and the rise time to the mass-averaged velocity is the same. The peak
luminosity is a few times $10^{40}$\,erg\,s$^{-1}$. This
value can increase by a factor $10$ if the ${}^{56}$Ni yield is on the higher end
of our estimates, $10^{-2}M_\odot$, coming closer to normal supernova luminosities.
A rough approximation to the light curve can be obtained by plotting $L_{\rm pk}(>v)$ versus $t_{\rm pk}(>v)$,
which is shown in Figure~\ref{f:lpk_tpk_ni56}. 

The short rise time suggests a connection to previously found rapidly-evolving
blue transients (e.g., \citealt{drout_2014,rest_2018,chen_2019}) but with much lower
luminosities. It is possible that the ejecta from the disk collides with
material previously ejected in a stellar wind by one or both of the progenitors 
of the WD and/or NS/BH, resulting in enhanced emission relative to our 
simple estimates based on radioactive heating. Another way to enhance the
luminosity above that from radioactive decay is through accretion power 
(e.g., \citealt{dexter_2013}, MM16). Extrapolating the 
accretion rate in Figure~\ref{f:mdot_LB} for model {\tt CO+NS(s)} to $t\simeq t_{\rm peak}\simeq 3$\,d,
yields $\sim 10^{43}$\,erg\,s$^{-1}$ for a $10\%$ thermalization efficiency.

In addition to powering a supernova-like transient from the unbound ejecta, we
speculate that the inner parts of the accretion flow (near the central NS or
BH) could generate a relativistic jet similar to those in gamma-ray bursts
(e.g., \citealt{fryer1999,King+07}).  We obtain peak accretion rates onto the central compact
object of $\sim 10^{-6}-10^{-3}M_\odot$\,s$^{-1}$, with a peak timescale of tens
to hundreds of seconds (Figure~\ref{f:mdot_LB}). Assuming a jet launching efficiency of $\epsilon_{\rm j} \lesssim 0.1$, 
the peak jet power could therefore be $\epsilon_j\dot{M}c^2 \lesssim 10^{47} - 10^{50}$\,erg\,s$^{-1}$.  
While these characteristic luminosities (timescales) are somewhat too
low (long) compared to the majority of long-duration gamma-ray bursts, they may
be compatible with other high energy transients.  For instance, \citet{xue_2019} recently
discovered an X-ray transient, CDF-S XT2, with {\it Chandra} with a peak
isotropic luminosity $L_{\rm X} \sim 3\times 10^{45}$\,erg\,s$^{-1}$ and peak duration $\sim 10^3$\,s.
The late-time decay of the X-ray luminosity, with a time exponent $2.16^{+26}_{-29}$, is in broad
agreement with the decay rate of the accretion rate in our models (Figure~\ref{f:mdot_LB}).
While peak accretion in our models occurs somewhat earlier, this peak time is tied to
how angular momentum transport is modeled.
The host galaxy and spatial offset of CDF-S XT2 from its host, while consistent with those of
NS-NS mergers, would plausibly also be consistent with the older stellar
populations that can host WD-NS/BH mergers.

WD-NS mergers have also been discussed as a possible formation channel of
pulsar planets \citep{Phinney&Hansen93, Podsiadlowski93, margalit_2017}.
Using a semi-analytic model extending the torus evolution to $\sim$kyr post
merger, \citet{margalit_2017} found that conditions conducive to formation
of planetary bodies consistent with the B1257+12 pulsar planets
\citep{Wolszczan&Frail92, Wolszczan94, Konacki&Wolszczan03} can be achieved for
sufficiently low values of the alpha viscosity parameter $\alpha$ and accretion
exponent $p$. The index of $p \sim 0.7$ we find in our current work 
(Figure~\ref{f:accretion_exponents_rns}) is somewhat
higher than that required to obtain significant mass at the location of the planets
and to spin-up the NS to millisecond periods, however this is subject to several
uncertainties. Simulations of radiatively-inefficient accretion disks typically
find $p \sim 0.4-0.8$ depending on the physics (hydrodynamic vs MHD
simulations), the value of the alpha viscosity parameter, and the initial
magnetic field \citep[e.g.][]{yuan2012,Yuan&Narayan14}, while observations of
Sgr A* suggest even lower values, $p \sim 0.3$ \citep{Yuan&Narayan14} (although
the physical accretion regime of Sgr A* is very different than the WD-NS merger
accretion disks considered here). Whether or not some of the matter expelled
from the disk midplane remains bound and eventually circulates back is also not
entirely resolved and can increase the remaining disk mass at late times,
increasing the viability of the WD-NS merger pulsar-planet formation scenario.

\section{Summary and Discussion}
\label{s:discussion}

We have carried out two-dimensional axisymmetric, time-dependent simulations of accretion 
disks formed during the (quasi-circular) merger of a CO or ONe WD by a NS or BH. Our models
include a physical equation of state, viscous angular momentum transport, self-gravity, and a coupled $19$-isotope 
nuclear reaction network. We studied both the long-term mass ejection 
from the disk, by excluding the innermost regions, and fully 
global models that resolve the compact object but which can only be evolved
for shorter than the viscous timescale of the disk. Our main results are the following:
\newline

\noindent 1. In all of the models we study, accretion and mass ejection proceed in a quasi-steady 
             manner on the viscous time, with no detonations. Nuclear energy generation is at most 
             comparable to viscous heating (Figures~\ref{f:timeave_profiles_LB}, \ref{f:timeave_profiles_resolution},
             \ref{f:timeave_profiles_rns}, and \ref{f:timeave_profiles_bh}).
             \newline

\noindent 2. The radiatively-inefficient character of the disk results in vigorous outflows.
             At least $50\%$ of the initial torus should be ejected in the wind (Figure~\ref{f:evolution_LB}
             and Table~\ref{t:results}). The velocity distribution of this outflow is broad,
             covering the range $10^2-10^4$\,km\,s$^{-1}$ (Figures~\ref{f:evolution_LB} 
             and \ref{f:hist_vel_def_LB-SB}). The outflow is concentrated within a
             cone of $\sim 40$\,deg from the rotation axis (Figure~\ref{f:evolution_LB} and
             \ref{f:hist_ang_def_LB-SB}).
             Energy losses due to photodissociation and thermal neutrino emission become important 
             only near the central compact object, with neutrino emission from electron/positron capture
             onto nucleons being sub-dominant (Figure~\ref{f:timeave_profiles_rns} 
             and \ref{f:timeave_profiles_bh}).
             \newline

\noindent 3. Our models can capture the burning of increasingly heavier elements, as
             accretion proceeds to large radii, all the way to the iron group elements
             and its subsequent photodissociation into ${}^{4}$He and nucleons 
             (Figures~\ref{f:timeave_abund_snapshots}, \ref{f:abund_profiles_SB},  
             \ref{f:abund_profiles_rns} and \ref{f:abund_profiles_bh}).
             The outflow composition is dominated by that of the initial WD,
             with burning products accounting for $10-30\%$ by mass (Table~\ref{t:results}).
             Based on the mass fractions of elements in the wind and the
             ejecta masses from large boundary models, we estimate that $10^{-3}-10^{-2}M_\odot$ 
             of ${}^{56}$Ni should be produced generically by these disk outflows. The
             wind composition is relatively robust to variations in the disk viscosity, 
             rotation rate of the NS or BH, and spatial resolution. No significant neutronization
             (and thus $r$-process production) is expected from our models.
             \newline

\noindent 4. Two predictions from our results are that (1) the average velocities of burning products
             generated at smaller radii are higher (i.e., helium and iron should be faster
             on average that Mg or Si; Figure~\ref{f:hist_vel_def_LB-SB}), 
             and that (2) these burning products 
             should be (on average) concentrated closer to the rotation axis than 
             lighter elements (Figure~\ref{f:hist_ang_def_LB-SB}).
             \newline

\noindent 5. Based on the ejecta mass and velocity, we estimate that the resulting transients
	     should rise to their peak brightness within a few days (Figures~\ref{f:rise_time_ni56}).
             When including only heating due to radioactive decay of ${}^{56}$Ni (and ${}^{56}$Co) generated
             in the outflow, we obtain peak bolometric luminosities in the range 
             $\sim 10^{40}-10^{41}$\,erg\,s$^{-1}$.
             This luminosity can be enhanced by circumstellar interaction or late-time accretion
             onto the central object (Figure~\ref{f:mdot_LB}), potentially accounting for 
             the properties of rapidly-evolving blue transients (\S\ref{s:observations}). 
             The generation of a relativistic jet by accretion onto the central object could also 
             account for X-ray transients such as CDF-S XT2.
             \newline

The main improvement to be made in our models is the replacement of a viscous
stress tensor for full magnetohydrodyanmic (MHD) modeling. Comparison between hydrodynamic 
and MHD models (with initial poloidal geometry) of accretion disk from NS-NS/BH mergers shows close similarity
in the ejection properties of the thermal outflows in the radiatively-inefficient phase; this
thermal component is the entirety of the
wind in hydrodynamics, but only a subset of the ejection when MHD is included \citep{fernandez2019}.
While in principle such magnetized disks can generate jets, the disruption of the
WD will leave a significant amount of material along the rotation axis, which can 
pose difficulties for launching relativistic outflows.

The second possible improvement is using realistic initial conditions obtained
from a self-consistent simulation of unstable Roche lobe overflow. Since the thermodynamics
of the disk become quickly dominated by heating from angular momentum transport
and nuclear reactions, it is not expected that the details of the initial disk
thermodynamics will have much incidence on the subsequent dynamics except if
(1) nuclear burning becomes important during the disruption process itself,
as expected for a ONe WD merging with a NS (e.g., \citealt{M12}) or (2)
if the magnetic field configuration post merger (which should be mostly toroidal, in analogy with NS-NS mergers)
generates significant deviations from the evolution obtained with viscous hydrodynamics.

The evolution of disks from He WDs around NS or BHs is expected to be more sensitive to the
choice of parameters such as the disk entropy, mass, and viscosity parameter (MM16). 
We therefore leave simulations of such systems for future work.

\section*{Acknowledgments}

We thank Craig Heinke for helpful discussions, and the anonymous referee for constructive comments.
RF acknowledges support from the National Science and Engineering Research Council (NSERC) 
of Canada and from the Faculty of Science at the University of Alberta. 
BM is supported by the U.S. National Aeronautics and Space Administration (NASA) through the 
NASA Hubble Fellowship grant $\#$HST-HF2-51412.001-A awarded by the Space Telescope Science Institute, 
which is operated by the Association of Universities for Research in Astronomy, Inc., for NASA, 
under contract NAS5-26555. BDM is supported in part by NASA through the Astrophysics Theory 
Program (grant number $\#$NNX17AK43G).
The software used in this work was in part developed by the DOE
NNSA-ASC OASCR Flash Center at the University of Chicago.
This research was enabled in part by support
provided by WestGrid (www.westgrid.ca), the Shared Hierarchical 
Academic Research Computing Network (SHARCNET, www.sharcnet.ca), and Compute Canada (www.computecanada.ca).
Computations were performed on \emph{Graham} and \emph{Cedar}. 
This research also used compute and stoage resources of the U.S. National Energy Research Scientific Computing
Center (NERSC), which is supported by the Office of Science of the U.S. Department of Energy
under Contract No. DE-AC02-05CH11231. Computations were performed in \emph{Edison} 
(repository m2058).
Graphics were developed with {\tt matplotlib} \citep{hunter2007}.


\appendix

\section[]{Implementation of Self-Gravity}
\label{s:self_gravity_appendix}

We implement self-gravity in spherical coordinates using the
multipole algorithm of \citet{MuellerSteinmetz1995}. While {\tt FLASH3} includes
a version of this algorithm, it is not optimized for non-uniform
axisymmetric spherical grids. Here we provide a brief description of our customized 
implementation and tests of it.

The truncated multipole expansion of the gravitational potential $\Phi$ 
in axisymmetry is
\begin{equation}
\label{eq:multipole_expansion}
\Phi(r,\theta) = -2\pi G\sum_{\ell=0}^{\ell_{\rm max}}\,P_\ell(\cos\theta)
\left[\frac{1}{r^{\ell+1}}C_\ell(r) + r^\ell D_\ell(r) \right],
\end{equation}
where $P_\ell$ is the Legendre polynomial of index $\ell$, and the radial
density moments are given by
\begin{eqnarray}
\label{eq:C_sg_moment_def}
C_\ell(r) & = & \int_0^\pi \sin\theta d\theta P_\ell(\cos\theta)\int_0^r dR\,R^{2+\ell}\rho(R,\theta)\\
\label{eq:D_sg_moment_def}
D_\ell(r) & = & \int_0^\pi \sin\theta d\theta P_\ell(\cos\theta)\int_r^\infty dR\,R^{1-\ell}\rho(R,\theta).
\end{eqnarray}
Equation~(\ref{eq:multipole_expansion}) is an exact solution of Poisson's equation (equation~\ref{eq:poisson}) when 
$\ell_{\rm max}\to\infty$. In practice, the sum needs to be truncated at some finite $\ell_{\rm max}$, 
the optimal value of which is
problem-dependent \citep{MuellerSteinmetz1995}. The main computational work involves calculation 
of the moments $C_\ell(r)$ and $D_\ell(r)$.

In the \citet{MuellerSteinmetz1995} algorithm, the integrals in 
equations~(\ref{eq:C_sg_moment_def})-(\ref{eq:D_sg_moment_def}) are first replaced by sums
of integrals inside each computational cell. One then assumes that
the density varies smoothly within a cell, thus decoupling the angular integral of Legendre
polynomials, the radial integral of the weight, and the density. The angular integral can be calculated exactly
from recursion relations of these polynomials, while the radial integral can be solved 
analytically. For a cell with indices $(i,j)$, the moments are thus
\begin{eqnarray}
\label{eq:C_sg_discrete}
C^{ij}_\ell & = & \sum_{q=i}^{i_{\rm max}}\sum_{j=1}^{j_{\rm max}} 
\left[\int_{\theta_{j-1/2}}^{\theta_{j+1/2}} \sin\theta d\theta P_\ell(\cos\theta)\right]\nonumber\\
            & &  \qquad\times\,\frac{\rho_{ij}}{3+\ell}\left(r_{q+1/2}^{3+\ell} - r_{q-1/2}^{3+\ell} \right)\\
\label{eq:D_sg_discrete}
D^{ij}_\ell & = & \sum_{q=1}^{i}\sum_{j=1}^{j_{\rm max}} 
\left[\int_{\theta_{j-1/2}}^{\theta_{j+1/2}} \sin\theta d\theta P_\ell(\cos\theta)\right]\nonumber\\
            & &  \times\rho_{ij}\left\{\begin{array}{l}
                 \left(r_{q+1/2}^{2-\ell} - r_{q-1/2}^{2-\ell} \right)/(2-\ell)\quad (\ell\ne 2)\\
		 \noalign{\smallskip}
                 \ln{\left(r_{q+1/2}/r_{q-1/2}\right)}\qquad\qquad\phantom{a}(\ell=2)
                 \end{array}\right.
\end{eqnarray}
where half-integer indices denote cell edges, and $\rho_{ij}$ is the volume-averaged density of the cell.

The angular and radial integrals are computed once at the beginning of the simulation. To improve
accuracy, \citet{MuellerSteinmetz1995} recommend computing the sum in equation~(\ref{eq:C_sg_discrete}) 
from small radii to large radii, and vice-versa for equation~(\ref{eq:D_sg_discrete}). In practice,
in our non-uniform grid implementation, the domain is spatially decomposed by compute core. The sums
are first computed locally within each core, and then each core broadcasts the total of the sum within itself
to the others. Finally, global cumulative sums are constructed locally with the information from 
all other cores. 
The final gravitational potential is computed by adding the contribution of 
the point mass from the neutron star (equation~\ref{eq:poisson}). Overall, the gravity solver
ads a cost of approximately $50\%$ that of the hydrodynamic solver. The latter is comparable
or smaller than the cost of the nuclear reaction network, therefore the inclusion of self-gravity
is only a moderate addition to the computational budget.

\begin{figure}
\includegraphics*[width=\columnwidth]{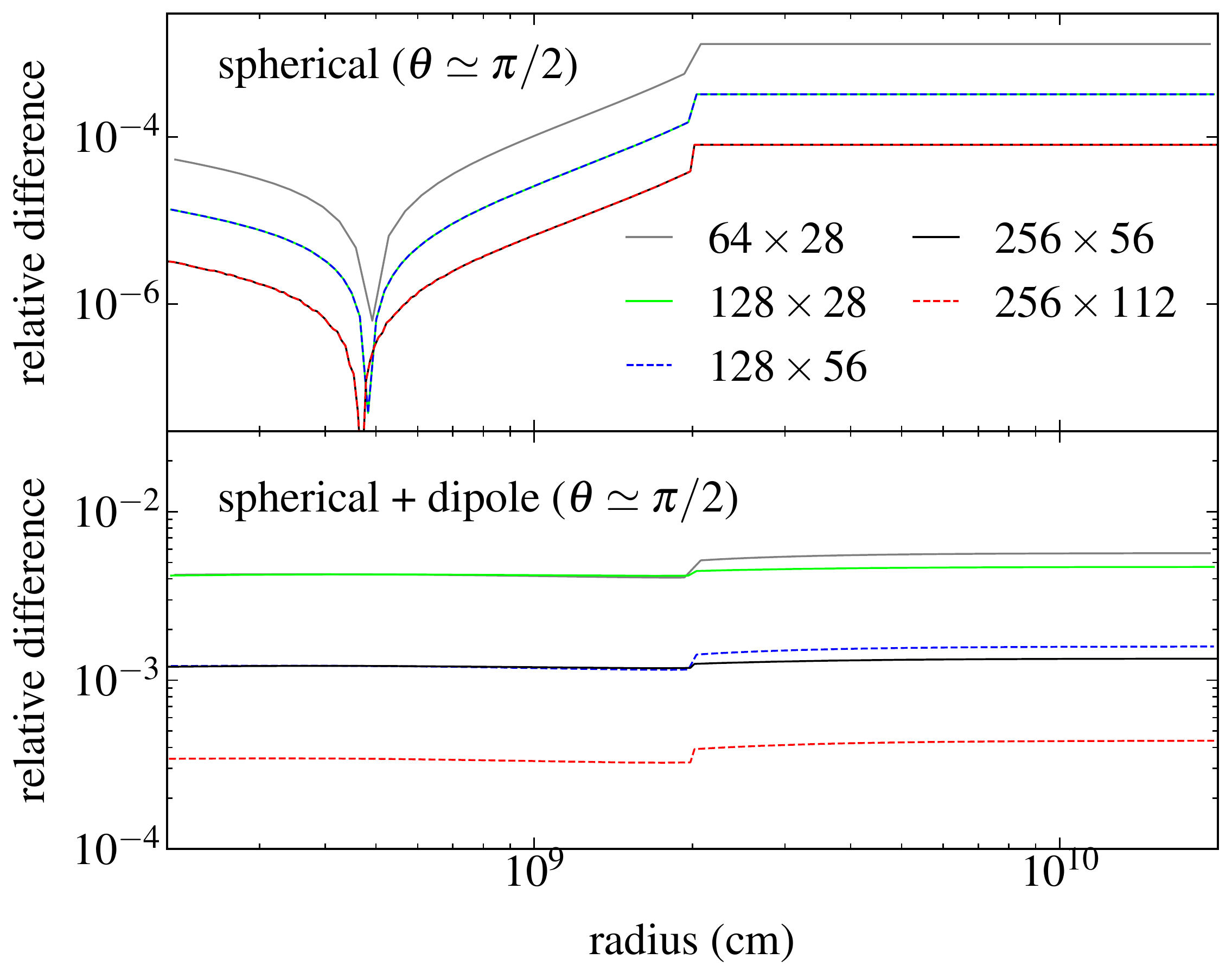}
\caption{Numerical test of the multipole solver. Shown as a function of radius 
is the fractional difference between the numerical solution and the analytic potential in 
equation~(\ref{eq:solution_sg-test}), when initializing the domain with the density 
profile of equation~(\ref{eq:density_profile_sg-test}). The top panel restricts the
density profile to the spherical component only ($P_2 = 0$), while the bottom panel
uses both spherical and dipolar components (the multipole solver allows $\ell \leq \ell_{\rm max}=12$).
The resolutions shown correspond to grids logarithmically spaced in radius and equispaced 
in $\cos\theta$, extending over the full range of polar angles, and an inner boundary
at $r_{\rm in}=2\times 10^8$\,cm and an outer boundary $100$ times larger. The parameters of the
analytic solution are $\rho_0 = 10^7$\,g\,cm$^{-3}$ and $R = 2\times 10^8$\,cm. Curves
are computed for a single angular direction (cell center adjacent to equatorial plane from above).}
\label{f:sg_moments_test}
\end{figure}

\begin{figure}
\includegraphics*[width=\columnwidth]{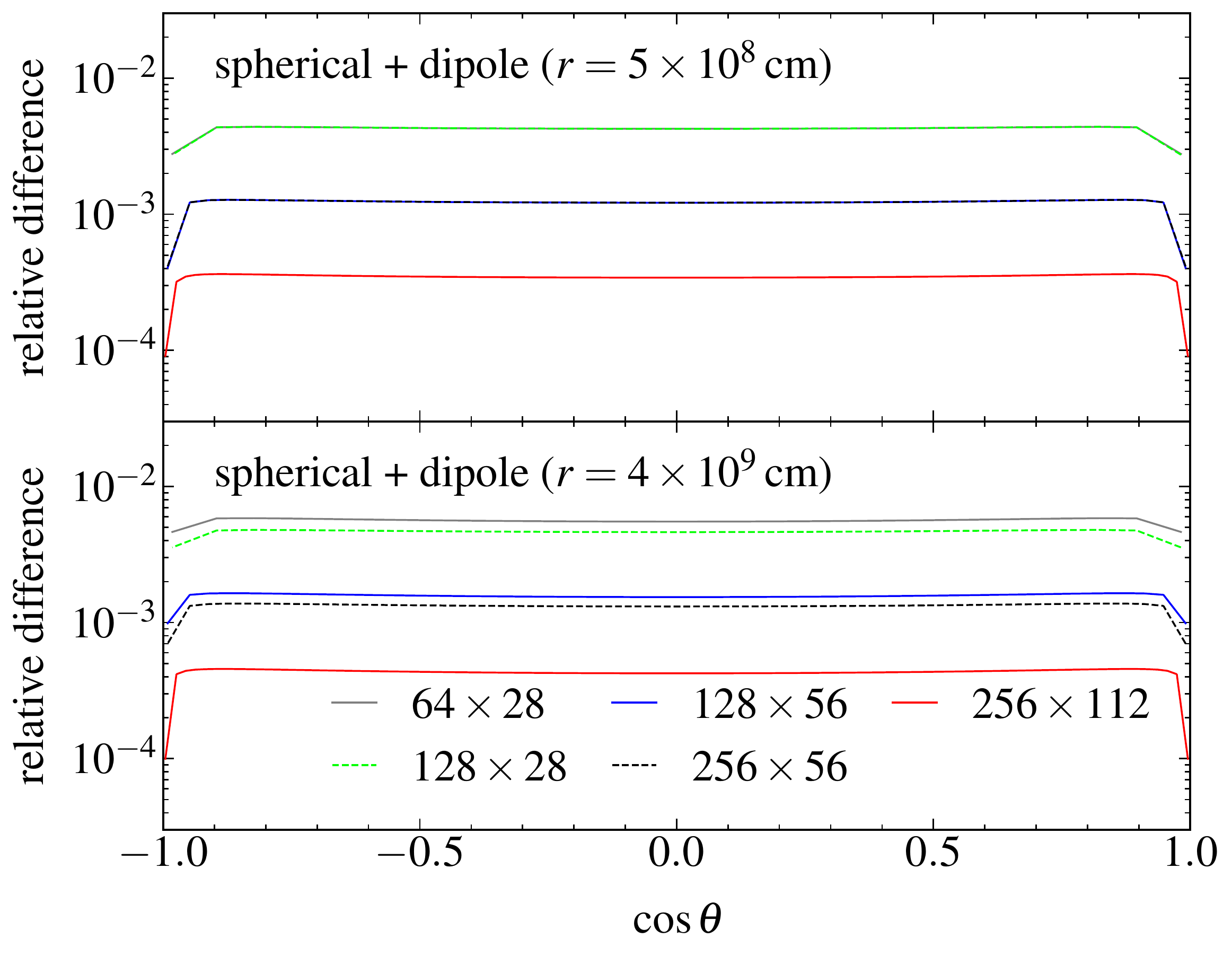}
\caption{Same as Figure~\ref{f:sg_moments_test}, but showing results as a function of
polar angle when including both spherical and dipolar components. Top and bottom panels
use radii in the interior ($r \leq R$) and exterior ($r>R$) parts of the solution, respectively.}
\label{f:sg_moments_angular}
\end{figure}

We test our implementation by comparing it against an analytic solution. Using
the density profile
\begin{equation}
\label{eq:density_profile_sg-test}
\rho(r,\theta) = 
\bigg\{
\begin{array}{lr}
\rho_0\left[ 1 + P_2(\cos\theta)\right] & r\leq R\\
\noalign{\smallskip}
0 & r > R
\end{array}
\end{equation}
with $\rho_0$ and $R$ constant, yields the following gravitational potential:
\begin{equation}
\label{eq:solution_sg-test}
\Phi = -2\pi G\rho_0\left[I_0(r) + I_2(r)P_2(\cos\theta)\right] 
\end{equation}
with
\begin{eqnarray}
I_0(r) & = &
\bigg\{\begin{array}{lr}
(2/3)(r^3 - r_{\rm in}^3)/r + (R^2 - r^2) & r \leq R\\
\noalign{\smallskip}
(2/3)(R^3 - r_{\rm in}^3)/r & r > R
\end{array}\\
\noalign{\smallskip}
\noalign{\smallskip}
I_2(r) & = &\frac{2}{5}
\bigg\{\begin{array}{lr} 
(r^5 - r_{\rm in}^5)/(5r^3) + r^2\ln (R/r) & r\leq R\\
\noalign{\smallskip}
(R^5 - r_{\rm in}^5)/(5r^3) & r > R
\end{array}
\end{eqnarray}
where $r_{\rm in}$ corresponds to the inner radial boundary.

For our tests, we use a computational domain extending from $r_{\rm in}=2\times 10^8$\,cm
to a radius $100$ times larger, covering all polar angles ($\theta \in [0,\pi]$). The density
normalization and transition radius are $\rho_0 = 10^7$\,g\,cm$^{-3}$ 
and $R = 2\times 10^8$\,cm, respectively. The multipole solver is run with $\ell_{\rm max}=12$.
The grid sizes used are $64\times 28$, $128\times 28$, $128\times 56$, $256\times 56$, 
and $256\times 112$ in radius and polar angle (logarithmic, and equispaced in $\cos\theta$, respectively). 
Figure~\ref{f:sg_moments_test} shows the fractional difference between the potential obtained from the 
multipole solver and that in the analytic solution. In all cases, increasing the spatial
resolution brings the numerical value closer to the analytic solution. Agreement is better
when restricting  the density profile to be spherical only 
($P_2 = 0$ in equations~\ref{eq:density_profile_sg-test} and \ref{eq:solution_sg-test})
than when using both spherical and dipole components. Note that agreement requires all
other moments (up to $\ell_{\rm max}=12$) to have vanishing amplitudes.

At our standard resolution ($128\times 56$), agreement is of the order of $10^{-3}$, with
a very weak radial dependence. The small bump at $r=R$ coincides with the transition from
interior to exterior solution in equation~\ref{eq:solution_sg-test}. 
Figure~\ref{f:sg_moments_angular} also shows that the fractional deviation is mostly
uniform with polar angle, both in the interior and exterior regions. The accuracy
of the $\ell=0$ moment is determined by the radial resolution only, while the 
angular resolution becomes more important when adding the dipole component, with
smaller changes introduced by the radial resolution.

\begin{figure}
\includegraphics*[width=\columnwidth]{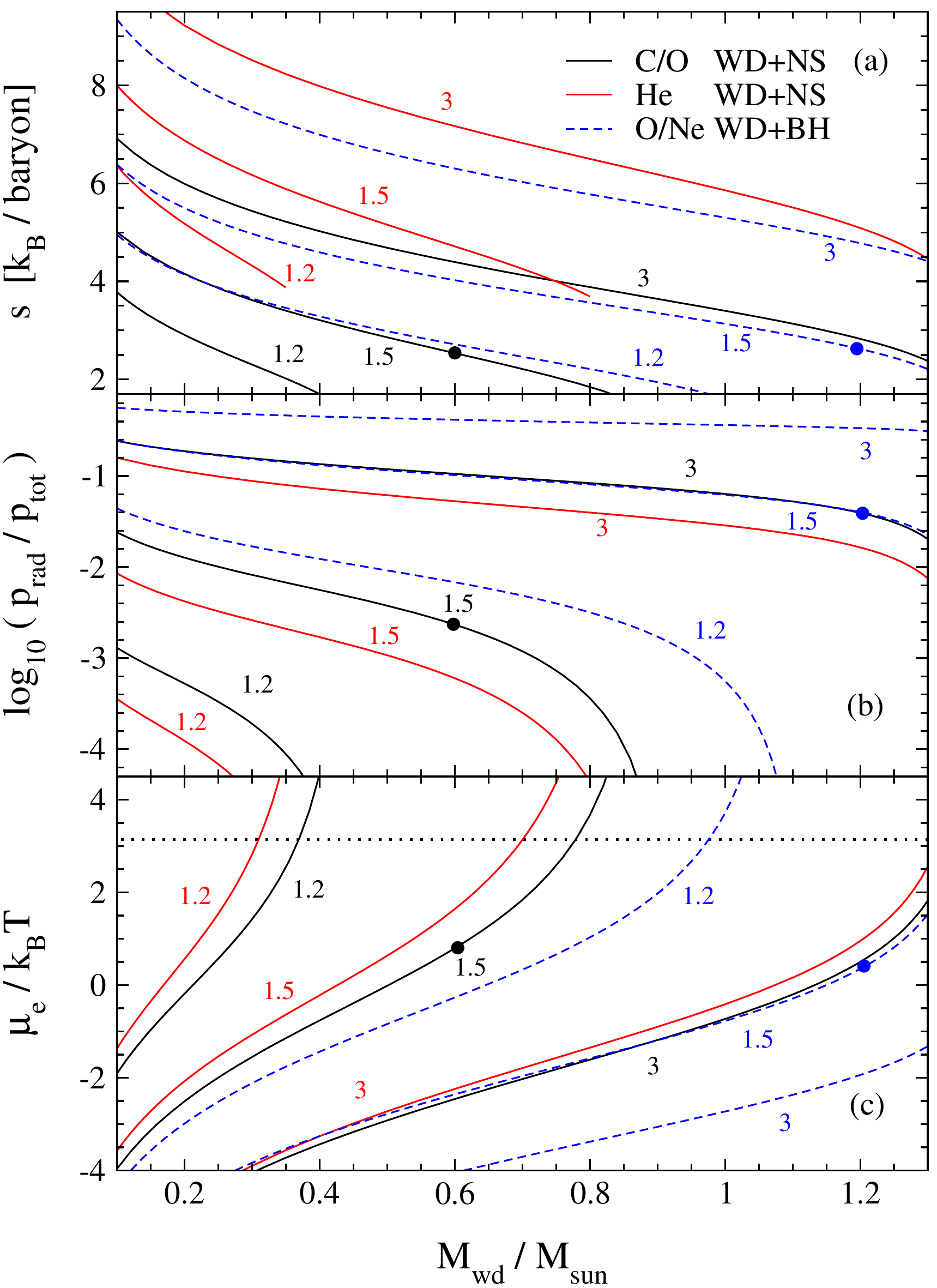}
\caption{Properties of equilibrium tori constructed with the Helmholtz equation of state
and point mass gravity, as a function of disk (WD) mass.
Shown are (a) entropy, (b) ratio of radiation to total pressure at density maximum,
and (c) degeneracy parameter at density maximum. Each curve is labeled by the value of
the torus distortion parameter (eq.~[\ref{eq:pp_general}]).
For reference, $d=\{1.2,1.5,3\}$ correspond to $e_{\rm int}/(GM_{\rm c}/R_0)\simeq\{5\%,10\%,20\%\}$
at pressure maximum, respectively, and to $H/R_0\sim \{0.4,0.6,1\}$, respectively.
Solid and dashed lines correspond to $M_{\rm c}=1.4M_\odot$ and $M_{\rm c}=5M_\sun$, respectively.
Colors label the composition: $\{X_{\rm C}=X_{\rm O}=0.5\}$ (black),
$X_{\rm He}=1$ (red), and $\{X_{\rm O}=0.6,X_{\rm Ne}=0.4\}$ (blue). The horizontal dotted
line in panel (c) marks the onset of degeneracy, $\mu_{\rm e} \ge \pi k_{\rm B} T$. The black and blue
dots correspond to our fiducial CO and ONe WDs (c.f. Table~\ref{t:models}).}
\label{f:pptorus_helmholtz}
\end{figure}

\begin{figure*}
\includegraphics*[width=\textwidth]{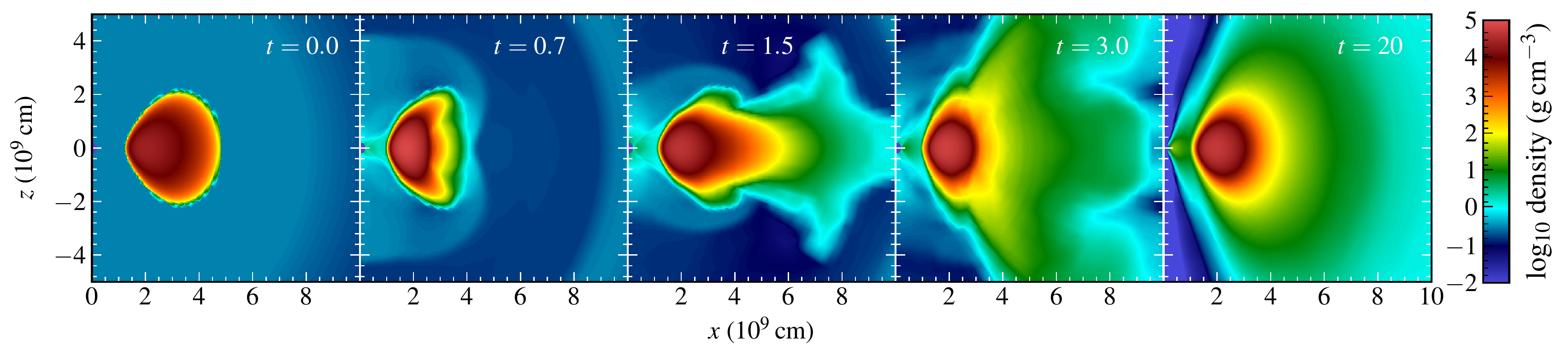}
\caption{Relaxation of the torus with self-gravity and no other source terms. 
Shown is the density of the fiducial model at various times in units of the
orbital time at the initial density peak (equation~\ref{eq:torb_def}).}
\label{f:relax_snapshots}
\end{figure*}

\begin{figure}
\includegraphics*[width=\columnwidth]{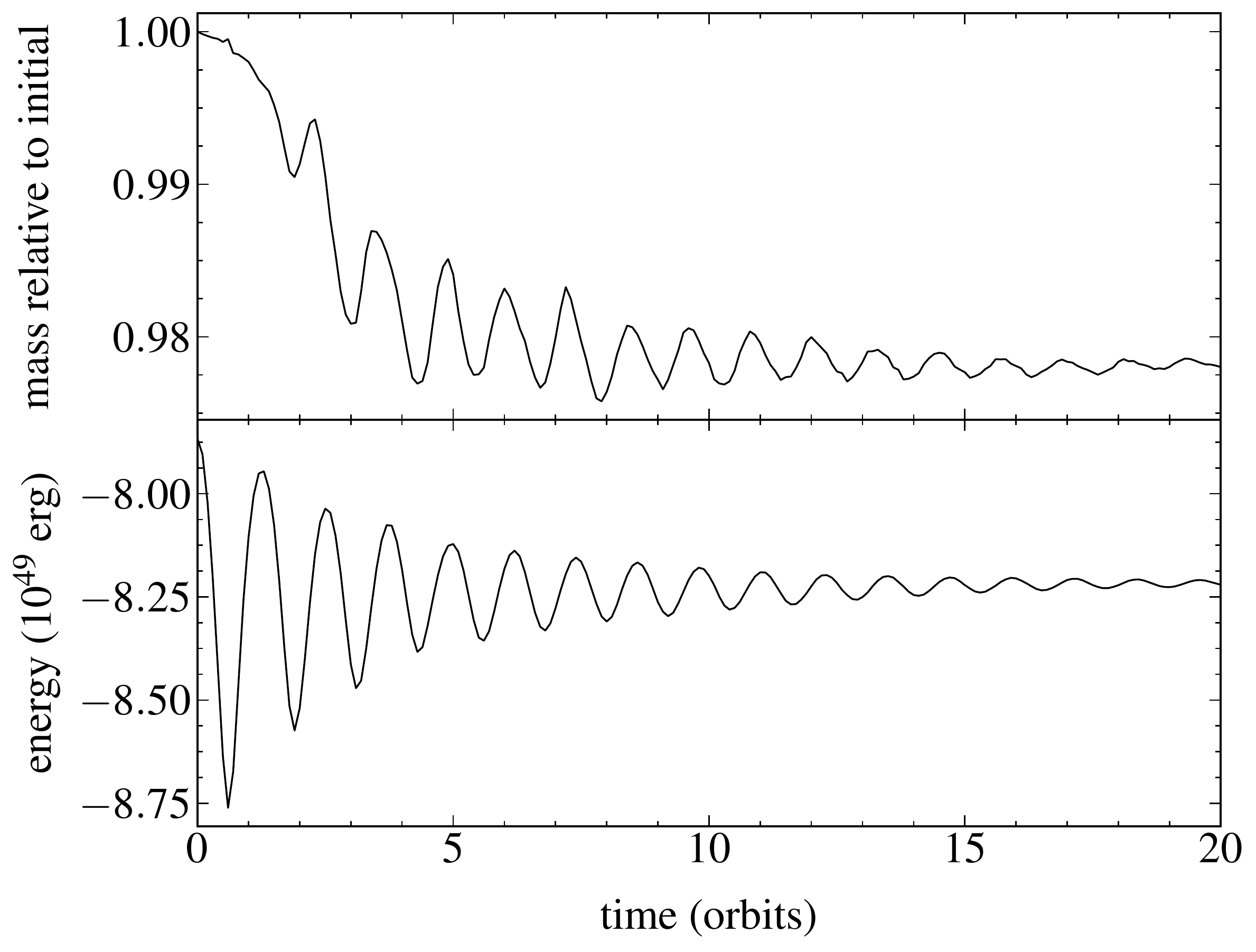}
\caption{Relaxation of the initial torus with self-gravity, starting from an initial
condition calculated with the gravity of the central object only. Shown are the torus 
mass relative to its initial value, restricted to densities higher than $10^{-3}$ times
the maximum (top), and the total torus energy (bottom). Parameters correspond to model {\tt CO+NS(l)}.}
\label{f:sg_e-mass}
\end{figure}

\begin{figure}
\includegraphics*[width=\columnwidth]{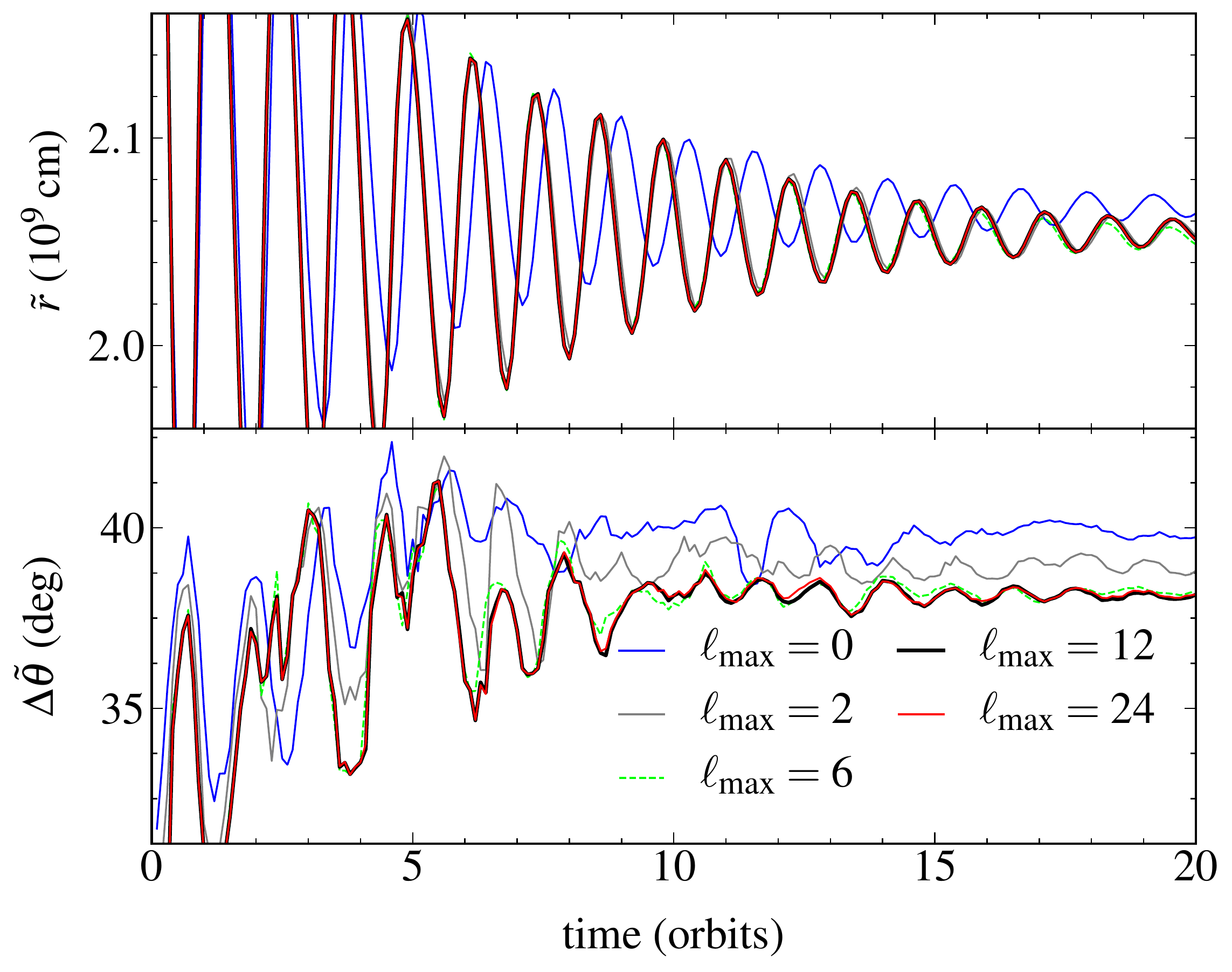}
\caption{Density weighted radius (top) and opening angle from
the equator (bottom) as a function of time, for different values of the
maximum Legendre index $\ell_{\rm max}$ in the 
multipole expansion (equation~\ref{eq:multipole_expansion}). Torus parameters
correspond to our fiducial model, relaxed in self-gravity without other source terms. 
See the text for the definition $\tilde r$ and $\Delta \tilde \theta$.}
\label{f:lmax_convergence}
\end{figure}

\section[]{Construction of initial torus}
\label{s:initial_condition_appendix}

\subsection{Equilibrium torus without self-gravity}
\label{s:initial_torus_ptmass}

As a starting point for the initial condition, we construct
an equilibrium torus with constant entropy $s$, angular momentum,
and composition $\mathbf{X}$. By solving the Bernoulli equation (e.g., \citealt{PP84}), 
we obtain an expression for the specific enthalpy of the torus as a
function of position, given a central mass $M_{\rm c}$, 
radius of density maximum in the torus $R_{\rm t}$ (set to the circularization
radius of the tidally-disrupted white dwarf; Paper I), and 
a dimensionless `distortion parameter' $d$ (which is a function of the torus 
entropy or $H/R$, see e.g., \citealt{stone1999})
\begin{equation}
\label{eq:pp_general}
w(r,\theta) = \frac{GM_{\rm c}}{R_{\rm t}}\left[
\frac{R_{\rm t}}{r}-\frac{1}{2}\frac{R_{\rm t}^2}{(r\sin\theta)^2} - \frac{1}{2d}\right],
\end{equation}
where $w = e_{\rm int} + p/\rho$ is the specific enthalpy of the fluid. 

For fixed entropy and composition, there is also a one-to-one thermodynamic mapping between the
enthalpy and density $w(\rho)|_{s,\mathbf{X}}$ from the equation of state. Inverting this
function in combination with equation~(\ref{eq:pp_general})
yields the mass of the torus after spatial integration. The limits of integration are obtained by 
setting the left-hand side to zero in equation~(\ref{eq:pp_general}). 
An iteration is required to find the distortion parameter $d$ that yields
the desired torus mass $M_{\rm t}$ [which amounts to solving for the function $d(s)$].
Note that the circularization radius (and thus $R_{\rm t}$) is a function of the 
torus mass and central object mass, hence $M_{\rm t}$ and $R_{\rm t}$ are not 
independent in this problem.

Figure~\ref{f:pptorus_helmholtz} shows properties of these tori as a function of 
mass, for three different compositions. Our fiducial C/O WD of mass $M_{\rm wd}=0.6M_\odot$
with distortion parameter $d=1.5$ has an entropy $3k_{\rm B}$ per baryon, has
very small degree of electron degeneracy, and a small contribution of radiation
to the total pressure. Helium WDs of the same mass and distortion parameter have
higher entropy, lower contribution of radiation pressure, and higher degeneracy.
Increasing the WD mass at constant distortion parameter decreases the entropy,
decreases the contribution of radiation pressure, and increases electron dengeneracy.
Our fiducial ONe WD has very similar entropy and degeneracy level compared to the
fiducial CO WD, but with a higher relative contribution from radiation to the total 
pressure.

\subsection{Relaxation with self-gravity}
\label{s:relax_sg}

We obtain a quiescent initial torus with self-gravity by evolving
the initial torus solution obtained without 
self-gravity (\S\ref{s:initial_torus_ptmass}) for $20$
orbits without any other source terms. The torus undergoes radial and vertical
oscillations as it adjusts to the new gravitational field, eventually reaching a new
equilibrium configuration. Figure~\ref{f:relax_snapshots} shows snapshots
in the evolution of the fiducial $0.6M_\odot$ CO WD, illustrating the amplitude
of these oscillations. The new radius of maximum density is $5\%$ smaller
than the original, and the maximum density is a factor $1.6$ higher.

The relaxation process results in the ejection of some mass to large radii.
Figure~\ref{f:sg_e-mass} shows that about $2\%$ of the mass contained
in material denser than $10^{-3}$ times the maximum density is redistributed
to larger radii. The frequency of the oscillations is approximately the
orbital frequency at the density maximum. Figure~\ref{f:sg_e-mass} also shows
the total energy of the torus, which undergoes oscillations of decreasing
amplitude, eventually settling into a new equilibrium value. By the
time we stop the relaxation, the amplitude of the oscillations has
decrease to about $1\%$.

We also use this torus relaxation process to find the optimal Legendre
index at which to truncate the multipole expansion (equation~\ref{eq:multipole_expansion}).
We perform the relaxation process over 20 orbits for our fiducial torus using
different values of the maximum Legendre index $\ell_{\rm max}$, with a
reference value of $12$ as recommended by \citet{MuellerSteinmetz1995}.
Convergence is quantified by the radial position of the torus and its opening
angle. We define these quantities as an average radius $\tilde r$, weighted by the
angle-averaged density profile, and the opening polar angle $\Delta \tilde\theta$
from the equator (at a constant radius $r=2\times 10^9$\,cm) at which the 
density drops to $10^{-3}$ of its maximum value in the simulation.
Figure~\ref{f:lmax_convergence} shows the evolution of these two metrics
as a function of time for different values of $\ell_{\rm max}$. The evolution is
essentially converged after $\ell_{\rm max}=6$, with $\ell_{\rm max}=12$ and
$24$ causing an indistinguishable change relative to each other. We therefore
adopt $\ell_{\rm max}=12$ for all of our simulations.

\begin{figure}
\includegraphics*[width=\columnwidth]{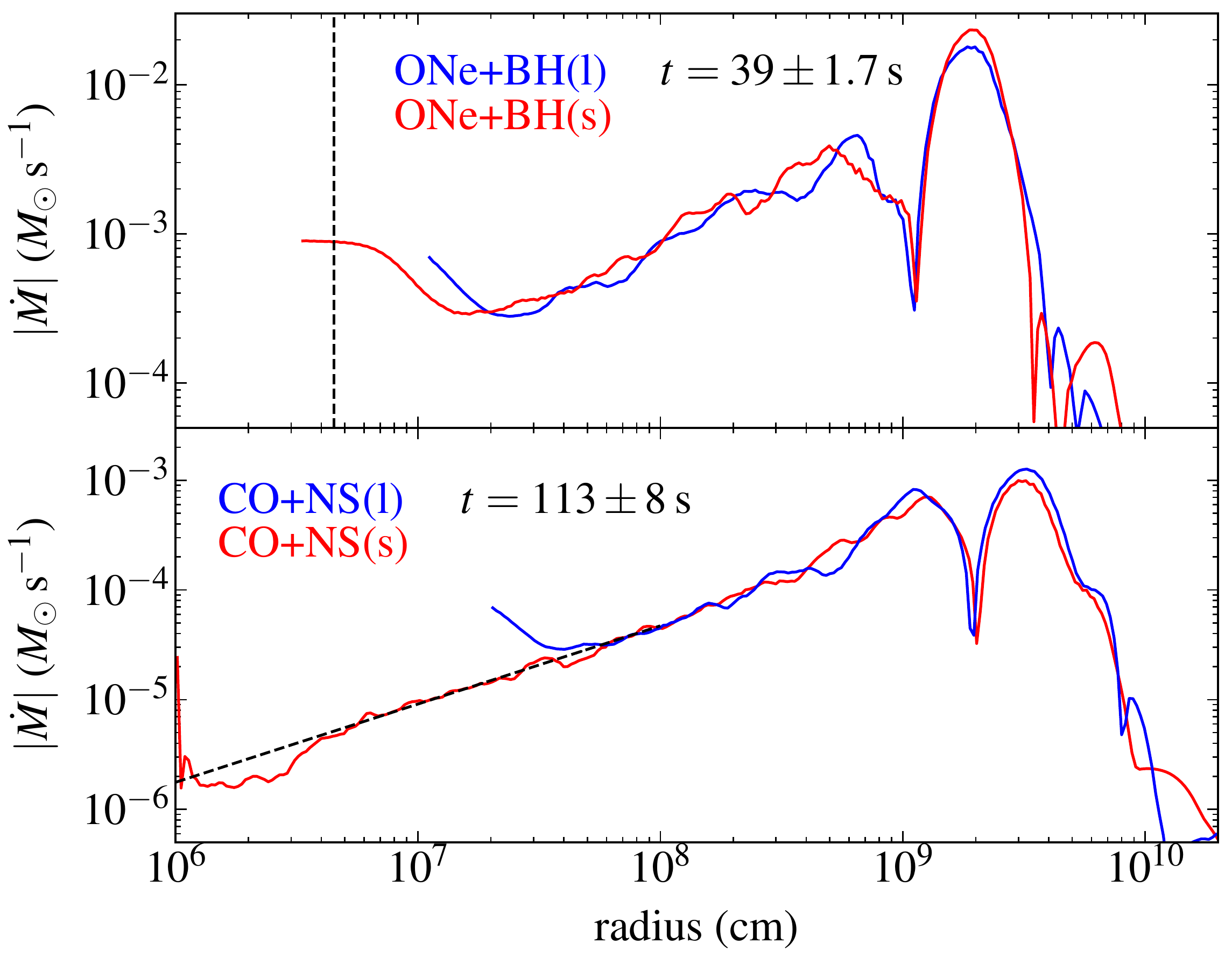}
\caption{Absolute value of the mass accretion rate as a function of radius,
averaged in time and within $30$\,deg of the equatorial plane, for
selected models, as labeled.
\emph{Top:} ONe WD model with non-spinning BH at the center (large boundary: blue, 
small boundary: red), the dashed line indicates the position of the ISCO radius. 
\emph{Bottom:} fiducial WD+NS model (large radius: blue, small radius: red). The
dashed line is the same power-law fit ($\propto r^{0.71}$) to the accretion rate 
as in Figure~\ref{f:accretion_exponents_rns}.}
\label{f:mdot_radius_appendix}
\end{figure}


\section[]{Long-term accretion rate at the central object}
\label{s:accretion_central}

Despite the fact that our fully global (``small boundary") models cannot be evolved 
for long enough to obtain a reliable long-term measure of the accretion rate at the central
object (to assess jet or fallback power, etc.), we can still estimate this quantity from our 
large boundary models by examining the radial behavior of the accretion rate, as shown in 
Figure~\ref{f:mdot_radius_appendix}.

A general feature of large boundary models is that the placement of an outflow
boundary condition at a radius when the flow is subsonic alters the behavior
compared to what it would be had that boundary not be there. Figure~\ref{f:mdot_radius_appendix}
shows that there is an increase in the accretion rate by a factor of a few
as this boundary is approached, deviating from the power-law behavior at
larger radii.

In the case of fully global BH models, the boundary is placed midway between
the ISCO and the horizon, where the flow is supersonic and thus causally
disconnected from that at larger radii. Figure~\ref{f:mdot_radius_appendix} shows
that the ISCO results in an increase in the accretion rate similar to that
obtained when placing the boundary further out, such that the value at the
ISCO is essentially the same as that measured at the innermost active radius
in the large boundary run. We therefore estimate the accretion rate onto
the black hole, for Figure~\ref{f:mdot_LB}, as simply the value of the accretion rate 
at the smallest radius in the large boundary run.

When a NS sits at the center, the discrepancy in the accretion rate at the smallest
radii between the large- and small boundary runs is more significant. Nevertheless,
we can estimate a reasonable value by measuring the accretion rate in the large
boundary model at a radius where the power-law behavior still holds, and then
extrapolating using the power-law exponent, as indicated in Figure~\ref{f:mdot_radius_appendix}.
In Figure~\ref{f:mdot_LB}, the accretion rate for the two NS models is obtained
by measuring it at $r=10^8$\,cm and applying a suppression factor $(10^{-2})^{0.7}$.
This assumes that the radial exponent of the accretion rate remains constant in
time, which is roughly satisfied.

\bibliographystyle{mn2e}
\bibliography{nudaf_mnras,apj-jour,rodrigo}

\begin{thebibliography}{}
{
\small
\bibitem[\protect\citeauthoryear{{Arnett}}{{Arnett}}{1979}]{arnett_1979}
{Arnett} W.~D.,  1979, ApJ, 230, L37

\bibitem[\protect\citeauthoryear{{Artemova}, {Bjoernsson} \&
  {Novikov}}{{Artemova} et~al.}{1996}]{artemova1996}
{Artemova} I.~V.,  {Bjoernsson} G.,    {Novikov} I.~D.,  1996, ApJ, 461, 565

\bibitem[\protect\citeauthoryear{{Bahramian} et~al.,}{{Bahramian}
  et~al.}{2017}]{bahramian_2017}
{Bahramian} A.,  et~al., 2017, MNRAS, 467, 2199

\bibitem[\protect\citeauthoryear{{Bobrick}, {Davies} \& {Church}}{{Bobrick}
  et~al.}{2017}]{bobrick_2017}
{Bobrick} A.,  {Davies} M.~B.,    {Church} R.~P.,  2017, MNRAS, 467, 3556

\bibitem[\protect\citeauthoryear{{Chen} et~al.,}{{Chen}
  et~al.}{2019}]{chen_2019}
{Chen} P.,  et~al., 2019, ApJL, submitted, arXiv:1905.02205

\bibitem[\protect\citeauthoryear{{Colella} \& {Woodward}}{{Colella} \&
  {Woodward}}{1984}]{colella84}
{Colella} P.,  {Woodward} P.~R.,  1984, JCP, 54, 174

\bibitem[\protect\citeauthoryear{{Dan}, {Rosswog}, {Br{\"u}ggen} \&
  {Podsiadlowski}}{{Dan} et~al.}{2014}]{dan_2014}
{Dan} M.,  {Rosswog} S.,  {Br{\"u}ggen} M.,    {Podsiadlowski} P.,  2014,
  MNRAS, 438, 14

\bibitem[\protect\citeauthoryear{{Datta} \& {Mukhopadhyay}}{{Datta} \&
  {Mukhopadhyay}}{2019}]{ranjan_2019}
{Datta} S.~R.,  {Mukhopadhyay} B.,  2019, MNRAS, 486, 1641

\bibitem[\protect\citeauthoryear{{Dexter} \& {Kasen}}{{Dexter} \&
  {Kasen}}{2013}]{dexter_2013}
{Dexter} J.,  {Kasen} D.,  2013, ApJ, 772, 30

\bibitem[\protect\citeauthoryear{{Drout} et~al.,}{{Drout}
  et~al.}{2014}]{drout_2014}
{Drout} M.~R.,  et~al., 2014, ApJ, 794, 23

\bibitem[\protect\citeauthoryear{Dubey, Antypas, Ganapathy, Reid, Riley,
  Sheeler, Siegel \& Weide}{Dubey et~al.}{2009}]{dubey2009}
Dubey A.,  Antypas K.,  Ganapathy M.~K.,  Reid L.~B.,  Riley K.,  Sheeler D.,
  Siegel A.,    Weide K.,  2009, J. Par. Comp., 35, 512

\bibitem[\protect\citeauthoryear{{Eggleton}}{{Eggleton}}{1983}]{eggleton1983}
{Eggleton} P.~P.,  1983, ApJ, 268, 368

\bibitem[\protect\citeauthoryear{{Fern{\'a}ndez}, {Kasen}, {Metzger} \&
  {Quataert}}{{Fern{\'a}ndez} et~al.}{2015}]{FKMQ14}
{Fern{\'a}ndez} R.,  {Kasen} D.,  {Metzger} B.~D.,    {Quataert} E.,  2015,
  MNRAS, 446, 750

\bibitem[\protect\citeauthoryear{{Fern{\'a}ndez} \& {Metzger}}{{Fern{\'a}ndez}
  \& {Metzger}}{2013a}]{FM13}
{Fern{\'a}ndez} R.,  {Metzger} B.~D.,  2013a, MNRAS, 435, 502

\bibitem[\protect\citeauthoryear{{Fern{\'a}ndez} \& {Metzger}}{{Fern{\'a}ndez}
  \& {Metzger}}{2013b}]{FM12}
{Fern{\'a}ndez} R.,  {Metzger} B.~D.,  2013b, ApJ, 763, 108

\bibitem[\protect\citeauthoryear{{Fern{\'a}ndez} \& {Metzger}}{{Fern{\'a}ndez}
  \& {Metzger}}{2016}]{FM16}
{Fern{\'a}ndez} R.,  {Metzger} B.~D.,  2016, ARNPS, 66, 23

\bibitem[\protect\citeauthoryear{{Fern{\'a}ndez}, {Tchekhovskoy}, {Quataert},
  {Foucart} \& {Kasen}}{{Fern{\'a}ndez} et~al.}{2019}]{fernandez2019}
{Fern{\'a}ndez} R.,  {Tchekhovskoy} A.,  {Quataert} E.,  {Foucart} F.,
  {Kasen} D.,  2019, MNRAS, 482, 3373

\bibitem[\protect\citeauthoryear{{Foley} et~al.,}{{Foley}
  et~al.}{2013}]{foley_2013}
{Foley} R.~J.,  et~al., 2013, ApJ, 767, 57

\bibitem[\protect\citeauthoryear{{Fryer}, {Woosley}, {Herant} \&
  {Davies}}{{Fryer} et~al.}{1999}]{fryer1999}
{Fryer} C.~L.,  {Woosley} S.~E.,  {Herant} M.,    {Davies} M.~B.,  1999, ApJ,
  520, 650

\bibitem[\protect\citeauthoryear{{Fryxell}, {Olson}, {Ricker}, {Timmes},
  {Zingale}, {Lamb}, {MacNeice}, {Rosner}, {Truran} \& {Tufo}}{{Fryxell}
  et~al.}{2000}]{fryxell00}
{Fryxell} B.,  {Olson} K.,  {Ricker} P.,  {Timmes} F.~X.,  {Zingale} M.,
  {Lamb} D.~Q.,  {MacNeice} P.,  {Rosner} R.,  {Truran} J.~W.,    {Tufo} H.,
  2000, ApJS, 131, 273

\bibitem[\protect\citeauthoryear{Hunter}{Hunter}{2007}]{hunter2007}
Hunter J.~D.,  2007, Computing In Science \& Engineering, 9, 90

\bibitem[\protect\citeauthoryear{Itoh, Hayashi, Nishikawa \& Kohyama}{Itoh
  et~al.}{1996}]{itoh1996}
Itoh N.,  Hayashi H.,  Nishikawa A.,    Kohyama Y.,  1996, ApJS, 102, 411

\bibitem[\protect\citeauthoryear{{Kasliwal} et~al.,}{{Kasliwal}
  et~al.}{2012}]{kasliwal_2012}
{Kasliwal} M.~M.,  et~al., 2012, ApJ, 755, 161

\bibitem[\protect\citeauthoryear{{Kawana}, {Tanikawa} \& {Yoshida}}{{Kawana}
  et~al.}{2018}]{kawana_2018}
{Kawana} K.,  {Tanikawa} A.,    {Yoshida} N.,  2018, MNRAS, 477, 3449

\bibitem[\protect\citeauthoryear{{Kim}, {Kalogera}, {Lorimer} \& {White}}{{Kim}
  et~al.}{2004}]{kim_2004}
{Kim} C.,  {Kalogera} V.,  {Lorimer} D.~R.,    {White} T.,  2004, ApJ, 616,
  1109

\bibitem[\protect\citeauthoryear{{King}, {Olsson} \& {Davies}}{{King}
  et~al.}{2007}]{King+07}
{King} A.,  {Olsson} E.,    {Davies} M.~B.,  2007, MNRAS, 374, L34

\bibitem[\protect\citeauthoryear{{Konacki} \& {Wolszczan}}{{Konacki} \&
  {Wolszczan}}{2003}]{Konacki&Wolszczan03}
{Konacki} M.,  {Wolszczan} A.,  2003, ApJ, 591, L147

\bibitem[\protect\citeauthoryear{{Kulkarni}}{{Kulkarni}}{2012}]{kulkarni_2012}
{Kulkarni} S.~R.,  2012, in {Griffin} E.,  {Hanisch} R.,   {Seaman} R.,  eds,
  New Horizons in Time Domain Astronomy Vol.~285 of IAU Symposium, {Cosmic
  Explosions (Optical)}.
pp 55--61

\bibitem[\protect\citeauthoryear{{Lorimer}}{{Lorimer}}{2008}]{lorimer_2008}
{Lorimer} D.~R.,  2008, Living Reviews in Relativity, 11, 8

\bibitem[\protect\citeauthoryear{{Luminet} \& {Pichon}}{{Luminet} \&
  {Pichon}}{1989}]{luminet_1989}
{Luminet} J.~P.,  {Pichon} B.,  1989, A\&A, 209, 103

\bibitem[\protect\citeauthoryear{{MacLeod}, {Guillochon}, {Ramirez-Ruiz},
  {Kasen} \& {Rosswog}}{{MacLeod} et~al.}{2016}]{macleod_2016}
{MacLeod} M.,  {Guillochon} J.,  {Ramirez-Ruiz} E.,  {Kasen} D.,    {Rosswog}
  S.,  2016, ApJ, 819, 3

\bibitem[\protect\citeauthoryear{{Margalit} \& {Metzger}}{{Margalit} \&
  {Metzger}}{2016}]{margalit_2016}
{Margalit} B.,  {Metzger} B.~D.,  2016, MNRAS, 461, 1154

\bibitem[\protect\citeauthoryear{{Margalit} \& {Metzger}}{{Margalit} \&
  {Metzger}}{2017}]{margalit_2017}
{Margalit} B.,  {Metzger} B.~D.,  2017, MNRAS, 465, 2790

\bibitem[\protect\citeauthoryear{{Metzger}}{{Metzger}}{2012}]{M12}
{Metzger} B.~D.,  2012, MNRAS, 419, 827

\bibitem[\protect\citeauthoryear{{M{\"u}ller} \& {Steinmetz}}{{M{\"u}ller} \&
  {Steinmetz}}{1995}]{MuellerSteinmetz1995}
{M{\"u}ller} E.,  {Steinmetz} M.,  1995, Computer Physics Communications, 89,
  45

\bibitem[\protect\citeauthoryear{{Nauenberg}}{{Nauenberg}}{1972}]{nauenberg1972}
{Nauenberg} M.,  1972, ApJ, 175, 417

\bibitem[\protect\citeauthoryear{{O'Shaughnessy} \& {Kim}}{{O'Shaughnessy} \&
  {Kim}}{2010}]{OShaughnessy&Kim10}
{O'Shaughnessy} R.,  {Kim} C.,  2010, ApJ, 715, 230

\bibitem[\protect\citeauthoryear{{Papaloizou} \& {Pringle}}{{Papaloizou} \&
  {Pringle}}{1984}]{PP84}
{Papaloizou} J.~C.~B.,  {Pringle} J.~E.,  1984, MNRAS, 208, 721

\bibitem[\protect\citeauthoryear{{Paschalidis}, {Liu}, {Etienne} \&
  {Shapiro}}{{Paschalidis} et~al.}{2011}]{Paschalidis+11}
{Paschalidis} V.,  {Liu} Y.~T.,  {Etienne} Z.,    {Shapiro} S.~L.,  2011, PRD,
  84, 104032

\bibitem[\protect\citeauthoryear{Perets et~al.,}{Perets
  et~al.}{2010}]{Perets+10}
Perets H.~B.,  et~al., 2010, Nature, 465, 322

\bibitem[\protect\citeauthoryear{{Phinney} \& {Hansen}}{{Phinney} \&
  {Hansen}}{1993}]{Phinney&Hansen93}
{Phinney} E.~S.,  {Hansen} B.~M.~S.,  1993, in {Phillips} J.~A.,  {Thorsett}
  S.~E.,   {Kulkarni} S.~R.,  eds, Planets Around Pulsars Vol.~36 of
  Astronomical Society of the Pacific Conference Series, {The pulsar planet
  production process.}.
pp 371--390

\bibitem[\protect\citeauthoryear{{Podsiadlowski}}{{Podsiadlowski}}{1993}]{Podsiadlowski93}
{Podsiadlowski} P.,  1993, in {Phillips} J.~A.,  {Thorsett} S.~E.,   {Kulkarni}
  S.~R.,  eds, Planets Around Pulsars Vol.~36 of Astronomical Society of the
  Pacific Conference Series, {Planet formation scenarios.}.
pp 149--165

\bibitem[\protect\citeauthoryear{{Rest} et~al.,}{{Rest}
  et~al.}{2018}]{rest_2018}
{Rest} A.,  et~al., 2018, Nature Astronomy, 2, 307

\bibitem[\protect\citeauthoryear{{Rosswog}, {Ramirez-Ruiz} \& {Hix}}{{Rosswog}
  et~al.}{2009}]{rosswog_2009}
{Rosswog} S.,  {Ramirez-Ruiz} E.,    {Hix} W.~R.,  2009, ApJ, 695, 404

\bibitem[\protect\citeauthoryear{{Shakura} \& {Sunyaev}}{{Shakura} \&
  {Sunyaev}}{1973}]{shakura1973}
{Shakura} N.~I.,  {Sunyaev} R.~A.,  1973, A\&A, 24, 337

\bibitem[\protect\citeauthoryear{{Stone}, {Pringle} \& {Begelman}}{{Stone}
  et~al.}{1999}]{stone1999}
{Stone} J.~M.,  {Pringle} J.~E.,    {Begelman} M.~C.,  1999, MNRAS, 310, 1002

\bibitem[\protect\citeauthoryear{{Timmes}}{{Timmes}}{1999}]{timmes1999}
{Timmes} F.~X.,  1999, ApJS, 124, 241

\bibitem[\protect\citeauthoryear{{Timmes} \& {Swesty}}{{Timmes} \&
  {Swesty}}{2000}]{timmes2000}
{Timmes} F.~X.,  {Swesty} F.~D.,  2000, ApJS, 126, 501

\bibitem[\protect\citeauthoryear{{Toonen}, {Perets}, {Igoshev}, {Michaely} \&
  {Zenati}}{{Toonen} et~al.}{2018}]{toonen_2018}
{Toonen} S.,  {Perets} H.~B.,  {Igoshev} A.~P.,  {Michaely} E.,    {Zenati} Y.,
   2018, A\&A, 619, A53

\bibitem[\protect\citeauthoryear{{van Kerkwijk}, {Bassa}, {Jacoby} \&
  {Jonker}}{{van Kerkwijk} et~al.}{2005}]{vankerkwijk_2005}
{van Kerkwijk} M.~H.,  {Bassa} C.~G.,  {Jacoby} B.~A.,    {Jonker} P.~G.,
  2005, in {Rasio} F.~A.,  {Stairs} I.~H.,  eds, Binary Radio Pulsars Vol.~328
  of Astronomical Society of the Pacific Conference Series, {Optical Studies of
  Companions to Millisecond Pulsars}.
p.~357

\bibitem[\protect\citeauthoryear{{Weaver}, {Zimmerman} \& {Woosley}}{{Weaver}
  et~al.}{1978}]{weaver1978}
{Weaver} T.~A.,  {Zimmerman} G.~B.,    {Woosley} S.~E.,  1978, ApJ, 225, 1021

\bibitem[\protect\citeauthoryear{{Wolszczan}}{{Wolszczan}}{1994}]{Wolszczan94}
{Wolszczan} A.,  1994, Science, 264, 538

\bibitem[\protect\citeauthoryear{{Wolszczan} \& {Frail}}{{Wolszczan} \&
  {Frail}}{1992}]{Wolszczan&Frail92}
{Wolszczan} A.,  {Frail} D.~A.,  1992, Nature, 355, 145

\bibitem[\protect\citeauthoryear{{Xue} et~al.,}{{Xue}  et~al.}{2019}]{xue_2019}
{Xue} Y.~Q.,  et~al., 2019, Nature, 568, 198

\bibitem[\protect\citeauthoryear{{Yuan} \& {Narayan}}{{Yuan} \&
  {Narayan}}{2014}]{Yuan&Narayan14}
{Yuan} F.,  {Narayan} R.,  2014, ARA\&A, 52, 529

\bibitem[\protect\citeauthoryear{{Yuan}, {Wu} \& {Bu}}{{Yuan}
  et~al.}{2012}]{yuan2012}
{Yuan} F.,  {Wu} M.,    {Bu} D.,  2012, ApJ, 761, 129

\bibitem[\protect\citeauthoryear{{Zenati}, {Perets} \& {Toonen}}{{Zenati}
  et~al.}{2019}]{zenati_2019}
{Zenati} Y.,  {Perets} H.~B.,    {Toonen} S.,  2019, MNRAS, 486, 1805
}
\end{thebibliography}


\label{lastpage}
\end{document}